\documentclass[aps,12pt]{revtex4-2}

\usepackage{amsmath,bm}
\usepackage{amssymb}
\usepackage{mathrsfs}
\usepackage{amsfonts}
\usepackage{graphicx}
\usepackage{longtable}
\usepackage{booktabs}
\usepackage{siunitx}
\usepackage[version=3]{mhchem}
\usepackage{xcolor}
\usepackage{listings}
\definecolor{lstcomment}{gray}{0.40}
\definecolor{lstkey}{rgb}{0.10,0.25,0.60}
\definecolor{lstbg}{gray}{0.97}
\lstdefinestyle{cbo}{
  basicstyle=\ttfamily\scriptsize,
  keywordstyle=\color{lstkey}\bfseries,
  commentstyle=\color{lstcomment}\itshape,
  stringstyle=\color{black},
  backgroundcolor=\color{lstbg},
  breaklines=true, showstringspaces=false, columns=fullflexible,
  frame=single, rulecolor=\color{lstcomment}, framesep=3pt,
  xleftmargin=4pt, xrightmargin=2pt, aboveskip=6pt, belowskip=4pt,
}
\lstdefinelanguage{yamlcbo}{
  keywords={task,method,output,molecule,cavity,lambda,omega,qpos,psi4_options,
            basis,dse,qopt,nstates,hessian_type,raman,nrt,verbosity,
            polarizability_type,excitation_level,optimization_options},
  sensitive=true, comment=[l]{\#}, morestring=[b]",
}
\usepackage[acronym]{glossaries}

\DeclareSIUnit\au{\text {au}}

\newcommand{\ke}[1]{\vert #1 \rangle}

\newcommand{\mel}[3]{\langle #1 \vert #2 \vert #3 \rangle}

\newacronym{cboa}{CBOA}{cavity Born-Oppenheimer approximation}
\newacronym{bo}{BOA}{Born-Oppenheimer approximation}
\newacronym{cbohf}{CBO-HF}{cavity Born-Oppenheimer Hartree-Fock}
\newacronym{scf}{SCF}{self-consistent field}
\newacronym{dse}{DSE}{dipole self-energy}
\newacronym{pes}{PES}{potential energy surface}
\newacronym{cpes}{cPES}{cavity potential energy surface}
\newacronym{vsc}{VSC}{vibrational strong coupling}
\newacronym{esc}{ESC}{electronic strong coupling}
\newacronym{ejcm}{EJCM}{extended molecular Jaynes-Cummings model}

\newacronym{sd}{SD}{steepest descent}
\newacronym{nr}{NR}{Newton-Raphson}
\newacronym{bf}{BFGS}{Broyden-Fletcher-Goldfarb-Shanno}
\newacronym{diis}{DIIS}{direct inversion in the iterative subspace}

\newacronym{ci}{CI}{configuration interaction}
\newacronym{qedhf}{QED-HF}{quantum-electrodynamical Hartree-Fock}
\newacronym{cboci}{CBO-CI}{cavity Born-Oppenheimer configuration interaction}
\newacronym{qedci}{QED-CI}{quantum-elec\-tro\-dy\-nam\-i\-cal configuration interaction}
\newacronym{qedfci}{QED-FCI}{quantum-electrodynamical full configuration interaction}
\newacronym{cbofci}{CBO-FCI}{cavity Born-Oppenheimer full configuration interaction}
\newacronym{ir}{IR}{infrared}
\newacronym{fwhm}{FWHM}{full width at half maximum}
\newacronym{dna}{DNA}{deoxyribonucleic acid}
\newacronym{lp}{LP}{lower polariton}
\newacronym{up}{UP}{upper polariton}
\newacronym{qed}{QED}{quantum electrodynamics}

\newacronym{cbodft}{CBO-DFT}{cavity Born-Oppenheimer density functional theory}

\newacronym{cphf}{CPHF}{coupled-perturbed Hartree-Fock}

\newacronym{api}{API}{application programming interface}
\newacronym{gga}{GGA}{generalized gradient approximation}
\newacronym{rms}{RMS}{root mean square}
\newacronym{jk}{JK}{Coulomb and exchange matrix}
\newacronym{blas}{BLAS}{basic linear algebra subprograms}

\begin{document}

\title{Enabling Polaritonic Electronic Structure Theory for Psi4:\\ The \texttt{psi4-cbo} package}

\author{Thomas Schnappinger}
\email{thomas.schnappinger@fysik.su.se}
\affiliation{Department of Physics, Stockholm University, AlbaNova University Center, SE-106 91 Stockholm, Sweden}

\author{Markus Kowalewski}
\affiliation{Department of Physics, Stockholm University, AlbaNova University Center, SE-106 91 Stockholm, Sweden}

\date{\today}%

\begin{abstract}
We present \texttt{psi4-cbo}, an open-source extension of the Psi4 program for simulations within the \gls{cboa}.
By treating the cavity photon mode on the same footing  as the nuclear coordinates, \texttt{psi4-cbo} incorporates light-matter coupling into standard electronic structure workflows.
These workflows include self-consistent field calculations, response theory, analytic gradients, joint Hessians, and geometry optimization.
The package is designed for reproducibility and is documented with working examples for its main workflows.
To evaluate the approximations inherent to the \gls{cboa} framework, we use an exact \gls{qedfci} reference.
Resolving the correlation energy into electron-electron and electron-photon channels enables us to precisely determine the point at which the \gls{cboa} becomes quantitatively sufficient.
Additionally, we highlight that, since self-consistent treatments dress the cavity modes, satisfying experimental resonance conditions is method-dependent.
As a practical application, we investigate the chromophore acrolein under vibrational strong coupling and detail how to locate resonances on the dressed photon modes rather than the bare molecular bands.
\end{abstract}

\maketitle
\glsresetall
\section{Introduction}
Tightly confined photon modes in optical or plasmonic cavities enable strong light-matter coupling.
When applied to molecules, electronic states or molecular vibrations hybridize with the photon modes and form polariton states~\cite{Ebbesen2023-fd, Fregoni2022-op}.
This phenomenon has been reported to alter reaction landscapes~\cite{Thomas2019-ve}, modify enzyme activity~\cite{Vergauwe2019-wt}, redirect stereoselectivity~\cite{Sau2021-vb}, and influence charge and ionic transport~\cite{Kumar2024-th,Thomas2025-we}.
There is increasing evidence that these observed effects are collective in nature, meaning that, rather than acting on isolated molecules, the cavity influences how molecules organize.
For example, \gls{vsc} can alter the supramolecular assembly of conjugated polymers~\cite{Joseph2021-sa}, poly(phenylene ethynylene)s~\cite{Sandeep2022-sa}, and porphyrins~\cite{Joseph2024-sp}, as well as drive the coassembly of DNA origami~\cite{Zhong2023-do}.
Similarly, \gls{esc} has been shown to modify ground-state intermolecular interactions in self-assembled chlorins~\cite{Biswas2025-in}.
Because these phenomena involve ensembles that share a single cavity mode, they place strict demands on theoretical descriptions.

Although \gls{vsc} and \gls{esc} offer promising new tools for chemistry, the underlying mechanisms are not yet fully understood, and experimental clarification of the underlying mechanisms remains challenging~\cite{Imperatore2021-rv,Wiesehan2021-yn}.
To explain these experimental observations, ab initio theory may be required that goes beyond minimal or phenomenological models and includes the light-matter coupling at the level of electronic structures.
The Pauli-Fierz Hamiltonian and its \gls{cboa} or \gls{qed} reductions provide this foundation~\cite{flick2017atoms,flick2017cavity,Tokatly2013-ho,Ruggenthaler2014-it}.
However, applying these methods to correlated wave functions, analytic derivatives, and large molecular ensembles remains challenging~\cite{Sidler2022-cg,Foley2023-pq,Sanchez-Barquilla2022-dq}.
Recent progress includes detailed analyses of the \gls{cboa}~\cite{Fiechter2024-vw,Fischer2023-ve}, its extension to perturbative electron-photon correlation~\cite{Fischer2024-mk}, and studies on how cavity-modified reactivity holds up under fully quantum treatments~\cite{Fischer2022-td,Ying2023-es,Ke2024-br}.

Efficient software packages have established a strong foundation for polaritonic electronic structure theory.
In particular, the Octopus code~\cite{Marques2003-gd,Castro2006-am,Andrade2015-zq,Tancogne-Dejean2020-ev} provides extensive capabilities with a focus on quantum electrodynamics and solid-state density functional theory, while the \texttt{$e^{T}$}~program package~\cite{Folkestad2020-cw,Folkestad2026-ui} offers comprehensive implementations for fully quantized photon methods~\cite{Haugland2020-xh,Angelico2023-jh} and \gls{cboa}~\cite{flick2017cavity,Schnappinger2023-cbohf} for molecular systems.
In this paper, we present a complementary tool, \texttt{psi4-cbo}, an open-source package built on Psi4~\cite{Smith2020-kq} and Psi4NumPy~\cite{Smith2018-tu}.
It is built to be easily extensible and allows for experimentation and analysis
of quantities, such as electronic densities and molecular properties that are typically not exposed in other program packages.
To achieve rapid prototyping of new theoretical features, \texttt{psi4-cbo} is written in Python.
Additionally, the \texttt{psi4-cbo} package specifically features the joint nuclear-photonic Hessian along with its associated analytical framework.
After developing and validating the mean-field ansatz, analytic gradients, and vibro-polaritonic spectra in previous studies~\cite{Schnappinger2023-cbohf,Schnappinger2023-wp,Schnappinger2024-vt,Schnappinger2025-fx,Schnappinger2024-rz}, the methods presented in these studies are available in \texttt{psi4-cbo}.
In addition, the program package also introduces post-mean field methods, diagnostic tools, and a comprehensive documentation.

The paper is organized as follows: Section~\ref{sec:theory} defines the underlying Pauli-Fierz Hamiltonian and its mean-field approaches \gls{cbohf} and \gls{qedhf}.
Section~\ref{sec:correlation} builds on this foundation by introducing both \gls{cboci} and \gls{qedci}, discussing their main differences, and building a suite of quantitative correlation diagnostics for mean-field solutions.
Section~\ref{sec:capabilities} outlines the complete technical capabilities of \texttt{psi4-cbo}, and Section~\ref{sec:usage} shows worked examples of the main workflows with the input needed to reproduce the results.
The two applications are as follows.
Section~\ref{sec:classical_photon} discusses the different sources of correlation, systematic diagnostics across different systems, and coupling strengths.
Moreover, the parameter regime where the \gls{cboa} remains reliable is discussed and where the quantized photon is required.
Section~\ref{sec:showcase_acrolein} applies \texttt{psi4-cbo} to acrolein, a prototypical chromophore used to quantitatively validate photon-mode dressing, and compares \gls{cbohf} against \gls{cbodft} for vibro-polaritonic \gls{ir} spectra.
Section~\ref{sec:limitations} then discusses its limitations, and Section~\ref{sec:availability} states how the package is obtained, installed, and tested.
The paper concludes with a summary of future developments.

\section{Theory}
\label{sec:theory}

\subsection{The Pauli-Fierz Hamiltonian and the cavity Born-Oppenheimer approximation}

We consider $N_{\rm mod}$ effective cavity modes, each of which is characterized by a frequency, $\omega_m$, and a coupling vector, $\boldsymbol\lambda_m$ along the polarization axis, $\hat{\mathbf e}_m$.
The coupling vector $\boldsymbol\lambda_m$ is defined as
\begin{equation}
\label{eq:lam}
\boldsymbol\lambda_m = \hat{\mathbf e}_m\,\lambda_m = \hat{\mathbf e}_m\sqrt{\frac{4\pi}{V_m}} ,
\end{equation}
with its magnitude, the coupling strength $\lambda_m$, is set by the effective mode volume $V_m$.
Atomic units are used throughout, unless otherwise stated.
Within the standard Born-Oppenheimer and long-wavelength approximation in the length gauge, the Pauli-Fierz Hamiltonian~\cite{flick2017atoms,flick2017cavity} can be written either in terms of Fock states or photon displacement coordinates.
In both representations, the nuclei are treated as classical point charges fixed at a fixed geometry $\mathbf R$.
The final Hamiltonian then consists of the electronic Hamiltonian $\hat H_{\rm el}(\mathbf R)$, the Hamiltonian of the cavity mode, and a light-matter coupling term.
The Pauli-Fierz Hamiltonian in the Fock state representation then reads:
\begin{equation}
\label{eq:pf_fock}
\hat H_{\rm PF} = \hat H_{\rm el}(\mathbf R) + \sum_{m=1}^{N_{\rm mod}}\Big[
  \omega_m\big(\hat b_m^\dagger\hat b_m+\tfrac12\big)
  - \sqrt{\tfrac{\omega_m}{2}}\,(\hat b_m+\hat b_m^\dagger)\,\hat d_m
  + \tfrac12\hat d_m^2\Big] ,
\end{equation}
where $\hat b_m^\dagger$ and $\hat b_m$ are the photon creation and annihilation operators of mode $m$, and $\omega_m$ is the frequency of the cavity mode.
$\hat d_m$ is the dressed dipole operator of mode $m$, which is given by the scalar product of the total (electronic and nuclear) molecular dipole operator $\hat{\boldsymbol\mu}$ and the coupling vector of Eq.~\eqref{eq:lam}:
\begin{equation}
\hat d_m \equiv \boldsymbol\lambda_m\cdot\hat{\boldsymbol\mu} = \lambda_m\,\hat{\mathbf e}_m\cdot\hat{\boldsymbol\mu}\,.
\end{equation}
The dressed dipole operator is used both in the linear light-matter coupling term and in the \gls{dse} term $\tfrac12\hat d_m^2$.
Alternatively, the Pauli-Fierz Hamiltonian can be expressed in terms of photon displacement coordinates $q_m$,
\begin{equation}
\label{eq:pf_cbo}
\hat H_{\rm PF} = \hat H_{\rm el}(\mathbf R) + \sum_{m=1}^{N_{\rm mod}}\Big[
  \tfrac12\hat p_{q_m}^2 + \tfrac12\omega_m^2\hat q_m^2 - \omega_m\hat q_m\hat d_m
  + \tfrac12\hat d_m^2\Big] \,,
\end{equation}
where we have used $\hat q_m=(\hat b_m+\hat b_m^\dagger)/\sqrt{2\omega_m}$ and $\hat p_{q_m}={\rm i}\sqrt{\omega_m/2}\,(\hat b_m^\dagger-\hat b_m)$ to rewrite Eq.\ (\ref{eq:pf_fock}).
Equations~\eqref{eq:pf_fock} and~\eqref{eq:pf_cbo} equivalently describe the identical joint electronic-photonic problem.

The \gls{cboa} \cite{flick2017cavity} applies the Born-Oppenheimer approximation to the photon displacement coordinates.
It assumes that the photon displacement coordinates change slowly relative to the electrons, analogous to the nuclear coordinates.
The \gls{cboa} is thus valid in the \gls{vsc} regime, when the electronic states and the eigenstates of the cavity modes are well separated.
The photon displacement coordinates can then be treated as classical parameters on the same footing as the nuclear coordinates $\mathbf R$.
The kinetic photon displacement term can then be separated from the electronic problem similarly to the nuclear kinetic energy.
This yields an effective electronic Hamiltonian
\begin{equation}
\label{eq:h_cbo}
\hat H_{\rm CBO}\big(\mathbf R;\{q_m\}\big) = \hat H_{\rm el}(\mathbf R)
  + \sum_{m=1}^{N_{\rm mod}}\Big[\tfrac12\omega_m^2 q_m^2 - \omega_m q_m\hat d_m
  + \tfrac12\hat d_m^2\Big].
\end{equation}
All \gls{cboa} methods discussed in this paper are based on Eq.~\eqref{eq:h_cbo}.
The set of photon displacement coordinates $\mathbf{q}\equiv\{q_m\}$ serves as variational parameters for the energy.
They can be optimized together with the molecular orbitals: their initial values only start the iteration, and in every \gls{scf} iteration, each $\mathbf{q}$ is updated to the stationary value of Eq.~\eqref{eq:qopt} for the current density, so that the converged density and $\mathbf{q}^{\rm opt}$ are mutually consistent.
Otherwise, $\mathbf{q}$ is treated as a constant.
When geometries are relaxed, it is also optimized alongside $\mathbf R$.
In contrast, the family of \gls{qed} methods~\cite{Haugland2020-xh} is based on Eq.~\eqref{eq:pf_fock} and retains a fully quantized photon field in terms of
Fock states.

\subsection{Mean-field Ans\"atze}
\label{sec:identity}

The two Hamiltonians presented in Eq.~\eqref{eq:pf_fock} and Eq.~\eqref{eq:h_cbo} support formally different mean-field ansätze.
In \gls{cbohf}~\cite{Schnappinger2023-cbohf}, a single electronic Slater determinant,
\begin{equation}
\Psi_{\rm \gls{cbohf}} = \Phi(\mathbf R;\mathbf{q}) ,
\end{equation}
is optimized self-consistently together with the photon displacement coordinates $\mathbf{q}$, which act as parameters of the energy functional.
Consequently, no photonic wavefunction is obtained using \gls{cbohf}.
Because \gls{qedhf}~\cite{Haugland2020-xh} is based on Eq.~\eqref{eq:pf_fock}, the electronic Slater determinant forms a tensor product with a coherent state of each cavity mode, that is, a displaced photon vacuum~\cite{Glauber1963-coh,Haugland2020-xh}:
\begin{equation}
\label{eq:qedhf_ansatz}
\ke{\Psi_{\rm \gls{qedhf}}} = \exp\Big[\sum_{m=1}^{N_{\rm mod}} z_m\big(\hat b_m^\dagger-\hat b_m\big)\Big]
  \,\ke{\Phi_{\rm HF}}\otimes\ke{\rm vac} ,
\end{equation}
with displacement amplitudes $z_m$ optimized jointly with orbitals~\cite{Haugland2020-xh,Foley2023-pq}.
The coherent-state transformation removes the coupling of the photon field to the mean molecular dipole~\cite{Haugland2020-xh,Liebenthal2023-ra}; therefore, the amplitudes vanish for a molecule without a static dipole component along $\boldsymbol\lambda_m$~\cite{Haugland2020-xh,Schnappinger2023-cbohf}.

Minimizing either energy with respect to its respective classical or quantum photon variable removes the net linear coupling between the photon and the mean dipole (zero-field condition).
Both yield the same stationarity condition, up to a normalization factor relating the two:
\begin{equation}
\label{eq:qopt}
z_m\sqrt{\frac{2}{\omega_m}} = q_m^{\rm opt} = \frac{\langle\hat d_m\rangle}{\omega_m} .
\end{equation}
Thus, each classical coordinate equals, mode by mode, the expectation value of the corresponding coherent state~\cite{Haugland2020-xh,Schnappinger2023-cbohf,Angelico2023-jh}.
Evaluating the energy of each theory at this stationary point yields identical results up to a constant shift~\cite{Angelico2023-jh}:
\begin{equation}
\label{eq:identity}
E_{\rm \gls{qedhf}} = E_{\rm \gls{cbohf}}\big(\mathbf{q}^{\rm opt}\big)
                      + \sum_m \tfrac12\omega_m .
\end{equation}
This offset corresponds to the zero-point energy of the photo modes.
For a given set of cavity frequencies $\{\omega_m\}$, this shift is a constant. Therefore, it does not alter gradients, vibrational frequencies, or densities.
For converged $E_{\rm \gls{cbohf}}(\mathbf{q}^{\rm opt})$, the corresponding density and the nuclear gradient do not depend on the mode frequencies $\{\omega_m\}$, as shown below.
The frequency dependence is confined to the photon coordinates and to derivatives with respect to them: $q_m^{\rm opt}\propto1/\omega_m$, the coherent-state amplitudes $z_m\propto\omega_m^{-1/2}$, the mixed nuclear-photon and photon-photon blocks of the joint Hessian, which scale with $\omega_m$ and $\omega_m\omega_n$, and the photon components of the dipole and polarizability derivatives, which scale with $\omega_m$ (see Eq.~\eqref{eq:hess_rescale} and Sec.~S7\,A of the supporting information).

The \gls{cboa} allows us to formulate a joint nuclear-photon Hessian, which includes $\mathbf{q}$~\cite{Bonini2021,Schnappinger2023-wp}.
In Eq.~\eqref{eq:h_cbo}, $q_m$ and $\omega_m$ enter only through $x_m=\omega_m q_m$, since $\tfrac12\omega_m^2q_m^2-\omega_m q_m\hat d_m=\tfrac12x_m^2-x_m\hat d_m$.
Any method that treats $q_m$ as a classical parameter therefore yields an energy $E(\mathbf R,\{x_m\})$ whose stationary point $x_m^{\rm opt}$, equal to $\langle\hat d_m\rangle$ at the mean-field level, does not depend on $\omega_m$.
By applying the chain rule, $\partial/\partial q_m=\omega_m\,\partial/\partial x_m$
it can be shown that every derivative with respect to $q_m$ contributes one factor $\omega_m$.
The derivatives with respect to $x_m$ at $x_m^{\rm opt}$ are thus frequency independent  (for details see Sec.~S7\,A of the supporting information).
A joint Hessian built at $q^{\rm opt}_m$ for reference frequencies $\omega_m^{\rm ref}$ thus yields the Hessian at any frequency $\omega_m$ in closed form:
\begin{equation}
\label{eq:hess_rescale}
\begin{aligned}
\mathbf H_{\mathbf R\mathbf R}(\omega) &= \mathbf H_{\mathbf R\mathbf R} , \\
\mathbf H_{\mathbf R q_m}(\omega) &= \frac{\omega_m}{\omega_m^{\rm ref}}\,\mathbf H_{\mathbf R q_m} , \\
H_{q_m q_n}(\omega) &= \frac{\omega_m\,\omega_n}{\omega_m^{\rm ref}\,\omega_n^{\rm ref}}\,H_{q_m q_n} ,
\end{aligned}
\end{equation}
that is, $\mathbf H(\omega)=\mathbf S\,\mathbf H(\omega^{\rm ref})\,\mathbf S$ with $\mathbf S={\rm diag}(\mathbf 1_{3N_{\rm at}},\omega_1/\omega_1^{\rm ref},\dots,\omega_{N_{\rm mod}}/\omega_{N_{\rm mod}}^{\rm ref})$, and a geometry relaxed jointly in $\mathbf R$ and $\mathbf{q}$ does not depend on the photon mode frequencies.
The rescaling introduces no approximation beyond those of Eq.~\eqref{eq:h_cbo}, which are namely the long-wavelength limit, classical photon displacement coordinates, and lossless modes.
It is exact for a fixed $\boldsymbol\lambda_m$ and a reference Hessian at $\mathbf{q}^{\rm opt}(\omega^{\rm ref})$, with and without the \gls{dse} and for any number of modes.
This assumption does not hold for $q_m\neq q_m^{\rm opt}$,
or for a coupling that changes with frequency (a fixed field amplitude gives $\lambda_m\propto\omega_m^{-1/2}$), or for a quantized photon beyond the mean field, whose energy depends on $\omega_m$ explicitly (Table~S8 validates the law against rebuilt Hessians).

We have verified Eq.~\eqref{eq:identity} numerically; the full results are given in Table~S1 of the supporting information.
Both methods were evaluated for \ce{HF} and \ce{H2O} using basis sets ranging from cc-pVDZ to cc-pVQZ at coupling strengths of $\lambda=0$, $0.05$ and $0.10$~a.u.
Comparing the \gls{cbohf} energy at its variationally optimized $\mathbf{q}^{\rm opt}$ with the energy of the independent coherent-state \gls{qedhf} \gls{scf} implemented in \lstinline|cbo_qedhf| module, the maximum residual across all thirty comparisons is $2.6\times10^{-13}$~$E_h$ (Table~S1), and the densities agree to $1.2\times10^{-9}$.

\section{Beyond mean field: Configuration-Interaction}
\label{sec:correlation}

\subsection{CI matrix structure: \glsentryshort{cboci} vs \glsentryshort{qedci}}

Both \gls{cboci} and \gls{qedci} are \gls{ci} methods that provide access to correlation effects and excited states.
The application of \gls{ci} methods to polaritonic problems, particularly in the context of \gls{esc} and the analysis of correlation effects, was pioneered by Koch and co-workers~\cite{Angelico2023-jh,Haugland2025-uk} and Foley and co-workers~\cite{Foley2023-pq,Vu2024-qh,McTague2022-xl}.
The two methods differ fundamentally in their definitions of the Hamiltonian and wavefunction, visible in their CI matrix structures.

\gls{cboci} diagonalizes Eq.~\eqref{eq:h_cbo} at a fixed photon displacement coordinate $q$ (usually $q=q^{\rm opt}$) within the basis of $N_{\rm det}$ pure electronic determinants $\{\Phi_I\}$. Since $q$ is treated as a parameter, it merely dresses the integrals.
The result is an ordinary, field-free electronic Hamiltonian with modified one-electron ($\tilde h$) and two-electron ($\tilde g$) integrals~\cite{Schnappinger2023-cbohf}:
\begin{equation}
\label{eq:cbo_dressed_integrals}
\tilde h = \hat h - \omega q\,\hat d + \tfrac12\big(\hat d^2\big)_{\rm 1-body} \quad \text{ and
} \quad
\tilde g = \hat g + \big(\hat d^2\big)_{\rm 2-body} ,
\end{equation}
where the \gls{dse} term $\tfrac12\hat d^2$ splits into one-electron and two-electron integrals.
The corresponding expressions in atomic orbitals are given in Eq.~(S2) of the supporting information.
Brillouin's theorem and the Slater-Condon rules then apply unchanged:
\begin{equation}
\label{eq:cboci_matrix}
\mel{\Phi_I}{\hat H_{\rm CBO}}{\Phi_J} =
\begin{cases}
\displaystyle\sum_{i}\tilde h_{ii} + \tfrac12\sum_{ij}\big(\tilde g_{iijj}-\tilde g_{ijji}\big)
   + E_{\rm c}
  & \Phi_I=\Phi_J , \\[10pt]
\displaystyle\tilde h_{pr} + \sum_{i}\big(\tilde g_{prii}-\tilde g_{piir}\big)
  & \text{single, } p\!\to\!r , \\[10pt]
\tilde g_{prqs} - \tilde g_{psqr}
  & \text{double, } pq\!\to\!rs , \\[6pt]
0 & \text{three or more,}
\end{cases}
\end{equation}
with $i,j$ running over all spin-orbitals occupied in $\Phi_I$.
This preserves the exact sparsity pattern of an ordinary \gls{ci} matrix: the cavity enters exclusively through $\tilde h$ and $\tilde g$, never altering the selection rules, while the constant $E_{\rm c}=\tfrac12\omega^2q^2-\omega q\,\boldsymbol\lambda\cdot\boldsymbol\mu_{\rm nuc}+\tfrac12(\boldsymbol\lambda\cdot\boldsymbol\mu_{\rm nuc})^2+E_{\rm nuc}$ shifts every diagonal element equally; its two terms with the nuclear dipole moment $\boldsymbol\mu_{\rm nuc}$ vanish with the origin at the center of nuclear charge.
Consequently, the matrix carries no photon index and remains strictly $N_{\rm det}\times N_{\rm det}$ in dimension.

In contrast, \gls{qedci} diagonalizes Eq.~\eqref{eq:pf_fock} in the product basis $\{\Phi_I\}\otimes\{n\}$, where $n=0,\ldots,n_{\max}$ represents the photon quanta.
For the sake of readability, the following discussion is restricted to a single cavity mode.
The Hamiltonian matrix can be partitioned based on changes in the photon number:
\begin{equation}
\label{eq:qedci_matrix}
\mel{\Phi_I,n}{\hat H_{\rm PF}}{\Phi_J,n'} =
\begin{cases}
\mel{\Phi_I}{\hat H_{\rm el}(\mathbf R)+\tfrac12\hat d^2}{\Phi_J} + \delta_{IJ}\,\omega(n+\tfrac12)
  & n'=n , \\[4pt]
-\sqrt{\dfrac{\omega}{2}}\sqrt{\max(n,n')}\,\mel{\Phi_I}{\hat d}{\Phi_J} & n'=n\pm1 , \\[8pt]
0 & |n-n'|\geq2 .
\end{cases}
\end{equation}
The diagonal $n'=n$ blocks are purely electronic and share the underlying structure of Eq.~\eqref{eq:cboci_matrix}.
The distinguishing feature of \gls{qedci} lies in the $n'=n\pm1$ blocks.
Since $\hat d$ is a one-electron operator, it connects only determinants that differ by no more than one spin-orbital.
Consequently, the off-diagonal blocks of these equations couple only adjacent photon numbers, weighted by the ladder factor $\sqrt{\max(n,n')}$.
These entries produce polaritonic states, which are mixtures of electronic and photonic excitations and by construction do not exist in \gls{cboci}.
Within the scope of this paper, we focus exclusively on the correlated ground state obtained from both ansätze.
A detailed investigation of electronic excited states under \gls{vsc} using \gls{cboci} will be presented in a follow-up work.

\subsection{Correlation energy: electron-electron and electron-photon correlation}
\label{sec:corr_channels}

The block structure of Eq.~\eqref{eq:qedci_matrix} allows us to split the correlation energy into two distinct terms, which can be isolated by comparing \gls{cbofci}, \gls{qedfci}, and \gls{cbohf}:
\begin{align}
\varepsilon_{ee} &\equiv E_{\rm \gls{cbofci}} - E_{\rm \gls{cbohf}} , \label{eq:eps_ee}\\
\varepsilon_{ep} &\equiv E_{\rm \gls{qedfci}} - E_{\rm \gls{cbofci}} - \tfrac12\omega_c .
\label{eq:eps_ep}
\end{align}
Here and in the following, $\omega_c$ and $\lambda$ denote the frequency and coupling strength of a single cavity mode.
$\varepsilon_{ee}$ represents the electron-electron correlation contained entirely within the $n'=n$ blocks.
This corresponds to the \gls{cboa} frame in which no photonic excitation is possible.
Nevertheless, it captures the beyond-mean-field dressing of the electronic problem induced by the classical photon field.
In contrast $\varepsilon_{ep}$ is the electron-photon correlation arising from the $n'=n\pm1$ blocks and the full quantum treatment of the photon mode.
In the weak-coupling limit ($\lambda\to0$), $\varepsilon_{ep}$ scales as $\varepsilon_{ep} \propto \lambda^2$, since these off-diagonal blocks scale linearly with $\lambda$ through $\hat d$.

The subtraction of $\tfrac12\omega_c$ in Eq.~\eqref{eq:eps_ep} removes the photon vacuum energy that the Fock-space calculation inherently carries, but the \gls{cboa} framework lacks.
This ensures that $\varepsilon_{ep}$ vanishes at zero coupling.
However, the remainder does not solely represent the pure electron-photon correlation, because \gls{dse} effectively screens $\omega_c$~\cite{Rokaj2018-ww,Schafer2020-cb,Schnappinger2023-wp,Sidler2024-tw}.
For clarity, restricting the discussion to a single cavity mode, the effective cavity frequency is given by:
\begin{equation}
\label{eq:omega_eff}
\omega_{\rm eff}^2 = \omega_c^2\big(1-\boldsymbol\lambda^\top\boldsymbol\alpha\boldsymbol\lambda\big) ,
\end{equation}
where $\boldsymbol\alpha$ is the static polarizability tensor of the molecule in the cavity, obtained from the coupled \gls{scf} solution including the \gls{dse}. For several modes, $\omega_{\rm eff}^2$ generalizes to the eigenvalues of $\omega_m\omega_n(\delta_{mn}-\boldsymbol\lambda_m^\top\boldsymbol\alpha\boldsymbol\lambda_n)$.
The resulting shift of the zero-point energy from its bare value accounts for most of $\varepsilon_{ep}$ through polarizability alone.
Subtracting this shift isolates the contribution beyond the mean field to $\varepsilon_{ep}$:
\begin{equation}
\label{eq:eps_ep_dressed}
\Delta_{\rm ZPE} = \tfrac12\omega_{\rm eff} - \tfrac12\omega_c , \qquad
\varepsilon_{ep}^{\rm dressed} \equiv \varepsilon_{ep} - \Delta_{\rm ZPE} .
\end{equation}

\subsection{Stability and Correlation Diagnostics for \gls{cbohf}}
\label{sec:diagnostics}

A converged \gls{scf} calculation only guarantees that the energy is stationary.
It does not guarantee that the solution is a true minimum, that it preserves the spatial symmetry, or that a single Slater determinant describes it sufficiently.
While this is generally true for standard electronic-structure mean-field methods, strong coupling introduces an additional layer of potential stability and correlation issues.
To address these issues, we introduce five diagnostics for \gls{cbohf}. The first three form an eigenvalue-stability triad.
Following standard Hartree-Fock stability analysis~\cite{Cizek1967-stability,Seeger1977-stability} and its Kohn-Sham extension~\cite{Bauernschmitt1996-stability}, the cavity introduces a third source of stability.
At a stationary point, the second derivative of the energy with respect to orbital rotations and photon optimization separates into electronic blocks of singlet and triplet character, plus a photon block.
Here, $T_1$, the lowest \emph{singlet} orbital Hessian eigenvalue, certifies a true minimum when positive and a saddle point when negative.
$T_2$, the lowest \emph{triplet} eigenvalue, signals an instability toward spin polarization (RHF$\to$UHF) when negative.
Neither $T_1$ nor $T_2$ involve the photon displacement coordinate.
Instead, $T_3$ is the photon sector analog and equals $\omega_{\rm eff}^2$ from Eq.~\eqref{eq:omega_eff} (the curvature along $q$ after electronic relaxation); thus, $T_3>0$ is the local stability condition of the cavity mode.
Note that these $T$ labels strictly denote stability criteria and are completely unrelated to the $T_1$ amplitude diagnostic used in coupled-cluster theory~\cite{Lee1989-t1diag}.
The remaining two criteria determine whether a single determinant was ever an adequate representation of the electronic wave function. $C_0^2$, the squared leading-configuration coefficient, typically indicates a significant multireference character when it falls roughly below $0.9$.
Because we employ two distinct \gls{ci} formulations (see Sec.~\ref{sec:correlation}), this metric takes two forms.
Evaluated using the \gls{cboci} reference wave function, $C_0^2=\vert{}c_0\vert{}^2$ is the reference weight strictly within the electronic-only $n'=n$ blocks.
This means that it is only reduced by electron-electron correlation.
In contrast, evaluated against the \gls{qedfci} reference wave function, the joint weight $(C_0^2)_{\rm joint}=\vert{}\langle{\Phi_{\rm HF},0}\vert{}{\Psi}\rangle\vert{}^2$ projects onto the Hartree-Fock determinant multiplied by the photon vacuum.
It therefore loses amplitude to the $n'=n\pm1$ blocks, responding to electron-electron and electron-photon correlation simultaneously.
This wavefunction-based distinction exactly parallels the energy partitioning defined in Eqs.~\eqref{eq:eps_ee}--\eqref{eq:eps_ep}.

\section{Software capabilities}
\label{sec:capabilities}
This section summarizes the key capabilities provided by \texttt{psi4-cbo}.
At the mean-field level, the package implements \gls{cbohf} for closed-shell systems~\cite{Schnappinger2023-cbohf,Sidler2024-tw} and CBO-UHF for open-shell and broken-symmetry cases, complete with spin-contamination diagnostics, as well as a standalone coherent-state \gls{qedhf} \gls{scf}.
It also features \gls{cbodft}, which pairs the exchange-correlation functionals of Psi4 with a \gls{cboa} description of the cavity field; the supported functional classes are listed in Sec.~\ref{sec:limitations}.
Because standard functionals lack explicit photon dependence, \gls{cbodft} functions as a DFT equivalent to \gls{cbohf} rather than a full quantum-electrodynamical DFT implementation~\cite{Tokatly2013-ho,Ruggenthaler2014-it}.
For excited states and correlation, CIS and CISD return permanent and transition dipoles, oscillator strengths, and natural transition orbitals for each root.
Since this CI framework diagonalizes all spin states simultaneously, each root is returned with its exact value $\langle S^2\rangle$, ensuring that triplet roots are properly identified rather than misidentified as weakly absorbing singlets.
To benchmark the \gls{cboa} (Sec.~\ref{sec:correlation}), the package includes an exact \gls{qedfci} and a complete active-space (CAS) CI implementation.
Both use a dense eigensolver by default; an optional matrix-free direct-CI solver (\lstinline|solver="matrix_free"|, for a given number of states) avoids the $\mathcal{O}(N_{\rm det}^2)$ storage bottleneck.

Geometry optimizations and vibrational spectrum calculations~\cite{Schnappinger2023-wp,Schnappinger2024-vt,Schnappinger2025-fx} are driven by analytic gradients and Hessians that treat the photon displacement coordinates $\mathrm{q}$ as variational parameters alongside the nuclei.
When scanning the cavity frequency $\omega_c$ at a fixed geometry and coupling strength, redundant Hessian calculations are bypassed because the mean-field energy, density, and nuclear gradient are independent of $\omega_c$ (Sec.~\ref{sec:identity}).
The joint Hessian is computed once and can be rescaled to any target frequency by Eq.~\eqref{eq:hess_rescale}.
For larger systems, a numerical Hessian with density fitting is used (see Table~S5).
To facilitate ab initio molecular dynamics, a force interface returns the energy along with the joint force vector of size $(3N_{\rm at}+N_{\rm mod})$ and a restart record.
This enables an external propagator to drive cavity Born-Oppenheimer molecular dynamics.

Finally, the software provides tools for physical interpretation and stability analysis.
An SCF-stability analysis evaluates singlet, triplet, and photon channels (Sec.~\ref{sec:diagnostics}) to identify per-system threshold behaviors, while symmetry tools provide point-group selection rules and quantify symmetry breaking in the cavity-dressed density (Table~S6).
Electron densities, quadrupole moments, and Mulliken, L\"owdin and electrostatic-potential-fitted atomic charges are evaluated directly from the cavity-dressed density matrix.
This allows charge redistribution under coupling to be examined directly rather than inferred from the dipole.
A complementary suite of analysis routines decomposes the total energy, separates the contributions to the correlation energy, and determines the molecular and photonic composition of each vibro-polaritonic mode.
Throughout these routines, efficiency and robustness are prioritized.
By default, the origin is placed at the center of nuclear charge, which also fixes the photon displacement and the energies at a fixed $\mathrm{q}$.
Meanwhile, density fitting~\cite{Whitten1973-df,Dunlap1979-df,Vahtras1993-ri,Weigend2002-rihf}, warm-started densities, and matrix-free conjugate-gradient solvers~\cite{Hestenes1952-cg} improve computational throughput (Table~S7).
\texttt{psi4-cbo} adds no parallel layer of its own: calculations run shared-memory parallel through the OpenMP-threaded integral and \gls{jk} engines of Psi4 and the threaded \gls{blas} behind NumPy (input key \texttt{nthreads}), without MPI or GPU support.

\subsection{Driving the package: calls, results and output}
\label{sec:usage}

The package provides two equivalent interfaces to run calculations.
An input file (\textsc{yaml} or \textsc{json}) specifies the computational tasks
(see Table~\ref{tab:tasks})
and the calculation is started via the command line \lstinline|python -m psi4_cbo input.yaml|.
The Python \gls{api} exposes the same routines as functions, for composed workflows, parameter scans and scripting.

A minimal \gls{cbohf} single point shows both variants for the same type of calculation.
As an input file:
\begin{lstlisting}[language=yamlcbo]
task:   scf
method: RHF
output: water_scf                # -> water_scf.out (report) + water_scf.h5 (arrays)
molecule: |
  O  0.000000  0.000000 -0.068516
  H  0.000000 -0.790690  0.543701
  H  0.000000  0.790690  0.543701
  symmetry c1
cavity:  {lambda: [[0.0, 0.0, 0.05]], omega: 0.5}  # lambda as a vector in the molecular frame (both in a.u.)
psi4_options: {basis: cc-pvdz, maxiter: 100, e_convergence: 1.0e-9, d_convergence: 1.0e-8}
qopt: true                       # relax the photon coordinate q
nthreads: 8                      # Psi4 thread count
\end{lstlisting}
run with \lstinline|python -m psi4_cbo water_scf.yaml|.
Alternatively, the same calculation can be run directly as a Python script through the \gls{api}:
\begin{lstlisting}[language=Python]
import numpy as np, psi4
from psi4_cbo import cbo_rhf, write_output
psi4.set_num_threads(8)          # Psi4 does NOT read OMP_NUM_THREADS
molecule = """
O  0.000000  0.000000 -0.068516
H  0.000000 -0.790690  0.543701
H  0.000000  0.790690  0.543701
symmetry c1
"""
cavity = {"lambda": np.array([[0.0, 0.0, 0.05]]), "omega": np.array([0.5]), "qpos": np.array([0.0])}
opts = {"basis": "cc-pvdz", "maxiter": 100, "e_convergence": 1e-9, "d_convergence": 1e-8}
wfn = cbo_rhf(cavity, molecule, opts, qopt=True)
write_output(wfn, "water_scf", cavity=cavity, opts=opts, molecule=molecule, method="RHF")
\end{lstlisting}
The input options for every task are documented in the repository (\texttt{docs/wiki/yaml-cli.md}), with example input files in \texttt{examples/}.
All subsequent examples use the Python \gls{api}.

Basis sets throughout this work are Dunning's correlation-consistent sets, with diffuse augmentation where the polarizability matters~\cite{Dunning1989-xc,Kendall1992-wu}; the minimal STO-3G basis~\cite{Hehre1969-xx} appears only for qualitative demonstrations, and each calculation below states the basis it uses.
The examples use the \gls{scf} thresholds shown above, $E_{\rm conv}=10^{-9}\,E_h$ and $D_{\rm conv}=10^{-8}$, with \texttt{maxiter}~$=100$.
The \gls{scf} is accelerated by standard \gls{diis}~\cite{Pulay1980-diis,Pulay1982-diis}.
Strong coupling makes convergence more challenging.
An optional fallback (\lstinline|"scf_fallback": True|) retries a failed \gls{scf} with damping and then with damping and a level shift.

All routines take a unified description of the cavity mode as an input:
a dictionary of coupling vectors \lstinline|"lambda"| of shape $(N_{\rm mod},3)$ (norm $\lambda$, direction = polarization), frequencies \lstinline|"omega"| and initial photon displacements \lstinline|"qpos"| (in atomic units).
\lstinline|setup_cavity(wc, pol_cavity, lam=..., q=...)| builds the cavity description from a frequency in \si{\per\centi\meter} and the coupling strength $\lambda$ in atomic units.
The coupling can also be given as a field amplitude $E_c$ in \si{\volt\per\nano\meter} (\lstinline|ec=|, $\lambda=E_c\sqrt{2/\omega}$) or as a mode volume $V$ in \si{\cubic\angstrom} (\lstinline|volume_a3=|, $\lambda=\sqrt{4\pi/V}$).
Since $\lambda\propto E_c/\sqrt{\omega}$, a scan of the cavity frequency must ensure $\lambda$ or $V$ is kept fixed (not $E_c$).
The number of modes is the leading dimension of the list \lstinline|"lambda"|; \lstinline|"omega"| and \lstinline|"qpos"| need one list entry per mode, which is checked on input, and \lstinline|setup_cavity| accepts a single frequency and displacement for all modes.

\begin{table}
\caption{Standard tasks accepted by the input file and their principal outputs. All tasks are equally accessible via the Python \gls{api}. Analytic derivatives are used by default where available; \gls{cbodft} currently falls back to numerical evaluation (analytic joint \gls{cbodft} Hessians are not yet implemented).}
\label{tab:tasks}
\resizebox{\columnwidth}{!}{%
\begin{tabular}{ll}
\toprule
task & returns \\
\midrule
\texttt{scf}            & energy, density, orbitals, dipole, $\partial E/\partial q_m$ \\
\texttt{gradient}       & nuclear and photon gradient, resolved by contribution \\
\texttt{hessian}        & joint $(\mathbf R,q)$ Hessian, $\boldsymbol\mu$ derivatives ($\alpha$ derivatives: numeric route) \\
\texttt{frequencies}    & vibro-polaritonic frequencies, modes, \gls{ir}, Raman (numeric Hessian) \\
\texttt{optimize}       & geometry optimization over nuclei and photon coordinate \\
\texttt{polarizability} & $\alpha(3,3)$ from \gls{cphf}/CPKS, analytic or numeric \\
\texttt{cis}, \texttt{cisd} & excitation energies, dipoles, $\langle S^2\rangle$ per root \\
\texttt{analysis}       & energy decomposition into fragment and pair contributions \\
\texttt{stability}      & singlet, triplet and photon \gls{scf} stability, $\omega_{\rm eff}^2$ \\
\texttt{properties}     & quadrupole moment, Mulliken, L\"owdin and ESP-fitted charges \\
\bottomrule
\end{tabular}}
\end{table}

The results are returned as dictionaries, and the same keys are used whether the calculation is run from a file or from a Python script.
\begin{lstlisting}[language=Python]
wfn = cbo_rhf(cavity, molecule, opts, qopt=True)   # dict, energies in Hartree
wfn["CBO-RHF ENERGY"]        # float
wfn["PHOTON GRADIENT"]       # (Nmod,)  dE/dq_m -- zero at the relaxed photon coordinate
wfn["QPOS"]                  # (Nmod,)  photon displacement used
grad = cbo_rhf_grad(cavity, molecule, wfn)
grad["Total"], grad["DSE"]   # (natoms, 3) each; contributions resolved separately
\end{lstlisting}
For example, \lstinline|compute_vibrational_analysis| returns a copy of the Hessian dictionary extended by \lstinline|"frequencies"|, \lstinline|"normal_modes"|, and \lstinline|"IR_intensity"| ($I_m\propto|\partial\boldsymbol\mu/\partial Q|^2$ per mode); \lstinline|cbo_cisd| returns \lstinline|"EXCITATION ENERGY"|, \lstinline|"PERMANENT DIPOLE"|, and \lstinline|"STATE S2"|  per state ($\langle \hat S^2\rangle$); and \lstinline|cbo_stability_analysis| returns \lstinline|"SINGLET STABLE"|, \lstinline|"TRIPLET STABLE"|, and \lstinline|"PHOTON STABLE"|, along with \lstinline|"OMEGA EFF SQUARED"| (the squared dressed photon frequencies of Eq.~\eqref{eq:omega_eff}, whose signs serve as the boundedness diagnostic).

Since the photon enters strictly as a parameter, the data structures scale naturally in size: every nuclear array is paired with a corresponding photon array of length $N_{\rm mod}$.
An input-file run, or \lstinline|write_output| in Python, writes a human-readable report and a \textsc{hdf5} archive.

The report breaks down the energy into the critical cavity contributions and explicitly resolves the \gls{dse} into its one-electron, Coulomb-like, exchange-like, nuclear, and electron-nuclear cross components; the last two vanish in the default center-of-nuclear-charge frame:
\begin{lstlisting}
  ENERGIES                                                    [Hartree]
    CBO-RHF ENERGY                                 -76.0163552466
    NUCLEAR REPULSION ENERGY                         8.8014639821
    1E ENERGY                                     -122.4541137378
    2E ENERGY                                       37.6312373968
    CBO DIPOLE ENERGY                               -0.0017136317
    CAVITY POTENTIAL                                 0.0008568158
  Dipole self-energy contributions:
    DSE-1E                                           0.0081506891
    DSE-2E-J                                         0.0008568158
    DSE-2E-K                                        -0.0030935769
    DSE-NUCLEAR                                      0.0000000000
    DSE-CROSS                                       -0.0000000000  <- electronic-nuclear cross terms
    DSE-TOTAL                                        0.0059139280
  PROPERTIES
    CBO DIPOLE MOMENT                    [+0.00000000  -0.00000000  +0.82792069]   |v| = 0.82792069  (2.1042 D)
    NUCLEAR DIPOLE MOMENT                [+0.00000000  +0.00000000  -0.00000000]   |v| = 0.00000000  (0.0000 D)
\end{lstlisting}
Finally, the accompanying \textsc{hdf5} file stores the result arrays under identical keys, alongside \lstinline|_cavity| and \lstinline|_molecule| metadata groups that record the exact coupling vectors, frequencies, photon coordinates, and geometry utilized; the density and orbitals are stored only on request (input key \texttt{hdf5\_wfn: true}).
Because a cavity calculation requires more parameters than a standard gas-phase run, archiving these states alongside the results ensures robust data provenance and reproducibility.

\subsection{A full workflow: structure, spectrum and detuning}

The results for acrolein shown in Sec.~\ref{sec:showcase_acrolein} are produced by chaining three tasks together.
This identical sequence can be used for any molecule.
Since the nuclei and the photon coordinates are relaxed together, the structure is first optimized \emph{inside} the cavity with \lstinline|opt_cbo|.
The joint Hessian at the relaxed structure gives the vibro-polaritonic spectrum.
For larger systems (Table~S5), \lstinline|cbo_hess_num| with \lstinline|density_fitting=True| replaces \lstinline|cbo_hess_analytic|; with \lstinline|raman={"numeric": False}| it also returns the polarizability derivatives (\lstinline|"alph_xx"|, \ldots), from which \lstinline|compute_vibrational_analysis| computes Raman activities. The analytic Hessian provides \gls{ir} intensities only.

The custom optimizer for joint geometry photon displacement coordinate optimization
is described in ef.~\cite{Schnappinger2024-vt}.
It uses line search, augmented-Hessian level shift, convergence tests, and checkpointing.
The photon coordinate and the light-matter coupling typically increase
the number of iterations required to converge.
Four types of step solvers are available:
\gls{sd},
\gls{nr} rebuilds the Hessian at every step, numerically or, with \lstinline|"hessian": "analytic"| (\gls{cbohf} only), analytically;
\gls{bf}~\cite{Broyden1970-bfgs,Fletcher1970-bfgs,Goldfarb1970-bfgs,Shanno1970-bfgs} is the default.
BFGS-H combines \gls{nr} and \gls{bf}, starting from a Hessian and updating it thereafter; with \lstinline|"hessian": "analytic"| that starting Hessian is the analytic joint one rather than a numerical approximation, which for a small molecule is the fastest route to a converged structure.
For the last three, an indefinite Hessian is shifted by its lowest eigenvalue $\varepsilon_{\min}$, $\mathbf H+(\mu-\varepsilon_{\min})\mathbf 1$, with $\mu$ increasing until the step falls within the trust radius; this keeps the step bounded along the flat photon directions, where the Hessian is often indefinite.
Therefore, the optimizer configuration addresses these specific challenges.
An example is given in the following (the option values are the defaults):
\begin{lstlisting}[language=Python]
options = {
    "method":      "BFGS-H",   # SD | NR | BFGS | BFGS-H
    "hessian":     "analytic", # Hessian of NR and BFGS-H (CBO-HF only)
    "preopt_q":    True,       # relax q before the first nuclear step
    "line_search": True,       # backtracking line search
    "aug_hess":    True,       # level-shift an indefinite Hessian
    "g_conv":      1e-5,       # max and RMS gradient
    "D_conv":      5e-5,       # max and RMS displacement
    "ckpt":        "opt.npz",  # geometry, q, energy after each accepted step
}
opts_df = dict(opts, scf_type="df")   # density_fitting=True needs a density-fitted SCF
res = opt_cbo(cavity, molecule, opts_df, options=options, density_fitting=True, restart=wfn_prev)
\end{lstlisting}

Strict convergence requires that all four criteria (maximum and \gls{rms} gradient, plus maximum and \gls{rms} displacement) be met simultaneously, and satisfying the gradient alone is insufficient.
Four of these settings warrant specific comment because they address intrinsic failures of the strongly coupled problem.
First, \lstinline|density_fitting=True| replaces the exact two-electron gradient, a NumPy contraction of the four-index derivative integrals, with the density-fitted gradient of Psi4 (\lstinline|scfgrad|, which requires \texttt{scf\_type df}).
For larger molecules this poses the largest contribution to computational cost in an optimization.
Second, \lstinline|preopt_q| relaxes the photon coordinate before the first nuclear gradient.
This is important because a structure optimized outside the cavity provides an inadequate starting point for $\mathrm{q}$ and thus would add unnecessary iteration cycles to the optimizations.
Third, \lstinline|restart=| warm-starts the first \gls{scf} from a previously converged density.
This is the standard remedy for strong-coupling convergence failures that appear at the beginning of an optimization.
 The gradual increase in $\lambda$ values aids in achieving convergence for calculations that would otherwise not converge from a cold start.
Finally, \lstinline|ckpt| writes the geometry, photon coordinate, and energy after every accepted step; the file is not read back automatically.
The option \lstinline|"hessian": "analytic"| is restricted to \gls{cbohf}, as no analytic joint Hessian currently exists for \gls{cbodft} (Sec.~\ref{sec:limitations}).
\gls{cbodft} optimizations with \gls{nr} or BFGS-H thus use the numerical Hessian.

In Python, these steps can be composed directly, making frequency scans highly efficient:
\begin{lstlisting}[language=Python]
mol  = psi4.geometry(molecule)
res  = opt_cbo(cavity, mol, opts, options=options)            # relax R and q together
mol.set_geometry(psi4.core.Matrix.from_array(res["last_geom"]))   # Bohr
cavity["qpos"] = res["last_q"]
wfn  = cbo_rhf(cavity, mol, opts, qopt=True)                   # SCF at q_opt
hess = cbo_hess_analytic(cavity, mol, wfn)                     # joint (R, q) Hessian
vib  = compute_vibrational_analysis(hess)                      # frequencies, modes, IR

# one Hessian build, then a closed-form rescale per cavity frequency
for w in omega_scan:                                           # (Nmod,) in a.u.
    vib_w = compute_vibrational_analysis(
        cbo_hess_rescale_omega(hess, cavity["omega"], w, mol.natom()))
    frac  = cbo_photon_fraction(vib_w, mol.natom())            # photon weight -> LP/UP
\end{lstlisting}
The functions in the loop apply Eq.~\eqref{eq:hess_rescale}.
Each invocation of \lstinline|compute_vibrational_analysis|
thus only rescales the Hessian instead of recalculating the second derivatives.
The \lstinline | qopt=True| \gls{scf} is required since Eq.~\eqref{eq:hess_rescale} holds only at $\mathrm{q}^{\rm opt}$.

\subsection{Diagnostics: stability of \gls{cboa} calculations}

The stability and reference-weight diagnostics introduced in Sec.~\ref{sec:diagnostics} require only a few additional lines of code once a converged wavefunction is obtained.
These diagnostics are the primary intended method for validating the  \gls{cboa} on any new system.
\begin{lstlisting}[language=Python]
st = cbo_stability_analysis(wfn, cavity, molecule, opts, channels=("singlet", "triplet", "photon"))
st["SINGLET STABLE"], st["TRIPLET STABLE"], st["PHOTON STABLE"]   # bool each
st["SINGLET LOWEST EIG"]    # T1: RHF -> RHF, is this a minimum
st["TRIPLET LOWEST EIG"]    # T2: RHF -> UHF, spin-symmetry breaking
st["OMEGA EFF SQUARED"]     # T3: eigenvalues of omega_m omega_n (delta_mn - lambda_m^T alpha lambda_n);
                            #     < 0 saddle point along q
st["PHOTON ALPHA"]          # alpha: SCF polarizability in the cavity, driving T3
st["ZPE SHIFT"]             # Delta_ZPE = 1/2 sum_m (omega_eff,m - omega_m)

# estimate before any coupled calculation, from the cavity-free polarizability alpha0
from psi4_cbo.cbo_stability import cbo_photon_dressing
est = cbo_photon_dressing(alpha0, cavity, dse=True)   # "OMEGA EFF SQUARED", "PHOTON STABLE", "ZPE SHIFT"

# correlation channels and reference weights
pf = cbo_fci_pf(cavity, molecule, opts, n_max=4, qopt_hf=True, classical_ref=True, return_vectors=True)
pf["EPS EE"], pf["EPS EP"]                                    # epsilon_ee, epsilon_ep
cbo_mr_diagnostic(pf)["JOINT C0^2"]                           # joint weight against QED-FCI
cbo_mr_diagnostic(cbo_cisd(cavity, molecule, opts))["C0^2"]   # electronic C0^2; < ~0.9 is a warning
\end{lstlisting}
The $T_3$ diagnostic is especially important when scaling up the coupling strength or ensemble size, and because it inherits the sensitivity of the polarizability to diffuse basis functions, it should be evaluated with an augmented basis set (Table~S4).
\lstinline|cbo_photon_dressing| estimates $T_3$ before any coupled calculation from the cavity-free polarizability $\boldsymbol\alpha_0$: with $\boldsymbol\Omega={\rm diag}(\omega_m)$ and the coupling vectors as rows of $\boldsymbol\Lambda$, $\omega_{\rm eff}^2$ follows from the eigenvalues of $\boldsymbol\Omega(\mathbf 1+\boldsymbol\Lambda\boldsymbol\alpha_0\boldsymbol\Lambda^{\!\top})^{-1}\boldsymbol\Omega$ with the \gls{dse}, which stay positive at any coupling strength.
The converged $T_3$ of \lstinline|cbo_stability_analysis| is the photon block of the Hessian after electronic relaxation; a negative value means that the energy decreases along $q$, so the reference is a saddle point rather than a minimum.

\subsection{Symmetry: modified selection rules for vibrational spectroscopy}

The package provides tools that address two distinct aspects of molecular symmetry under strong coupling.
Symmetry in cavity \gls{qed} has drawn wider attention, from cavity-induced symmetry lowering to symmetry-reduced polaritonic coupled cluster~\cite{Barlini2025,Schnappinger2024-vt,Monzel2026-qedcc-sym}.
First, the point-group symmetry of the bare molecule dictates which vibrations a specific cavity polarization can couple to.
This selection rule can be evaluated before any coupled calculation is performed.
\begin{lstlisting}[language=Python]
lab = cbo_symmetry_labels(molecule, hess_dict, cavity=cavity)
# irrep label per bare vibration, and which modes the cavity polarization may couple to
\end{lstlisting}
Second, the continuous symmetry measure (CSM)~\cite{Zabrodsky1992-csm}, adapted to a one-particle density matrix, quantifies how close the cavity-polarized density $D$ is to each symmetry operation $R$ of the molecule,
\begin{equation}
\label{eq:csm}
\mathrm{CSM}(R) = 1 - \frac{\lVert D'-D\rVert_F}{\lVert D'\rVert_F+\lVert D\rVert_F} , \qquad
D' = T(R)\,D\,T(R)^{\!\top} ,
\end{equation}
where $T(R)$ is the AO representation of $R$ and $\lVert\cdot\rVert_F$ the Frobenius norm.
$\mathrm{CSM}(R)\in[0,1]$ is scale-invariant in $D$ and equals 1 when $D$ respects $R$ exactly:
\begin{lstlisting}[language=Python]
csm = cbo_symmetry_csm(molecule, wfn["CBO-RHF DENSITY"], basis)   # spherical-harmonic basis
csm["csm"]                  # {operation: CSM}; 1 = symmetric, < 1 = broken
\end{lstlisting}
These tools are designed to be used in tandem.
The irreducible representation labels establish which couplings are formally allowed in the uncoupled limit.
Meanwhile, the CSM reveals how much of that strict classification survives once the cavity field dresses the density.
In the context of molecular ensembles, this metric distinguishes a strict selection rule from a mere tendency.

\section{Evaluating the validity of the \gls{cboa}}
\label{sec:classical_photon}

\begin{figure}
\centering
\includegraphics[width=0.9\linewidth]{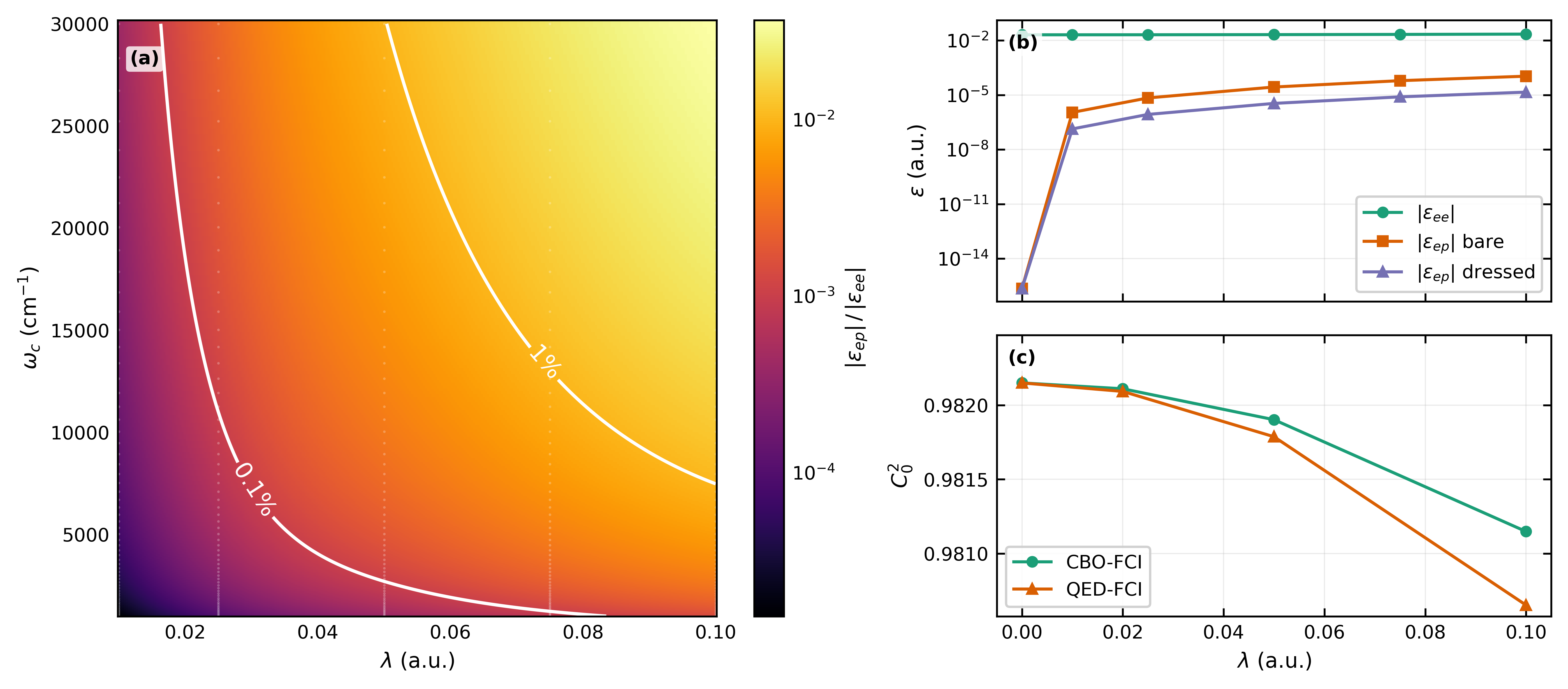}
\caption{Correlation map for \ce{H2}/STO-3G at its equilibrium geometry, coupled to a single $z$-polarized cavity mode, derived from \gls{cbohf}, \gls{cbofci}, and \gls{qedfci} energies. The scanned range spans the \gls{vsc} and \gls{esc} regimes. \textbf{(a)} Ratio $\vert{}\varepsilon_{ep}\vert{}/\vert{}\varepsilon_{ee}\vert{}$ as a function of cavity frequency $\omega_c$ and coupling $\lambda$. White contours mark the $0.1\%$ and $1\%$ levels; $\lambda=0$ (where $\varepsilon_{ep}$ vanishes) is omitted. \textbf{(b)} $\vert{}\varepsilon_{ee}\vert{}$, bare $\vert{}\varepsilon_{ep}\vert{}$, and dressed $\vert{}\varepsilon_{ep}\vert{}$ as functions of $\lambda$ at $\omega_c=\SI{3555}{\per\centi\meter}$ (\gls{vsc} condition; see Eqs.~\eqref{eq:eps_ee}--\eqref{eq:eps_ep_dressed} for definitions). \textbf{(c)} Reference weight $C_0^2$ against \gls{cbofci} and joint weight $(C_0^2)_{\rm joint}$ against \gls{qedfci}, as a function of $\lambda$ at $\omega_c=\SI{3555}{\per\centi\meter}$.}
\label{fig:correlation_map}
\end{figure}

To determine when explicit photon quantization becomes necessary, we evaluate electron-electron and  electron-photon correlation energies along with  the introduced correlation diagnostics across a comprehensive grid of cavity frequencies and coupling strengths.
Analyzing this two-dimensional parameter space allows us to systematically map the boundary where the \gls{cboa} begins to break down.
Figure~\ref{fig:correlation_map} applies both diagnostics to the test case of \ce{H2}/STO-3G coupled to a single $z$-polarized cavity mode.
This system is computationally inexpensive enough to compute densely in both cavity frequency $\omega_c$ and coupling strength $\lambda$, while remaining exactly solvable by \gls{qedfci}.

Figure~\ref{fig:correlation_map}(a) shows the ratio $\vert{}\varepsilon_{ep}\vert{}/\vert{}\varepsilon_{ee}\vert{}$ as a function of cavity frequency (\SIrange{1000}{30000}{\per\centi\meter}) and coupling strength $\lambda$ (up to $0.1$~a.u.).
In the low-frequency, weak-coupling regime (lower-left quadrant), the electron-photon correlation is four to five orders of magnitude below the electron-electron correlation at $\lambda=0.01$ and stays more than two orders below it up to $\lambda=0.1$ throughout the \gls{vsc} range.
This makes the  \gls{cboa} an excellent approximation.
The $0.1\%$ and $1\%$ ratios (white contour lines) mark where this approximation begins to worsen.
This threshold is reached by pushing simultaneously toward electronic-transition frequencies and strong coupling (upper-right quadrant).
At the extreme corner of the grid ($\lambda=0.1$, $\omega_c=\SI{30000}{\per\centi\meter}$), the ratio is $3.6\%$.
Thus, even in this limit, photon quantization primarily corrects the correlation energy rather than qualitatively changing the ground state.

Figures~\ref{fig:correlation_map}~(b) and~(c) show vertical cuts through the correlation map at $\omega_c=\SI{3555}{\per\centi\meter}$, thus scanning coupling strengths in the \gls{vsc} regime.
While $\varepsilon_{ee}$ remains constant, the bare and dressed $\varepsilon_{ep}$ rise steeply from zero at $\lambda=0$, though they remain orders of magnitude smaller than $\varepsilon_{ee}$.
The polarizability-dressed $\varepsilon_{ep}$ is about an order of magnitude smaller than the bare value and of opposite sign, demonstrating that most of the bare electron-photon correlation is already accounted for by dressing the photon frequency with the molecular polarizability (see Eq.~\eqref{eq:omega_eff}).
Figure~\ref{fig:correlation_map}(c) shows that both reference weights decrease with increasing $\lambda$, with the joint weight decreasing faster because it also loses amplitude to the $n'=n\pm1$ blocks.
The energy-based (b) and amplitude-based (c) diagnostics completely agree on where the coupling begins to significantly perturb the system.

The same structure appears for \ce{LiH}, \ce{HF} and \ce{H2O} and across five basis sets from STO-3G to cc-pVTZ (see Tables~S2 and~S3 of the supporting information): $\varepsilon_{ee}$ is essentially independent of $\lambda$, while $\varepsilon_{ep}$ grows as $\lambda^2$.
The ratio itself does not transfer between systems, because the two quantities depend on different properties: $\varepsilon_{ee}$ grows with the number of correlated electron pairs and with the basis set size, while $\varepsilon_{ep}$ is dominated by the polarizability dressing of the cavity mode, $\varepsilon_{ep}\approx\tfrac12(\omega_{\rm eff}-\omega_c)\approx-\tfrac14\omega_c\boldsymbol\lambda^{\!\top}\boldsymbol\alpha\boldsymbol\lambda$, and saturates once $\boldsymbol\alpha$ converges in the basis set size.
At $\omega_c=\SI{3550}{\per\centi\meter}$ the ratio spans $7\times10^{-5}$ to $1.2\times10^{-2}$ and differs by a factor of 7 to 11 between \ce{LiH}, which combines a large polarizability with few correlated electrons, and \ce{H2O}; for \ce{H2} it is largest with 6-31G and decreases towards larger basis sets, where $\varepsilon_{ep}$ is converged but $\varepsilon_{ee}$ keeps growing.
Fig.~\ref{fig:correlation_map} thus shows where the electron-photon correlation becomes relevant for one system. It is not a system-independent threshold.

\section{Acrolein under vibrational strong coupling}
\label{sec:showcase_acrolein}

All vibrational frequencies quoted here are harmonic values without any empirical documented Hartree Fock scaling factors. At the HF/aug-cc-pVDZ level these overestimate experiment by approximately $10$--$15\%$ ($14\%$ for the \ce{C=O} stretch of acrolein). The \ce{C=O} stretch of acrolein appears near \SI{1965.5}{\per\centi\meter} compared to the experimental value $\approx\SI{1720}{\per\centi\meter}$.

We performed geometry optimizations under strong coupling, evaluated analytic and numeric joint Hessians and computed vibro-polaritonic \gls{ir} spectra for acrolein (\ce{C3H4O}, aug-cc-pVDZ).
While these individual capabilities have been previously demonstrated for smaller systems~\cite{Schnappinger2023-cbohf,Schnappinger2025-fx}, the advance here lies in deploying them simultaneously at this scale and integrating the stability diagnostics (Sec.~\ref{sec:diagnostics}) concurrently with the full simulations.
Acrolein lies in the $xy$ plane, with the \ce{C=O} bond oriented along $y$, and is coupled to two cavity modes with the same coupling strength $\lambda=0.05$ and the same frequency $\omega_c$: one polarized along $x$, in the molecular plane and perpendicular to the \ce{C=O} bond, and one along $z$, perpendicular to the plane (coordinates in Sec.~S9 of the supporting information).
The field-free dipole moment, $\boldsymbol\mu=(0.234,-1.406,0)$~a.u.\ ($|\boldsymbol\mu|=1.425$~a.u.), points almost along $-y$, so that it projects only weakly onto the $x$-polarized mode and not at all onto the $z$-polarized one.
The molecular orientation with respect to the cavity modes is preserved during the optimizations.
The cavity is tuned to the \ce{C=O} stretch mode at \SI{1965.5}{\per\centi\meter},
which is a strong \gls{ir}-active band carrying.
The \ce{C=C} stretch mode is redshifted by \SI{148}{\per\centi\meter}
and ins 153 times weaker in intensity compared to \ce{C=O} stretch mode.
In a fully self-consistent ab initio treatment, tuning the cavity to resonance is no longer trivial.
Setting $\omega_c$ to the bare molecular frequency does not automatically put the cavity on resonance, as the \gls{dse} dresses the photon to $\omega_c\sqrt{1-\boldsymbol\lambda^{\!\top}\boldsymbol\alpha\boldsymbol\lambda}$ due to the presence of the molecule itself: setting $\omega_c=\SI{1965.5}{\per\centi\meter}$ puts the dressed photon at \SI{1885}{\per\centi\meter}, \SI{80}{\per\centi\meter} below the free-molecule \ce{C=O} stretch and \SI{90}{\per\centi\meter} below that of the cavity-relaxed geometry (\SI{1974.6}{\per\centi\meter}).
At the field-free  resonance frequency, the two cavity modes still carry photon fractions of $0.976$ and $0.997$ and there is no significant hybridization with the vibrational modes.
Consequently, the resonance condition of the coupled system must be evaluated for the dressed photon: $\omega_c\approx\omega_{\rm matter}/\sqrt{1-\boldsymbol\lambda^{\!\top}\boldsymbol\alpha\boldsymbol\lambda}$.

\begin{figure}
\centering
\includegraphics[width=0.78\linewidth]{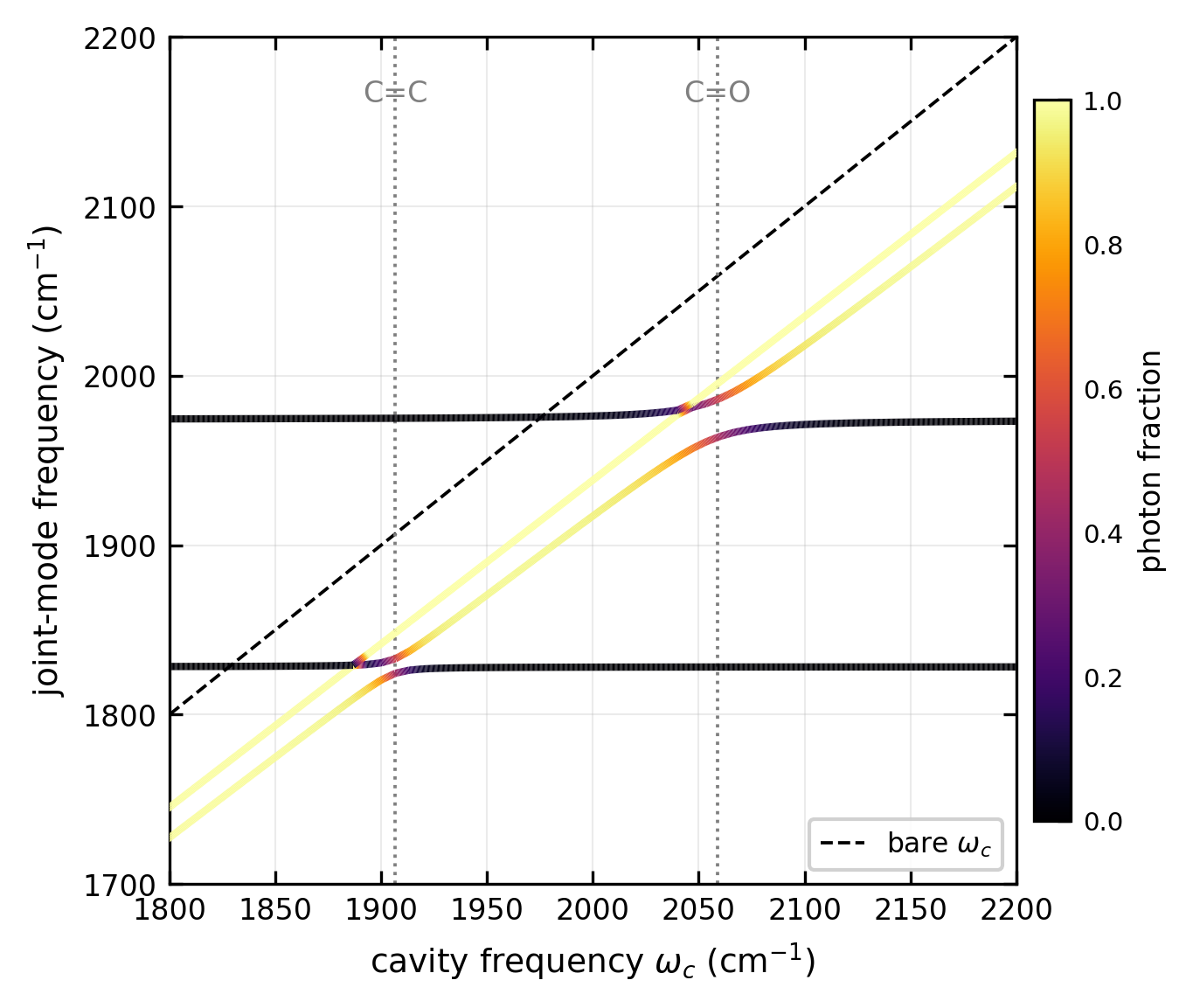}
\caption{Joint-mode frequencies of acrolein as functions of the cavity frequency $\omega_c$, colored by their photon fraction, with the bare $\omega_c$ dashed black. \gls{cbohf}, aug-cc-pVDZ, two cavity modes polarized along $x$ and $z$ at $\lambda=0.05$, scanned across the \gls{vsc} range through the \ce{C=C} and \ce{C=O} resonances. Two photon branches appear, one per polarization, split by the anisotropy of $\boldsymbol\alpha$. Each of the $91$ points is a closed-form rescaling of a single joint Hessian (Eq.~\eqref{eq:hess_rescale}).}
\label{fig:acrolein_dressing}
\end{figure}

One practical way to locate that resonance is to scan $\omega$.
Every geometry was re-optimized a fixed value of $\lambda$ in the presence of both cavity modes.
The detuning scan sweeps $\omega_c$ at fixed geometry, where $\mathrm{q}^{\rm opt}=\langle\hat d\rangle/\omega_c$ follows without a new \gls{scf}.
No joint Hessian here carries an imaginary vibro-polariton frequency.
This is possible because the cavity-relaxed geometry and the \gls{cbohf} density at $\mathrm{q}^{\rm opt}$ do not depend on $\omega_c$, thus the whole scan follows in closed form from one Hessian build (Eq.~\eqref{eq:hess_rescale}).
Rebuilding the Hessian at selected points reproduces the rescaled frequencies to \SI{2e-9}{\per\centi\meter}, and Table~S8 validates the rescaling across different systems.
The $91$ data points of Fig.~\ref{fig:acrolein_dressing} therefore require only
one full evaluation of the Hessian.
This makes such a scan computationally tractable and well resolved to locate the avoided crossing.

Figure~\ref{fig:acrolein_dressing} shows the resulting joint-mode frequencies as functions of $\omega_c$.
The photon-like branches are proportional to $\omega_c$, red-shifted by the \gls{dse} dressing.
They hybridize with every vibrational mode at their point of resonance.
Two avoided crossings are resolved in the window shown, the \ce{C=C} stretch at $\omega_c=\SI{1907}{\per\centi\meter}$ and the \ce{C=O} stretch at \SI{2060}{\per\centi\meter}.
Both lie well above the bare vibrational frequencies involved, \num{1828} and \SI{1975}{\per\centi\meter}, which is the dressing of Eq.~\eqref{eq:omega_eff} made directly visible.
There are two photon branches rather than one because the dressing is anisotropic: the same bare $\omega_c=\SI{2058.8}{\per\centi\meter}$ is shifted to \num{1974.6} and \SI{1993.1}{\per\centi\meter} by $\alpha_{xx}=32.06$ and $\alpha_{zz}=25.12$~a.u., a separation of \SI{18.5}{\per\centi\meter} produced entirely by the \gls{dse} dressing.

\begin{figure}
\centering
\includegraphics[width=0.9\linewidth]{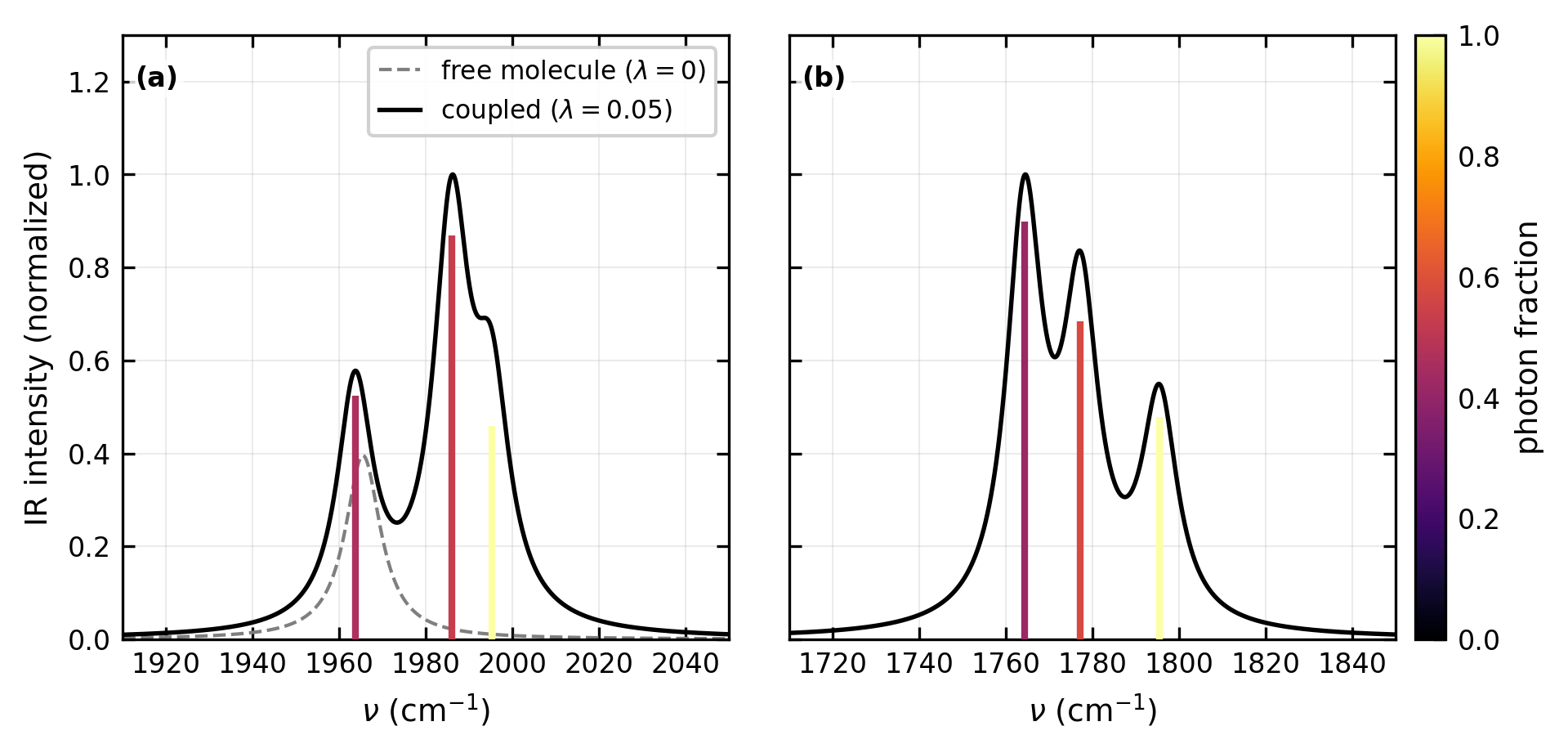}
\caption{Vibro-polaritonic \gls{ir} spectra of acrolein based on (a) \gls{cbohf} and (b) \gls{cbodft}/B3LYP, both
using aug-cc-pVDZ with two cavity modes polarized along $x$ and $z$ at $\lambda=0.05$, each tuned to
its own dressed \ce{C=O} resonance in the \gls{vsc} regime ($\omega_c=\num{2058.8}$ and
\SI{1853.9}{\per\centi\meter}). Lorentzian broadening (\SI{10}{\per\centi\meter} \gls{fwhm}) over
sticks colored by photon fraction. Each panel spans its own \SI{140}{\per\centi\meter} window, of
equal width so that the two splittings compare directly. The gray trace is the cavity-free RHF
spectrum, shown only in the \gls{cbohf} panel (a).}
\label{fig:acrolein_spectra}
\end{figure}

The band of interest is the \ce{C=O} stretch, at \SI{1974.6}{\per\centi\meter} for the cavity-relaxed geometry.
Mixing with the $x$-polarized photon mode splits it into a lower and an upper polariton (\acrshort{lp}, \acrshort{up})\glsunset{lp}\glsunset{up}.
Their separation is smallest, \SI{22.4}{\per\centi\meter}, at $\omega_c=\SI{2060}{\per\centi\meter}$, where both carry about half photon character ($0.441$ and $0.555$).
This minimum approximates the vacuum Rabi splitting $2g$ to be \SI{22.3}{\per\centi\meter} at the dressed resonance of Fig.~\ref{fig:acrolein_spectra}.
Away from resonance the branches separate further only because the photon is detuned, and they revert to one photon-like and one matter-like mode.
A larger splitting, therefore, requires a larger $\lambda$, not a differently tuned cavity.
The dressed-resonance estimate $\omega_c\approx\omega_{\rm matter}/\sqrt{1-\boldsymbol\lambda^{\!\top}\boldsymbol\alpha\boldsymbol\lambda}$ yields \SI{2058.8}{\per\centi\meter}, within one scan grid point (\SI{6.7}{\per\centi\meter}) of this minimum.
The same scan contains a second, much narrower avoided crossing, between the photon and the \ce{C=C} stretch (\SI{1828}{\per\centi\meter}) near $\omega_c=\SI{1907}{\per\centi\meter}$, where the branches are separated by only about \SI{9}{\per\centi\meter}.
Because the grid point at \SI{1906.7}{\per\centi\meter} almost coincides with this crossing, it shares the photon character slightly more evenly ($0.443$ and $0.547$) than any point near the \ce{C=O} crossing.
Choosing $\omega_c$ automatically as the most evenly mixed crossing would therefore select the \ce{C=C} band.
At that frequency the \ce{C=O} band keeps a photon fraction of only $0.005$ and does not couple.

Figure~\ref{fig:acrolein_spectra} compares the two electronic-structure methods \gls{cbohf} and \gls{cbodft}/B3LYP for $\lambda=0.05$, each tuned to its own dressed resonance as described above.
The resonance of the \ce{C=O} stretch mode is at $\omega_c=\SI{2058.8}{\per\centi\meter}$ for \gls{cbohf} and at \SI{1853.9}{\per\centi\meter} for \gls{cbodft}/B3LYP.
Both produce a polariton doublet around the bare \ce{C=O} mode, a \gls{lp} and a \gls{up}: \gls{lp} at \SI{1963.6}{\per\centi\meter} and \gls{up} at \SI{1986.0}{\per\centi\meter}, with photon fractions $0.467$ and $0.530$, for \gls{cbohf}, and \gls{lp} at \SI{1764.3}{\per\centi\meter} and \gls{up} at \SI{1777.2}{\per\centi\meter}, at $0.415$ and $0.577$, for \gls{cbodft} with B3LYP.
Both pairs divide the photon nearly evenly between the two branches, which is what makes the \gls{lp}/\gls{up} labels meaningful here: a single vibrational mode carries essentially all of the coupling, so the cavity mode has exactly one partner and the simple dressed state picture is complete.
The splitting is \SI{22.3}{\per\centi\meter} at the \gls{cbohf} level compared to \SI{12.9}{\per\centi\meter} for B3LYP.
Thus, the two methods disagree on the size of Rabi splitting by nearly a factor of two for identical $\lambda$.
They also disagree on the resonance frequency, by \SI{205}{\per\centi\meter} (\num{2058.8} against \SI{1853.9}{\per\centi\meter}), since both the band position and the polarizability that dresses the mode are method-dependent.

In addition to each pair \gls{lp} / \gls{up} for each vibration mode, there is a third, essentially pure photon mode with a photon fraction of $0.997$: the $z$-polarized cavity mode couples to no vibrational mode in this energy range.
Since acrolein is planar in the $xy$ plane and every relevant vibrational transition dipole coincides with that plane.
This photonic spectator mode carries no matter character, and yet it is not dark.
Its intensity can only be explained by the dipole derivative of the photon coordinate itself,
\begin{equation}
\label{eq:photon_ir}
\frac{\partial\boldsymbol\mu}{\partial q_m} \;=\; \omega_m\,\boldsymbol\alpha\,\boldsymbol\lambda_m ,
\end{equation}
where $\boldsymbol\alpha$ is the static polarizability of the coupled system, see more details in Sec.~S7\,A of the supporting information.
Displacing $q_m$ acts on the molecule like a static electric field $\omega_m\boldsymbol\lambda_m\,\delta q_m$, and the molecule responds with an induced dipole.
The derivative with respect to $q_m$ is linear in coupling and is determined by polarizability rather than by any vibrational transition dipole moment.
For a photon mode that does not mix with any vibrational mode, the normal coordinate is $q_m$ itself, and its infrared intensity, $I\propto|\partial\boldsymbol\mu/\partial Q|^2=\omega_m^2\,|\boldsymbol\alpha\boldsymbol\lambda_m|^2$, grows as $\lambda^2$: the mode is bright although no molecular vibration is excited.
This quadratic scaling law holds as long as $\boldsymbol\alpha$ does not change with coupling.
In a single water molecule, for example, and a far red-detuned infrared cavity, $I/\lambda^2$ changes by approximately $1\%$ up to $\lambda=0.02$ and is reduced by $41\%$ at $\lambda=0.15$ (Table~S9), as \gls{dse} screens the polarizability of the coupled molecule: $|\boldsymbol\alpha\boldsymbol\lambda_m|^2/\lambda^2$ is reduced by $40\%$, and mixing with vibrations contributes less than $2\%$ overall.
The molecular intensity is, meanwhile, conserved: over the \SIrange{1910}{2050}{\per\centi\meter} window of the \gls{cbohf} panel of Fig.~\ref{fig:acrolein_spectra}, weighting each joint mode by its matter character recovers $\sum_i(1-f^{\rm ph}_i)I_i=0.405$ against the free molecular intensity of $0.395$, while the photon-weighted part of the same window contributes a further $0.63$.
The tall, purely photonic, peaks in Fig.~\ref{fig:acrolein_spectra} are therefore a consequence of deliberately amplified coupling rather than molecular absorption peaks.
At experimentally realizable couplings, the photon contribution shrinks quadratically and the spectrum converges to a vibrational spectrum  exhibiting the expected  polariton splitting.

\section{Known limitations}
\label{sec:limitations}

There are several restrictions in current implementation that should be noted.
First, \gls{cbodft} supports LDA, \gls{gga}, global hybrid and range-separated hybrid functionals.
Meta-GGA, double-hybrid, VV10 and dispersion-corrected functionals
do currently not work and raise an error rather than silently omitting the corresponding energy term.
Analytic derivatives are not available for range-separated functionals. Calculations using these functionals raise an error, and \glspl{gga} or global hybrids must be used for analytic gradients.
Similarly, \gls{cbodft} lacks the implementation of an analytic joint Hessian, such
that frequency calculations must use numerical second derivatives based on analytic gradients.
Moreover,  density fitting is not available for \gls{cbodft} gradients and
\gls{cbodft} linear response calculations.
The stability analysis is also restricted to the singlet and photon stability calculations, since the spin-flip exchange-correlation kernel is not implemented.
Second, CBO-UHF provides the \gls{scf} and the analytic gradient only.
Hessians, polarizabilities, stability analysis, geometry optimization, \gls{qedhf} and the CI methods require a closed-shell reference.
Third, the matrix-free \gls{qedfci} solver is used only on request and needs the number of states, and the continuous symmetry measure requires a spherical-harmonic basis set.

\section{Availability, installation and dependencies}
\label{sec:availability}

\texttt{psi4-cbo} requires Python~3.9 or newer.
The reference environment and the continuous integration tests use Python~3.13.
Beyond Psi4 itself, it depends only on NumPy, SciPy, h5py, PyYAML.
The package is distributed under the GPL-3.0-only license.
The repository contains a test suite of more than 1300 tests,
which are run with pytest.
The continuous integration runs the tests and requires a line coverage of at least $90\%$.
Since Psi4 is not distributed via PyPI, it must be provisioned independently, leading to three supported installation routes.
First, the package can be installed via \texttt{pip} into an existing environment that already contains Psi4.
Second, users can deploy a Conda environment from the specification file provided in the repository (\lstinline|conda env create -f environment.yml|), which explicitly pins every dependency, including Psi4.
Third, users can utilize the provided Nix flake, which supplies Psi4 alongside the complete Python stack.

We used this third method for all calculations reported in this work because it guarantees strict reproducibility.
The flake pins not only Psi4, but all build-time dependencies and all
run-time dependencies.
Consequently, a later rebuild uses the same package versions and build recipes rather than relying on the future state of dynamic package repositories.
The reference environment for the reported data comprises Psi4~1.10, Python~3.13.9, NumPy~2.3.4, SciPy~1.16.3, h5py~3.14.0, and OpenBLAS; the Conda specification file pins NumPy~2.4.3.
Concretely, every result in this paper was produced inside a Nix flake pinning NixOS-QChem~\cite{Kowalewski2022-nix} release-25.11 (NixOS-QChem revision \texttt{42055bb} of 2026-01-09, nixpkgs revision \texttt{c8cfcd6} of 2026-01-07), which fixes the Psi4 version and all numerical dependency.
The code and the flake are available at \url{https://gitlab.fysik.su.se/thomas.schnappinger/psi4-cbo}.
This paper describes release \texttt{psi4-cbo-v1.1.0} (2026-09-11).

\section{Summary and Conclusion}
\label{sec:summary}

We have presented \texttt{psi4-cbo}, an open-source package built on PSI4 and Psi4NumPy~\cite{Smith2018-tu,Smith2020-kq}.
It implements the \gls{cboa} treatment of the Pauli-Fierz Hamiltonian~\cite{flick2017atoms,flick2017cavity,Tokatly2013-ho,Ruggenthaler2014-it}, in which the photon displacement coordinate is a variational degree of freedom on the same footing as the nuclear coordinates.
The cavity photon mode then enters the electronic-structure calculation through a harmonic photon potential, a linear light-matter coupling, and the \gls{dse}, allowing for the calculation of energies, forces, force constants, and linear response properties under strong coupling.
The package covers mean-field methods for closed- and open-shell systems, a density-functional variant, configuration-interaction methods for correlation and excited states, and an exact quantum-electrodynamical full-CI reference wherever the system size allows one.
We have implemented analytic gradients, joint nuclear-photon Hessians and geometry optimization under coupling for the closed-shell methods, vibro-polaritonic \gls{ir} and Raman spectra, and diagnostics for stability, symmetry and mode composition.

With the help of \texttt{psi4-cbo} the conditions under which the
\gls{cboa} is valid have been analyzed.
Since \gls{cbohf} and \gls{qedhf} are completely equivalent at the mean-field level, the only deficiency of the \gls{cboa} beyond mean-field is the missing electron-photon correlation ($\varepsilon_{ep}$).
We show that $\varepsilon_{ep}$ remains a small correction to standard electron-electron correlation ($\varepsilon_{ee}$) throughout the \gls{vsc} regime for the systems studied and becomes significant only when electronic-transition frequencies and strong coupling are approached together.
Therefore, quantizing the photon mode merely provides a quantitative correction rather than a qualitative change to the ground state.
To allow routine verification of this limit, we provide built-in stability and wavefunction diagnostics that can assess the \gls{cboa} on a per-system basis.
Moreover, the package provides a flexible interface that allows for further
analysis of the \gls{dse} term contributions.

The acrolein showcase demonstrates the practical capabilities of \texttt{psi4-cbo} and highlights the physical nuance of cavity tuning.
The induced dipole contribution shifts the photon mode its empty resonance frequency and cavity needs to be re-tuned to match the vibrational
transitions.
By using a detuning scan—which only require a full evaluation of a single joint Hessian—we show that tuning is highly method-dependent.
\gls{cbohf} and \gls{cbodft} require different cavity frequencies to achieve resonance in \ce{C=O} band, since the vibrational frequencies are method
dependent. Moreover, method dependent polarizabilities lead to
different shifts in the cavity resonance.
Additionally, the spectra reveal that for high coupling strengths even cavity modes uncoupled to any vibration become infrared-active, as the molecular polarizability gives the photon coordinate its own induced dipole derivative.

The above-mentioned analysis required the \gls{cboa}
and its parametric inclusion of the photon displacement coordinates.
Treating the photon degrees of freedom on the same footing as the nuclear degrees of freedom enabled the re-tooling of well established tools such vibrational
frequency analysis \cite{Schnappinger2023-cbohf,Schnappinger2023-wp,Schnappinger2024-rz,Schnappinger2024-vt,Schnappinger2025-fx}.
The analytic joint Hessian serves as the computational tool that finally brings vibro-polaritonic spectra from of highly idealized toy models to more realistic molecular systems.
As the energy, density, and gradient at the optimal photon coordinate do not depend on the bare cavity frequency, it yields entire dispersion curves from a single Hessian build and a closed-form rescaling.
Paired with continuous symmetry analysis~\cite{Zabrodsky1992-csm} and matrix-free response solvers~\cite{Hestenes1952-cg}, this implementation makes routine \textit{ab initio} calculations of cavity-coupled molecules computationally viable.
Several methodological extensions naturally follow from the foundation laid here, and from the boundaries recorded in Sections~\ref{sec:scaling} and~\ref{sec:limitations}.
By leaving the mean-field description, follow-up investigations of electronic excited states using \gls{cboci} or extensions towards cavity Born-Oppenheimer coupled cluster~\cite{Angelico2023-jh,Fischer2025} are planned next steps.
Further developments must also realize analytic joint Hessians for the density-functional route (\gls{cbodft}) and investigate possibilities to incorporate cavity dissipation.
Since every cavity modelled here is lossless, extending the framework to include finite-$Q$ effects~\cite{Fetherolf2026-dc} is the essential prerequisite for directly comparing computed spectral line shapes with experimental line shapes.
Finally, the defining challenge of the field remains scale.
The calculations presented here use artificially large per-molecule couplings to enhance the physical effects and study them.
Real \gls{vsc} experiments~\cite{Ebbesen2023-fd,Fregoni2022-op}, however, operate in the opposite regime: macroscopic ensembles, minuscule single-molecule couplings, and dominant collective responses.
Bridging this gap requires a polarizable QM/MM embedding model.
Because the \gls{dse} strictly requires polarizability to respond to the field, standard fixed-charge environments will inevitably fail.
An advanced embedding scheme would allow a fully \textit{ab initio} quantum region to feel the collective macroscopic coupling of the surrounding ensemble without paying an impossible computational cost.
Delivering analytical gradients for such an embedded setup will allow for
ab initio dynamics for the condensed-phase systems where polaritonic chemistry actually happens~\cite{Thomas2019-ve,Vergauwe2019-wt,Gu2023-uq,Thomas2025-we}.

By proving that the photon coordinate can be seamlessly and rigorously integrated into standard analytic derivative and matrix-free response frameworks, the implementation presented here not only solves the isolated molecule problem, but provides the exact computational scaffold needed to finally bring correlated, embedded, and dynamic polaritonic chemistry to the macroscopic scale.

\section*{Acknowledgement}
The authors thank Eric Fischer and Michael Ruggenthaler for inspiring discussions and helpful comments. This project has received funding from the Swedish Research Council (Grant No.~VR 2024-04299).

\section*{Supporting Information}
Tables S1--S9, Eq.~(S2), and Secs.~S7\,A and S9 cited in the main text are available in the supporting information. This document provides comprehensive theoretical derivations, algorithmic details, and extended numerical analyses. Specifically, it includes the numerical verification of the \gls{cbohf} and \gls{qedhf} identity, the atomic-orbital (AO) level working equations, and the proof and numerical validation of the Hessian rescaling law.


\begin{thebibliography}{76}%
\makeatletter
\providecommand \@ifxundefined [1]{%
 \@ifx{#1\undefined}
}%
\providecommand \@ifnum [1]{%
 \ifnum #1\expandafter \@firstoftwo
 \else \expandafter \@secondoftwo
 \fi
}%
\providecommand \@ifx [1]{%
 \ifx #1\expandafter \@firstoftwo
 \else \expandafter \@secondoftwo
 \fi
}%
\providecommand \natexlab [1]{#1}%
\providecommand \enquote  [1]{``#1''}%
\providecommand \bibnamefont  [1]{#1}%
\providecommand \bibfnamefont [1]{#1}%
\providecommand \citenamefont [1]{#1}%
\providecommand \href@noop [0]{\@secondoftwo}%
\providecommand \href [0]{\begingroup \@sanitize@url \@href}%
\providecommand \@href[1]{\@@startlink{#1}\@@href}%
\providecommand \@@href[1]{\endgroup#1\@@endlink}%
\providecommand \@sanitize@url [0]{\catcode `\\12\catcode `\$12\catcode
  `\&12\catcode `\#12\catcode `\^12\catcode `\_12\catcode `\%12\relax}%
\providecommand \@@startlink[1]{}%
\providecommand \@@endlink[0]{}%
\providecommand \url  [0]{\begingroup\@sanitize@url \@url }%
\providecommand \@url [1]{\endgroup\@href {#1}{\urlprefix }}%
\providecommand \urlprefix  [0]{URL }%
\providecommand \Eprint [0]{\href }%
\providecommand \doibase [0]{https://doi.org/}%
\providecommand \selectlanguage [0]{\@gobble}%
\providecommand \bibinfo  [0]{\@secondoftwo}%
\providecommand \bibfield  [0]{\@secondoftwo}%
\providecommand \translation [1]{[#1]}%
\providecommand \BibitemOpen [0]{}%
\providecommand \bibitemStop [0]{}%
\providecommand \bibitemNoStop [0]{.\EOS\space}%
\providecommand \EOS [0]{\spacefactor3000\relax}%
\providecommand \BibitemShut  [1]{\csname bibitem#1\endcsname}%
\let\auto@bib@innerbib\@empty
\bibitem [{\citenamefont {Ebbesen}\ \emph {et~al.}(2023)\citenamefont
  {Ebbesen}, \citenamefont {Rubio},\ and\ \citenamefont
  {Scholes}}]{Ebbesen2023-fd}%
  \BibitemOpen
  \bibfield  {author} {\bibinfo {author} {\bibfnamefont {T.~W.}\ \bibnamefont
  {Ebbesen}}, \bibinfo {author} {\bibfnamefont {A.}~\bibnamefont {Rubio}},\
  and\ \bibinfo {author} {\bibfnamefont {G.~D.}\ \bibnamefont {Scholes}},\
  }\href {https://doi.org/10.1021/acs.chemrev.3c00637} {\bibfield  {journal}
  {\bibinfo  {journal} {Chem. Rev.}\ }\textbf {\bibinfo {volume} {123}},\
  \bibinfo {pages} {12037} (\bibinfo {year} {2023})}\BibitemShut {NoStop}%
\bibitem [{\citenamefont {Fregoni}\ \emph {et~al.}(2022)\citenamefont
  {Fregoni}, \citenamefont {Garcia-Vidal},\ and\ \citenamefont
  {Feist}}]{Fregoni2022-op}%
  \BibitemOpen
  \bibfield  {author} {\bibinfo {author} {\bibfnamefont {J.}~\bibnamefont
  {Fregoni}}, \bibinfo {author} {\bibfnamefont {F.~J.}\ \bibnamefont
  {Garcia-Vidal}},\ and\ \bibinfo {author} {\bibfnamefont {J.}~\bibnamefont
  {Feist}},\ }\href {https://doi.org/10.1021/acsphotonics.1c01749} {\bibfield
  {journal} {\bibinfo  {journal} {ACS Photonics}\ }\textbf {\bibinfo {volume}
  {9}},\ \bibinfo {pages} {1096} (\bibinfo {year} {2022})}\BibitemShut
  {NoStop}%
\bibitem [{\citenamefont {Thomas}\ \emph {et~al.}(2019)\citenamefont {Thomas},
  \citenamefont {Lethuillier-Karl}, \citenamefont {Nagarajan}, \citenamefont
  {Vergauwe}, \citenamefont {George}, \citenamefont {Chervy}, \citenamefont
  {Shalabney}, \citenamefont {Devaux}, \citenamefont {Genet}, \citenamefont
  {Moran},\ and\ \citenamefont {Ebbesen}}]{Thomas2019-ve}%
  \BibitemOpen
  \bibfield  {author} {\bibinfo {author} {\bibfnamefont {A.}~\bibnamefont
  {Thomas}}, \bibinfo {author} {\bibfnamefont {L.}~\bibnamefont
  {Lethuillier-Karl}}, \bibinfo {author} {\bibfnamefont {K.}~\bibnamefont
  {Nagarajan}}, \bibinfo {author} {\bibfnamefont {R.~M.~A.}\ \bibnamefont
  {Vergauwe}}, \bibinfo {author} {\bibfnamefont {J.}~\bibnamefont {George}},
  \bibinfo {author} {\bibfnamefont {T.}~\bibnamefont {Chervy}}, \bibinfo
  {author} {\bibfnamefont {A.}~\bibnamefont {Shalabney}}, \bibinfo {author}
  {\bibfnamefont {E.}~\bibnamefont {Devaux}}, \bibinfo {author} {\bibfnamefont
  {C.}~\bibnamefont {Genet}}, \bibinfo {author} {\bibfnamefont
  {J.}~\bibnamefont {Moran}},\ and\ \bibinfo {author} {\bibfnamefont {T.~W.}\
  \bibnamefont {Ebbesen}},\ }\href {https://doi.org/10.1126/science.aau7742}
  {\bibfield  {journal} {\bibinfo  {journal} {Science}\ }\textbf {\bibinfo
  {volume} {363}},\ \bibinfo {pages} {615} (\bibinfo {year}
  {2019})}\BibitemShut {NoStop}%
\bibitem [{\citenamefont {Vergauwe}\ \emph {et~al.}(2019)\citenamefont
  {Vergauwe}, \citenamefont {Thomas}, \citenamefont {Nagarajan}, \citenamefont
  {Shalabney}, \citenamefont {George}, \citenamefont {Chervy}, \citenamefont
  {Seidel}, \citenamefont {Devaux}, \citenamefont {Torbeev},\ and\
  \citenamefont {Ebbesen}}]{Vergauwe2019-wt}%
  \BibitemOpen
  \bibfield  {author} {\bibinfo {author} {\bibfnamefont {R.~M.~A.}\
  \bibnamefont {Vergauwe}}, \bibinfo {author} {\bibfnamefont {A.}~\bibnamefont
  {Thomas}}, \bibinfo {author} {\bibfnamefont {K.}~\bibnamefont {Nagarajan}},
  \bibinfo {author} {\bibfnamefont {A.}~\bibnamefont {Shalabney}}, \bibinfo
  {author} {\bibfnamefont {J.}~\bibnamefont {George}}, \bibinfo {author}
  {\bibfnamefont {T.}~\bibnamefont {Chervy}}, \bibinfo {author} {\bibfnamefont
  {M.}~\bibnamefont {Seidel}}, \bibinfo {author} {\bibfnamefont
  {E.}~\bibnamefont {Devaux}}, \bibinfo {author} {\bibfnamefont
  {V.}~\bibnamefont {Torbeev}},\ and\ \bibinfo {author} {\bibfnamefont {T.~W.}\
  \bibnamefont {Ebbesen}},\ }\href {https://doi.org/10.1002/anie.201908876}
  {\bibfield  {journal} {\bibinfo  {journal} {Angew. Chem. Int. Ed Engl.}\
  }\textbf {\bibinfo {volume} {58}},\ \bibinfo {pages} {15324} (\bibinfo {year}
  {2019})}\BibitemShut {NoStop}%
\bibitem [{\citenamefont {Sau}\ \emph {et~al.}(2021)\citenamefont {Sau},
  \citenamefont {Nagarajan}, \citenamefont {Patrahau}, \citenamefont
  {Lethuillier-Karl}, \citenamefont {Vergauwe}, \citenamefont {Thomas},
  \citenamefont {Moran}, \citenamefont {Genet},\ and\ \citenamefont
  {Ebbesen}}]{Sau2021-vb}%
  \BibitemOpen
  \bibfield  {author} {\bibinfo {author} {\bibfnamefont {A.}~\bibnamefont
  {Sau}}, \bibinfo {author} {\bibfnamefont {K.}~\bibnamefont {Nagarajan}},
  \bibinfo {author} {\bibfnamefont {B.}~\bibnamefont {Patrahau}}, \bibinfo
  {author} {\bibfnamefont {L.}~\bibnamefont {Lethuillier-Karl}}, \bibinfo
  {author} {\bibfnamefont {R.~M.~A.}\ \bibnamefont {Vergauwe}}, \bibinfo
  {author} {\bibfnamefont {A.}~\bibnamefont {Thomas}}, \bibinfo {author}
  {\bibfnamefont {J.}~\bibnamefont {Moran}}, \bibinfo {author} {\bibfnamefont
  {C.}~\bibnamefont {Genet}},\ and\ \bibinfo {author} {\bibfnamefont {T.~W.}\
  \bibnamefont {Ebbesen}},\ }\href {https://doi.org/10.1002/anie.202013465}
  {\bibfield  {journal} {\bibinfo  {journal} {Angew. Chem. Int. Ed Engl.}\
  }\textbf {\bibinfo {volume} {60}},\ \bibinfo {pages} {5712} (\bibinfo {year}
  {2021})}\BibitemShut {NoStop}%
\bibitem [{\citenamefont {Kumar}\ \emph {et~al.}(2024)\citenamefont {Kumar},
  \citenamefont {Biswas}, \citenamefont {Rashid}, \citenamefont {Mony},
  \citenamefont {Chandrasekharan}, \citenamefont {Mattiotti}, \citenamefont
  {Vergauwe}, \citenamefont {Hagenmuller}, \citenamefont {Kaliginedi},\ and\
  \citenamefont {Thomas}}]{Kumar2024-th}%
  \BibitemOpen
  \bibfield  {author} {\bibinfo {author} {\bibfnamefont {S.}~\bibnamefont
  {Kumar}}, \bibinfo {author} {\bibfnamefont {S.}~\bibnamefont {Biswas}},
  \bibinfo {author} {\bibfnamefont {U.}~\bibnamefont {Rashid}}, \bibinfo
  {author} {\bibfnamefont {K.~S.}\ \bibnamefont {Mony}}, \bibinfo {author}
  {\bibfnamefont {G.}~\bibnamefont {Chandrasekharan}}, \bibinfo {author}
  {\bibfnamefont {F.}~\bibnamefont {Mattiotti}}, \bibinfo {author}
  {\bibfnamefont {R.~M.~A.}\ \bibnamefont {Vergauwe}}, \bibinfo {author}
  {\bibfnamefont {D.}~\bibnamefont {Hagenmuller}}, \bibinfo {author}
  {\bibfnamefont {V.}~\bibnamefont {Kaliginedi}},\ and\ \bibinfo {author}
  {\bibfnamefont {A.}~\bibnamefont {Thomas}},\ }\bibfield  {journal} {\bibinfo
  {journal} {J. Am. Chem. Soc.}\ }\href {https://doi.org/10.1021/jacs.4c03016}
  {10.1021/jacs.4c03016} (\bibinfo {year} {2024})\BibitemShut {NoStop}%
\bibitem [{\citenamefont {Thomas}\ \emph {et~al.}(2025)\citenamefont {Thomas},
  \citenamefont {Devaux}, \citenamefont {Nagarajan}, \citenamefont {Chervy},
  \citenamefont {Seidel}, \citenamefont {Rogez}, \citenamefont {Robert},
  \citenamefont {Drillon}, \citenamefont {Ruan}, \citenamefont
  {Schlittenhardt}, \citenamefont {Ruben}, \citenamefont {Hagenmüller},
  \citenamefont {Schütz}, \citenamefont {Schachenmayer}, \citenamefont
  {Genet}, \citenamefont {Pupillo},\ and\ \citenamefont
  {Ebbesen}}]{Thomas2025-we}%
  \BibitemOpen
  \bibfield  {author} {\bibinfo {author} {\bibfnamefont {A.}~\bibnamefont
  {Thomas}}, \bibinfo {author} {\bibfnamefont {E.}~\bibnamefont {Devaux}},
  \bibinfo {author} {\bibfnamefont {K.}~\bibnamefont {Nagarajan}}, \bibinfo
  {author} {\bibfnamefont {T.}~\bibnamefont {Chervy}}, \bibinfo {author}
  {\bibfnamefont {M.}~\bibnamefont {Seidel}}, \bibinfo {author} {\bibfnamefont
  {G.}~\bibnamefont {Rogez}}, \bibinfo {author} {\bibfnamefont
  {J.}~\bibnamefont {Robert}}, \bibinfo {author} {\bibfnamefont
  {M.}~\bibnamefont {Drillon}}, \bibinfo {author} {\bibfnamefont {T.~T.}\
  \bibnamefont {Ruan}}, \bibinfo {author} {\bibfnamefont {S.}~\bibnamefont
  {Schlittenhardt}}, \bibinfo {author} {\bibfnamefont {M.}~\bibnamefont
  {Ruben}}, \bibinfo {author} {\bibfnamefont {D.}~\bibnamefont {Hagenmüller}},
  \bibinfo {author} {\bibfnamefont {S.}~\bibnamefont {Schütz}}, \bibinfo
  {author} {\bibfnamefont {J.}~\bibnamefont {Schachenmayer}}, \bibinfo {author}
  {\bibfnamefont {C.}~\bibnamefont {Genet}}, \bibinfo {author} {\bibfnamefont
  {G.}~\bibnamefont {Pupillo}},\ and\ \bibinfo {author} {\bibfnamefont {T.~W.}\
  \bibnamefont {Ebbesen}},\ }\href {https://doi.org/10.1063/5.0231202}
  {\bibfield  {journal} {\bibinfo  {journal} {J. Chem. Phys.}\ }\textbf
  {\bibinfo {volume} {162}},\ \bibinfo {pages} {134701} (\bibinfo {year}
  {2025})}\BibitemShut {NoStop}%
\bibitem [{\citenamefont {Joseph}\ \emph {et~al.}(2021)\citenamefont {Joseph},
  \citenamefont {Kushida}, \citenamefont {Smarsly}, \citenamefont {Ihiawakrim},
  \citenamefont {Thomas}, \citenamefont {Paravicini-Bagliani}, \citenamefont
  {Nagarajan}, \citenamefont {Vergauwe}, \citenamefont {Devaux}, \citenamefont
  {Ersen}, \citenamefont {Bunz},\ and\ \citenamefont
  {Ebbesen}}]{Joseph2021-sa}%
  \BibitemOpen
  \bibfield  {author} {\bibinfo {author} {\bibfnamefont {K.}~\bibnamefont
  {Joseph}}, \bibinfo {author} {\bibfnamefont {S.}~\bibnamefont {Kushida}},
  \bibinfo {author} {\bibfnamefont {E.}~\bibnamefont {Smarsly}}, \bibinfo
  {author} {\bibfnamefont {D.}~\bibnamefont {Ihiawakrim}}, \bibinfo {author}
  {\bibfnamefont {A.}~\bibnamefont {Thomas}}, \bibinfo {author} {\bibfnamefont
  {G.~L.}\ \bibnamefont {Paravicini-Bagliani}}, \bibinfo {author}
  {\bibfnamefont {K.}~\bibnamefont {Nagarajan}}, \bibinfo {author}
  {\bibfnamefont {R.}~\bibnamefont {Vergauwe}}, \bibinfo {author}
  {\bibfnamefont {E.}~\bibnamefont {Devaux}}, \bibinfo {author} {\bibfnamefont
  {O.}~\bibnamefont {Ersen}}, \bibinfo {author} {\bibfnamefont {U.~H.~F.}\
  \bibnamefont {Bunz}},\ and\ \bibinfo {author} {\bibfnamefont {T.~W.}\
  \bibnamefont {Ebbesen}},\ }\href@noop {} {\bibfield  {journal} {\bibinfo
  {journal} {Angew. Chem. Int. Ed.}\ }\textbf {\bibinfo {volume} {60}},\
  \bibinfo {pages} {19665} (\bibinfo {year} {2021})}\BibitemShut {NoStop}%
\bibitem [{\citenamefont {Sandeep}\ \emph {et~al.}(2022)\citenamefont
  {Sandeep}, \citenamefont {Joseph}, \citenamefont {Gautier}, \citenamefont
  {Nagarajan}, \citenamefont {Sujith}, \citenamefont {Thomas},\ and\
  \citenamefont {Ebbesen}}]{Sandeep2022-sa}%
  \BibitemOpen
  \bibfield  {author} {\bibinfo {author} {\bibfnamefont {K.}~\bibnamefont
  {Sandeep}}, \bibinfo {author} {\bibfnamefont {K.}~\bibnamefont {Joseph}},
  \bibinfo {author} {\bibfnamefont {J.}~\bibnamefont {Gautier}}, \bibinfo
  {author} {\bibfnamefont {K.}~\bibnamefont {Nagarajan}}, \bibinfo {author}
  {\bibfnamefont {M.}~\bibnamefont {Sujith}}, \bibinfo {author} {\bibfnamefont
  {K.~G.}\ \bibnamefont {Thomas}},\ and\ \bibinfo {author} {\bibfnamefont
  {T.~W.}\ \bibnamefont {Ebbesen}},\ }\href@noop {} {\bibfield  {journal}
  {\bibinfo  {journal} {J. Phys. Chem. Lett.}\ }\textbf {\bibinfo {volume}
  {13}},\ \bibinfo {pages} {1209} (\bibinfo {year} {2022})}\BibitemShut
  {NoStop}%
\bibitem [{\citenamefont {Joseph}\ \emph {et~al.}(2024)\citenamefont {Joseph},
  \citenamefont {De~Waal}, \citenamefont {Jansen}, \citenamefont {Van Der~Tol},
  \citenamefont {Vantomme},\ and\ \citenamefont {Meijer}}]{Joseph2024-sp}%
  \BibitemOpen
  \bibfield  {author} {\bibinfo {author} {\bibfnamefont {K.}~\bibnamefont
  {Joseph}}, \bibinfo {author} {\bibfnamefont {B.}~\bibnamefont {De~Waal}},
  \bibinfo {author} {\bibfnamefont {S.~A.~H.}\ \bibnamefont {Jansen}}, \bibinfo
  {author} {\bibfnamefont {J.~J.~B.}\ \bibnamefont {Van Der~Tol}}, \bibinfo
  {author} {\bibfnamefont {G.}~\bibnamefont {Vantomme}},\ and\ \bibinfo
  {author} {\bibfnamefont {E.~W.}\ \bibnamefont {Meijer}},\ }\href@noop {}
  {\bibfield  {journal} {\bibinfo  {journal} {J. Am. Chem. Soc.}\ }\textbf
  {\bibinfo {volume} {146}},\ \bibinfo {pages} {12130} (\bibinfo {year}
  {2024})}\BibitemShut {NoStop}%
\bibitem [{\citenamefont {Zhong}\ \emph {et~al.}(2023)\citenamefont {Zhong},
  \citenamefont {Hou}, \citenamefont {Zhao}, \citenamefont {Bai}, \citenamefont
  {Wang}, \citenamefont {Gao}, \citenamefont {Guo},\ and\ \citenamefont
  {Zhang}}]{Zhong2023-do}%
  \BibitemOpen
  \bibfield  {author} {\bibinfo {author} {\bibfnamefont {C.}~\bibnamefont
  {Zhong}}, \bibinfo {author} {\bibfnamefont {S.}~\bibnamefont {Hou}}, \bibinfo
  {author} {\bibfnamefont {X.}~\bibnamefont {Zhao}}, \bibinfo {author}
  {\bibfnamefont {J.}~\bibnamefont {Bai}}, \bibinfo {author} {\bibfnamefont
  {Z.}~\bibnamefont {Wang}}, \bibinfo {author} {\bibfnamefont {F.}~\bibnamefont
  {Gao}}, \bibinfo {author} {\bibfnamefont {J.}~\bibnamefont {Guo}},\ and\
  \bibinfo {author} {\bibfnamefont {F.}~\bibnamefont {Zhang}},\ }\href@noop {}
  {\bibfield  {journal} {\bibinfo  {journal} {ACS Photonics}\ }\textbf
  {\bibinfo {volume} {10}},\ \bibinfo {pages} {1618} (\bibinfo {year}
  {2023})}\BibitemShut {NoStop}%
\bibitem [{\citenamefont {Biswas}\ \emph {et~al.}(2025)\citenamefont {Biswas},
  \citenamefont {Mondal}, \citenamefont {Chandrasekharan}, \citenamefont
  {Mony}, \citenamefont {Singh},\ and\ \citenamefont {Thomas}}]{Biswas2025-in}%
  \BibitemOpen
  \bibfield  {author} {\bibinfo {author} {\bibfnamefont {S.}~\bibnamefont
  {Biswas}}, \bibinfo {author} {\bibfnamefont {M.}~\bibnamefont {Mondal}},
  \bibinfo {author} {\bibfnamefont {G.}~\bibnamefont {Chandrasekharan}},
  \bibinfo {author} {\bibfnamefont {K.~S.}\ \bibnamefont {Mony}}, \bibinfo
  {author} {\bibfnamefont {A.}~\bibnamefont {Singh}},\ and\ \bibinfo {author}
  {\bibfnamefont {A.}~\bibnamefont {Thomas}},\ }\href@noop {} {\bibfield
  {journal} {\bibinfo  {journal} {Nat. Commun.}\ }\textbf {\bibinfo {volume}
  {16}},\ \bibinfo {pages} {5115} (\bibinfo {year} {2025})}\BibitemShut
  {NoStop}%
\bibitem [{\citenamefont {Imperatore}\ \emph {et~al.}(2021)\citenamefont
  {Imperatore}, \citenamefont {Asbury},\ and\ \citenamefont
  {Giebink}}]{Imperatore2021-rv}%
  \BibitemOpen
  \bibfield  {author} {\bibinfo {author} {\bibfnamefont {M.~V.}\ \bibnamefont
  {Imperatore}}, \bibinfo {author} {\bibfnamefont {J.~B.}\ \bibnamefont
  {Asbury}},\ and\ \bibinfo {author} {\bibfnamefont {N.~C.}\ \bibnamefont
  {Giebink}},\ }\href {https://doi.org/10.1063/5.0046307} {\bibfield  {journal}
  {\bibinfo  {journal} {J. Chem. Phys.}\ }\textbf {\bibinfo {volume} {154}},\
  \bibinfo {pages} {191103} (\bibinfo {year} {2021})}\BibitemShut {NoStop}%
\bibitem [{\citenamefont {Wiesehan}\ and\ \citenamefont
  {Xiong}(2021)}]{Wiesehan2021-yn}%
  \BibitemOpen
  \bibfield  {author} {\bibinfo {author} {\bibfnamefont {G.~D.}\ \bibnamefont
  {Wiesehan}}\ and\ \bibinfo {author} {\bibfnamefont {W.}~\bibnamefont
  {Xiong}},\ }\href {https://doi.org/10.1063/5.0077549} {\bibfield  {journal}
  {\bibinfo  {journal} {J. Chem. Phys.}\ }\textbf {\bibinfo {volume} {155}},\
  \bibinfo {pages} {241103} (\bibinfo {year} {2021})}\BibitemShut {NoStop}%
\bibitem [{\citenamefont {Flick}\ \emph
  {et~al.}(2017{\natexlab{a}})\citenamefont {Flick}, \citenamefont
  {Ruggenthaler}, \citenamefont {Appel},\ and\ \citenamefont
  {Rubio}}]{flick2017atoms}%
  \BibitemOpen
  \bibfield  {author} {\bibinfo {author} {\bibfnamefont {J.}~\bibnamefont
  {Flick}}, \bibinfo {author} {\bibfnamefont {M.}~\bibnamefont {Ruggenthaler}},
  \bibinfo {author} {\bibfnamefont {H.}~\bibnamefont {Appel}},\ and\ \bibinfo
  {author} {\bibfnamefont {A.}~\bibnamefont {Rubio}},\ }\href@noop {}
  {\bibfield  {journal} {\bibinfo  {journal} {Proc. Natl. Acad. Sci. U.S.A.}\
  }\textbf {\bibinfo {volume} {114}},\ \bibinfo {pages} {3026} (\bibinfo {year}
  {2017}{\natexlab{a}})}\BibitemShut {NoStop}%
\bibitem [{\citenamefont {Flick}\ \emph
  {et~al.}(2017{\natexlab{b}})\citenamefont {Flick}, \citenamefont {Appel},
  \citenamefont {Ruggenthaler},\ and\ \citenamefont {Rubio}}]{flick2017cavity}%
  \BibitemOpen
  \bibfield  {author} {\bibinfo {author} {\bibfnamefont {J.}~\bibnamefont
  {Flick}}, \bibinfo {author} {\bibfnamefont {H.}~\bibnamefont {Appel}},
  \bibinfo {author} {\bibfnamefont {M.}~\bibnamefont {Ruggenthaler}},\ and\
  \bibinfo {author} {\bibfnamefont {A.}~\bibnamefont {Rubio}},\ }\href@noop {}
  {\bibfield  {journal} {\bibinfo  {journal} {J. Chem. Theory Comput.}\
  }\textbf {\bibinfo {volume} {13}},\ \bibinfo {pages} {1616} (\bibinfo {year}
  {2017}{\natexlab{b}})}\BibitemShut {NoStop}%
\bibitem [{\citenamefont {Tokatly}(2013)}]{Tokatly2013-ho}%
  \BibitemOpen
  \bibfield  {author} {\bibinfo {author} {\bibfnamefont {I.~V.}\ \bibnamefont
  {Tokatly}},\ }\href {https://doi.org/10.1103/PhysRevLett.110.233001}
  {\bibfield  {journal} {\bibinfo  {journal} {Phys. Rev. Lett.}\ }\textbf
  {\bibinfo {volume} {110}},\ \bibinfo {pages} {233001} (\bibinfo {year}
  {2013})}\BibitemShut {NoStop}%
\bibitem [{\citenamefont {Ruggenthaler}\ \emph {et~al.}(2014)\citenamefont
  {Ruggenthaler}, \citenamefont {Flick}, \citenamefont {Pellegrini},
  \citenamefont {Appel}, \citenamefont {Tokatly},\ and\ \citenamefont
  {Rubio}}]{Ruggenthaler2014-it}%
  \BibitemOpen
  \bibfield  {author} {\bibinfo {author} {\bibfnamefont {M.}~\bibnamefont
  {Ruggenthaler}}, \bibinfo {author} {\bibfnamefont {J.}~\bibnamefont {Flick}},
  \bibinfo {author} {\bibfnamefont {C.}~\bibnamefont {Pellegrini}}, \bibinfo
  {author} {\bibfnamefont {H.}~\bibnamefont {Appel}}, \bibinfo {author}
  {\bibfnamefont {I.~V.}\ \bibnamefont {Tokatly}},\ and\ \bibinfo {author}
  {\bibfnamefont {A.}~\bibnamefont {Rubio}},\ }\href
  {https://doi.org/10.1103/PhysRevA.90.012508} {\bibfield  {journal} {\bibinfo
  {journal} {Phys. Rev. A}\ }\textbf {\bibinfo {volume} {90}},\ \bibinfo
  {pages} {012508} (\bibinfo {year} {2014})}\BibitemShut {NoStop}%
\bibitem [{\citenamefont {Sidler}\ \emph {et~al.}(2022)\citenamefont {Sidler},
  \citenamefont {Ruggenthaler}, \citenamefont {Sch{\"a}fer}, \citenamefont
  {Ronca},\ and\ \citenamefont {Rubio}}]{Sidler2022-cg}%
  \BibitemOpen
  \bibfield  {author} {\bibinfo {author} {\bibfnamefont {D.}~\bibnamefont
  {Sidler}}, \bibinfo {author} {\bibfnamefont {M.}~\bibnamefont
  {Ruggenthaler}}, \bibinfo {author} {\bibfnamefont {C.}~\bibnamefont
  {Sch{\"a}fer}}, \bibinfo {author} {\bibfnamefont {E.}~\bibnamefont {Ronca}},\
  and\ \bibinfo {author} {\bibfnamefont {A.}~\bibnamefont {Rubio}},\ }\href
  {https://doi.org/10.1063/5.0094956} {\bibfield  {journal} {\bibinfo
  {journal} {J. Chem. Phys.}\ }\textbf {\bibinfo {volume} {156}},\ \bibinfo
  {pages} {230901} (\bibinfo {year} {2022})}\BibitemShut {NoStop}%
\bibitem [{\citenamefont {Foley}\ \emph {et~al.}(2023)\citenamefont {Foley},
  \citenamefont {McTague},\ and\ \citenamefont {DePrince}}]{Foley2023-pq}%
  \BibitemOpen
  \bibfield  {author} {\bibinfo {author} {\bibfnamefont {J.~J.}\ \bibnamefont
  {Foley}}, \bibinfo {author} {\bibfnamefont {J.~F.}\ \bibnamefont {McTague}},\
  and\ \bibinfo {author} {\bibfnamefont {A.~E.}\ \bibnamefont {DePrince}},\
  }\bibfield  {journal} {\bibinfo  {journal} {Chem. Phys. Rev.}\ }\textbf
  {\bibinfo {volume} {4}},\ \href {https://doi.org/10.1063/5.0167243}
  {10.1063/5.0167243} (\bibinfo {year} {2023})\BibitemShut {NoStop}%
\bibitem [{\citenamefont {Sánchez-Barquilla}\ \emph
  {et~al.}(2022)\citenamefont {Sánchez-Barquilla}, \citenamefont
  {Fernández-Domínguez}, \citenamefont {Feist},\ and\ \citenamefont
  {García-Vidal}}]{Sanchez-Barquilla2022-dq}%
  \BibitemOpen
  \bibfield  {author} {\bibinfo {author} {\bibfnamefont {M.}~\bibnamefont
  {Sánchez-Barquilla}}, \bibinfo {author} {\bibfnamefont {A.~I.}\ \bibnamefont
  {Fernández-Domínguez}}, \bibinfo {author} {\bibfnamefont {J.}~\bibnamefont
  {Feist}},\ and\ \bibinfo {author} {\bibfnamefont {F.~J.}\ \bibnamefont
  {García-Vidal}},\ }\bibfield  {journal} {\bibinfo  {journal} {ACS
  Photonics}\ }\href {https://doi.org/10.1021/acsphotonics.2c00048}
  {10.1021/acsphotonics.2c00048} (\bibinfo {year} {2022})\BibitemShut {NoStop}%
\bibitem [{\citenamefont {Fiechter}\ and\ \citenamefont
  {Richardson}(2024)}]{Fiechter2024-vw}%
  \BibitemOpen
  \bibfield  {author} {\bibinfo {author} {\bibfnamefont {M.~R.}\ \bibnamefont
  {Fiechter}}\ and\ \bibinfo {author} {\bibfnamefont {J.~O.}\ \bibnamefont
  {Richardson}},\ }\bibfield  {journal} {\bibinfo  {journal} {J. Chem. Phys.}\
  }\textbf {\bibinfo {volume} {160}},\ \href
  {https://doi.org/10.1063/5.0197248} {10.1063/5.0197248} (\bibinfo {year}
  {2024})\BibitemShut {NoStop}%
\bibitem [{\citenamefont {Fischer}\ and\ \citenamefont
  {Saalfrank}(2023)}]{Fischer2023-ve}%
  \BibitemOpen
  \bibfield  {author} {\bibinfo {author} {\bibfnamefont {E.~W.}\ \bibnamefont
  {Fischer}}\ and\ \bibinfo {author} {\bibfnamefont {P.}~\bibnamefont
  {Saalfrank}},\ }\href {https://doi.org/10.1021/acs.jctc.3c00708} {\bibfield
  {journal} {\bibinfo  {journal} {J. Chem. Theory Comput.}\ }\textbf {\bibinfo
  {volume} {19}},\ \bibinfo {pages} {7215} (\bibinfo {year}
  {2023})}\BibitemShut {NoStop}%
\bibitem [{\citenamefont {Fischer}\ \emph {et~al.}(2024)\citenamefont
  {Fischer}, \citenamefont {Syska},\ and\ \citenamefont
  {Saalfrank}}]{Fischer2024-mk}%
  \BibitemOpen
  \bibfield  {author} {\bibinfo {author} {\bibfnamefont {E.~W.}\ \bibnamefont
  {Fischer}}, \bibinfo {author} {\bibfnamefont {J.~A.}\ \bibnamefont {Syska}},\
  and\ \bibinfo {author} {\bibfnamefont {P.}~\bibnamefont {Saalfrank}},\ }\href
  {https://doi.org/10.1021/acs.jpclett.4c00105} {\bibfield  {journal} {\bibinfo
   {journal} {J. Phys. Chem. Lett.}\ ,\ \bibinfo {pages} {2262}} (\bibinfo
  {year} {2024})}\BibitemShut {NoStop}%
\bibitem [{\citenamefont {Fischer}\ \emph {et~al.}(2022)\citenamefont
  {Fischer}, \citenamefont {Anders},\ and\ \citenamefont
  {Saalfrank}}]{Fischer2022-td}%
  \BibitemOpen
  \bibfield  {author} {\bibinfo {author} {\bibfnamefont {E.~W.}\ \bibnamefont
  {Fischer}}, \bibinfo {author} {\bibfnamefont {J.}~\bibnamefont {Anders}},\
  and\ \bibinfo {author} {\bibfnamefont {P.}~\bibnamefont {Saalfrank}},\ }\href
  {https://doi.org/10.1063/5.0076434} {\bibfield  {journal} {\bibinfo
  {journal} {J. Chem. Phys.}\ }\textbf {\bibinfo {volume} {156}},\ \bibinfo
  {pages} {154305} (\bibinfo {year} {2022})}\BibitemShut {NoStop}%
\bibitem [{\citenamefont {Ying}\ and\ \citenamefont {Huo}(2023)}]{Ying2023-es}%
  \BibitemOpen
  \bibfield  {author} {\bibinfo {author} {\bibfnamefont {W.}~\bibnamefont
  {Ying}}\ and\ \bibinfo {author} {\bibfnamefont {P.}~\bibnamefont {Huo}},\
  }\bibfield  {journal} {\bibinfo  {journal} {J. Chem. Phys.}\ }\textbf
  {\bibinfo {volume} {159}},\ \href {https://doi.org/10.1063/5.0159791}
  {10.1063/5.0159791} (\bibinfo {year} {2023})\BibitemShut {NoStop}%
\bibitem [{\citenamefont {Ke}\ and\ \citenamefont
  {Richardson}(2024)}]{Ke2024-br}%
  \BibitemOpen
  \bibfield  {author} {\bibinfo {author} {\bibfnamefont {Y.}~\bibnamefont
  {Ke}}\ and\ \bibinfo {author} {\bibfnamefont {J.~O.}\ \bibnamefont
  {Richardson}},\ }\href {https://doi.org/10.1063/5.0220908} {\bibfield
  {journal} {\bibinfo  {journal} {J. Chem. Phys.}\ }\textbf {\bibinfo {volume}
  {161}},\ \bibinfo {pages} {054104} (\bibinfo {year} {2024})}\BibitemShut
  {NoStop}%
\bibitem [{\citenamefont {Marques}(2003)}]{Marques2003-gd}%
  \BibitemOpen
  \bibfield  {author} {\bibinfo {author} {\bibfnamefont {M.}~\bibnamefont
  {Marques}},\ }\href {https://doi.org/10.1016/s0010-4655(02)00686-0}
  {\bibfield  {journal} {\bibinfo  {journal} {Comput. Phys. Commun.}\ }\textbf
  {\bibinfo {volume} {151}},\ \bibinfo {pages} {60} (\bibinfo {year}
  {2003})}\BibitemShut {NoStop}%
\bibitem [{\citenamefont {Castro}\ \emph {et~al.}(2006)\citenamefont {Castro},
  \citenamefont {Appel}, \citenamefont {Oliveira}, \citenamefont {Rozzi},
  \citenamefont {Andrade}, \citenamefont {Lorenzen}, \citenamefont {Marques},
  \citenamefont {Gross},\ and\ \citenamefont {Rubio}}]{Castro2006-am}%
  \BibitemOpen
  \bibfield  {author} {\bibinfo {author} {\bibfnamefont {A.}~\bibnamefont
  {Castro}}, \bibinfo {author} {\bibfnamefont {H.}~\bibnamefont {Appel}},
  \bibinfo {author} {\bibfnamefont {M.}~\bibnamefont {Oliveira}}, \bibinfo
  {author} {\bibfnamefont {C.~A.}\ \bibnamefont {Rozzi}}, \bibinfo {author}
  {\bibfnamefont {X.}~\bibnamefont {Andrade}}, \bibinfo {author} {\bibfnamefont
  {F.}~\bibnamefont {Lorenzen}}, \bibinfo {author} {\bibfnamefont {M.~A.~L.}\
  \bibnamefont {Marques}}, \bibinfo {author} {\bibfnamefont {E.~K.~U.}\
  \bibnamefont {Gross}},\ and\ \bibinfo {author} {\bibfnamefont
  {A.}~\bibnamefont {Rubio}},\ }\href {https://doi.org/10.1002/pssb.200642067}
  {\bibfield  {journal} {\bibinfo  {journal} {Phys. Status Solidi B Basic
  Res.}\ }\textbf {\bibinfo {volume} {243}},\ \bibinfo {pages} {2465} (\bibinfo
  {year} {2006})}\BibitemShut {NoStop}%
\bibitem [{\citenamefont {Andrade}\ \emph {et~al.}(2015)\citenamefont
  {Andrade}, \citenamefont {Strubbe}, \citenamefont {De~Giovannini},
  \citenamefont {Larsen}, \citenamefont {Oliveira}, \citenamefont
  {Alberdi-Rodriguez}, \citenamefont {Varas}, \citenamefont {Theophilou},
  \citenamefont {Helbig}, \citenamefont {Verstraete}, \citenamefont {Stella},
  \citenamefont {Nogueira}, \citenamefont {Aspuru-Guzik}, \citenamefont
  {Castro}, \citenamefont {Marques},\ and\ \citenamefont
  {Rubio}}]{Andrade2015-zq}%
  \BibitemOpen
  \bibfield  {author} {\bibinfo {author} {\bibfnamefont {X.}~\bibnamefont
  {Andrade}}, \bibinfo {author} {\bibfnamefont {D.}~\bibnamefont {Strubbe}},
  \bibinfo {author} {\bibfnamefont {U.}~\bibnamefont {De~Giovannini}}, \bibinfo
  {author} {\bibfnamefont {A.~H.}\ \bibnamefont {Larsen}}, \bibinfo {author}
  {\bibfnamefont {M.~J.~T.}\ \bibnamefont {Oliveira}}, \bibinfo {author}
  {\bibfnamefont {J.}~\bibnamefont {Alberdi-Rodriguez}}, \bibinfo {author}
  {\bibfnamefont {A.}~\bibnamefont {Varas}}, \bibinfo {author} {\bibfnamefont
  {I.}~\bibnamefont {Theophilou}}, \bibinfo {author} {\bibfnamefont
  {N.}~\bibnamefont {Helbig}}, \bibinfo {author} {\bibfnamefont {M.~J.}\
  \bibnamefont {Verstraete}}, \bibinfo {author} {\bibfnamefont
  {L.}~\bibnamefont {Stella}}, \bibinfo {author} {\bibfnamefont
  {F.}~\bibnamefont {Nogueira}}, \bibinfo {author} {\bibfnamefont
  {A.}~\bibnamefont {Aspuru-Guzik}}, \bibinfo {author} {\bibfnamefont
  {A.}~\bibnamefont {Castro}}, \bibinfo {author} {\bibfnamefont {M.~A.~L.}\
  \bibnamefont {Marques}},\ and\ \bibinfo {author} {\bibfnamefont
  {A.}~\bibnamefont {Rubio}},\ }\href {https://doi.org/10.1039/c5cp00351b}
  {\bibfield  {journal} {\bibinfo  {journal} {Phys. Chem. Chem. Phys.}\
  }\textbf {\bibinfo {volume} {17}},\ \bibinfo {pages} {31371} (\bibinfo {year}
  {2015})}\BibitemShut {NoStop}%
\bibitem [{\citenamefont {Tancogne-Dejean}\ \emph {et~al.}(2020)\citenamefont
  {Tancogne-Dejean}, \citenamefont {Oliveira}, \citenamefont {Andrade},
  \citenamefont {Appel}, \citenamefont {Borca}, \citenamefont {Le~Breton},
  \citenamefont {Buchholz}, \citenamefont {Castro}, \citenamefont {Corni},
  \citenamefont {Correa}, \citenamefont {De~Giovannini}, \citenamefont
  {Delgado}, \citenamefont {Eich}, \citenamefont {Flick}, \citenamefont {Gil},
  \citenamefont {Gomez}, \citenamefont {Helbig}, \citenamefont {Hübener},
  \citenamefont {Jestädt}, \citenamefont {Jornet-Somoza}, \citenamefont
  {Larsen}, \citenamefont {Lebedeva}, \citenamefont {Lüders}, \citenamefont
  {Marques}, \citenamefont {Ohlmann}, \citenamefont {Pipolo}, \citenamefont
  {Rampp}, \citenamefont {Rozzi}, \citenamefont {Strubbe}, \citenamefont
  {Sato}, \citenamefont {Schäfer}, \citenamefont {Theophilou}, \citenamefont
  {Welden},\ and\ \citenamefont {Rubio}}]{Tancogne-Dejean2020-ev}%
  \BibitemOpen
  \bibfield  {author} {\bibinfo {author} {\bibfnamefont {N.}~\bibnamefont
  {Tancogne-Dejean}}, \bibinfo {author} {\bibfnamefont {M.~J.~T.}\ \bibnamefont
  {Oliveira}}, \bibinfo {author} {\bibfnamefont {X.}~\bibnamefont {Andrade}},
  \bibinfo {author} {\bibfnamefont {H.}~\bibnamefont {Appel}}, \bibinfo
  {author} {\bibfnamefont {C.~H.}\ \bibnamefont {Borca}}, \bibinfo {author}
  {\bibfnamefont {G.}~\bibnamefont {Le~Breton}}, \bibinfo {author}
  {\bibfnamefont {F.}~\bibnamefont {Buchholz}}, \bibinfo {author}
  {\bibfnamefont {A.}~\bibnamefont {Castro}}, \bibinfo {author} {\bibfnamefont
  {S.}~\bibnamefont {Corni}}, \bibinfo {author} {\bibfnamefont {A.~A.}\
  \bibnamefont {Correa}}, \bibinfo {author} {\bibfnamefont {U.}~\bibnamefont
  {De~Giovannini}}, \bibinfo {author} {\bibfnamefont {A.}~\bibnamefont
  {Delgado}}, \bibinfo {author} {\bibfnamefont {F.~G.}\ \bibnamefont {Eich}},
  \bibinfo {author} {\bibfnamefont {J.}~\bibnamefont {Flick}}, \bibinfo
  {author} {\bibfnamefont {G.}~\bibnamefont {Gil}}, \bibinfo {author}
  {\bibfnamefont {A.}~\bibnamefont {Gomez}}, \bibinfo {author} {\bibfnamefont
  {N.}~\bibnamefont {Helbig}}, \bibinfo {author} {\bibfnamefont
  {H.}~\bibnamefont {Hübener}}, \bibinfo {author} {\bibfnamefont
  {R.}~\bibnamefont {Jestädt}}, \bibinfo {author} {\bibfnamefont
  {J.}~\bibnamefont {Jornet-Somoza}}, \bibinfo {author} {\bibfnamefont {A.~H.}\
  \bibnamefont {Larsen}}, \bibinfo {author} {\bibfnamefont {I.~V.}\
  \bibnamefont {Lebedeva}}, \bibinfo {author} {\bibfnamefont {M.}~\bibnamefont
  {Lüders}}, \bibinfo {author} {\bibfnamefont {M.~A.~L.}\ \bibnamefont
  {Marques}}, \bibinfo {author} {\bibfnamefont {S.~T.}\ \bibnamefont
  {Ohlmann}}, \bibinfo {author} {\bibfnamefont {S.}~\bibnamefont {Pipolo}},
  \bibinfo {author} {\bibfnamefont {M.}~\bibnamefont {Rampp}}, \bibinfo
  {author} {\bibfnamefont {C.~A.}\ \bibnamefont {Rozzi}}, \bibinfo {author}
  {\bibfnamefont {D.~A.}\ \bibnamefont {Strubbe}}, \bibinfo {author}
  {\bibfnamefont {S.~A.}\ \bibnamefont {Sato}}, \bibinfo {author}
  {\bibfnamefont {C.}~\bibnamefont {Schäfer}}, \bibinfo {author}
  {\bibfnamefont {I.}~\bibnamefont {Theophilou}}, \bibinfo {author}
  {\bibfnamefont {A.}~\bibnamefont {Welden}},\ and\ \bibinfo {author}
  {\bibfnamefont {A.}~\bibnamefont {Rubio}},\ }\href
  {https://doi.org/10.1063/1.5142502} {\bibfield  {journal} {\bibinfo
  {journal} {J. Chem. Phys.}\ }\textbf {\bibinfo {volume} {152}},\ \bibinfo
  {pages} {124119} (\bibinfo {year} {2020})}\BibitemShut {NoStop}%
\bibitem [{\citenamefont {Folkestad}\ \emph {et~al.}(2020)\citenamefont
  {Folkestad}, \citenamefont {Kj{\o}nstad}, \citenamefont {Myhre},
  \citenamefont {Andersen}, \citenamefont {Balbi}, \citenamefont {Coriani},
  \citenamefont {Giovannini}, \citenamefont {Goletto}, \citenamefont
  {Haugland}, \citenamefont {Hutcheson}, \citenamefont {H{\o}yvik},
  \citenamefont {Moitra}, \citenamefont {Paul}, \citenamefont {Scavino},
  \citenamefont {Skeidsvoll}, \citenamefont {Tveten},\ and\ \citenamefont
  {Koch}}]{Folkestad2020-cw}%
  \BibitemOpen
  \bibfield  {author} {\bibinfo {author} {\bibfnamefont {S.~D.}\ \bibnamefont
  {Folkestad}}, \bibinfo {author} {\bibfnamefont {E.~F.}\ \bibnamefont
  {Kj{\o}nstad}}, \bibinfo {author} {\bibfnamefont {R.~H.}\ \bibnamefont
  {Myhre}}, \bibinfo {author} {\bibfnamefont {J.~H.}\ \bibnamefont {Andersen}},
  \bibinfo {author} {\bibfnamefont {A.}~\bibnamefont {Balbi}}, \bibinfo
  {author} {\bibfnamefont {S.}~\bibnamefont {Coriani}}, \bibinfo {author}
  {\bibfnamefont {T.}~\bibnamefont {Giovannini}}, \bibinfo {author}
  {\bibfnamefont {L.}~\bibnamefont {Goletto}}, \bibinfo {author} {\bibfnamefont
  {T.~S.}\ \bibnamefont {Haugland}}, \bibinfo {author} {\bibfnamefont
  {A.}~\bibnamefont {Hutcheson}}, \bibinfo {author} {\bibfnamefont {I.-M.}\
  \bibnamefont {H{\o}yvik}}, \bibinfo {author} {\bibfnamefont {T.}~\bibnamefont
  {Moitra}}, \bibinfo {author} {\bibfnamefont {A.~C.}\ \bibnamefont {Paul}},
  \bibinfo {author} {\bibfnamefont {M.}~\bibnamefont {Scavino}}, \bibinfo
  {author} {\bibfnamefont {A.~S.}\ \bibnamefont {Skeidsvoll}}, \bibinfo
  {author} {\bibfnamefont {{\AA}.~H.}\ \bibnamefont {Tveten}},\ and\ \bibinfo
  {author} {\bibfnamefont {H.}~\bibnamefont {Koch}},\ }\href
  {https://doi.org/10.1063/5.0004713} {\bibfield  {journal} {\bibinfo
  {journal} {J. Chem. Phys.}\ }\textbf {\bibinfo {volume} {152}},\ \bibinfo
  {pages} {184103} (\bibinfo {year} {2020})}\BibitemShut {NoStop}%
\bibitem [{\citenamefont {Folkestad}\ \emph {et~al.}(2026)\citenamefont
  {Folkestad}, \citenamefont {Kjønstad}, \citenamefont {Paul}, \citenamefont
  {Myhre}, \citenamefont {Alessandro}, \citenamefont {Angelico}, \citenamefont
  {Balbi}, \citenamefont {Barlini}, \citenamefont {Bianchi}, \citenamefont
  {Cappelli}, \citenamefont {Castagnola}, \citenamefont {Coriani},
  \citenamefont {El~Moutaoukal}, \citenamefont {Giovannini}, \citenamefont
  {Goletto}, \citenamefont {Haugland}, \citenamefont {Hollas}, \citenamefont
  {Høyvik}, \citenamefont {Lexander}, \citenamefont {Lipovec}, \citenamefont
  {Marrazzini}, \citenamefont {Moitra}, \citenamefont {Os}, \citenamefont
  {Paul}, \citenamefont {Pedersen}, \citenamefont {Rinaldi}, \citenamefont
  {Riso}, \citenamefont {Roet}, \citenamefont {Ronca}, \citenamefont {Rossi},
  \citenamefont {Sannes}, \citenamefont {Schnack-Petersen}, \citenamefont
  {Skeidsvoll}, \citenamefont {Stoll}, \citenamefont {Thiam}, \citenamefont
  {Trabski},\ and\ \citenamefont {Koch}}]{Folkestad2026-ui}%
  \BibitemOpen
  \bibfield  {author} {\bibinfo {author} {\bibfnamefont {S.~D.}\ \bibnamefont
  {Folkestad}}, \bibinfo {author} {\bibfnamefont {E.~F.}\ \bibnamefont
  {Kjønstad}}, \bibinfo {author} {\bibfnamefont {A.~C.}\ \bibnamefont {Paul}},
  \bibinfo {author} {\bibfnamefont {R.~H.}\ \bibnamefont {Myhre}}, \bibinfo
  {author} {\bibfnamefont {R.}~\bibnamefont {Alessandro}}, \bibinfo {author}
  {\bibfnamefont {S.}~\bibnamefont {Angelico}}, \bibinfo {author}
  {\bibfnamefont {A.}~\bibnamefont {Balbi}}, \bibinfo {author} {\bibfnamefont
  {A.}~\bibnamefont {Barlini}}, \bibinfo {author} {\bibfnamefont
  {A.}~\bibnamefont {Bianchi}}, \bibinfo {author} {\bibfnamefont
  {C.}~\bibnamefont {Cappelli}}, \bibinfo {author} {\bibfnamefont
  {M.}~\bibnamefont {Castagnola}}, \bibinfo {author} {\bibfnamefont
  {S.}~\bibnamefont {Coriani}}, \bibinfo {author} {\bibfnamefont
  {Y.}~\bibnamefont {El~Moutaoukal}}, \bibinfo {author} {\bibfnamefont
  {T.}~\bibnamefont {Giovannini}}, \bibinfo {author} {\bibfnamefont
  {L.}~\bibnamefont {Goletto}}, \bibinfo {author} {\bibfnamefont {T.~S.}\
  \bibnamefont {Haugland}}, \bibinfo {author} {\bibfnamefont {D.}~\bibnamefont
  {Hollas}}, \bibinfo {author} {\bibfnamefont {I.-M.}\ \bibnamefont {Høyvik}},
  \bibinfo {author} {\bibfnamefont {M.~T.}\ \bibnamefont {Lexander}}, \bibinfo
  {author} {\bibfnamefont {D.}~\bibnamefont {Lipovec}}, \bibinfo {author}
  {\bibfnamefont {G.}~\bibnamefont {Marrazzini}}, \bibinfo {author}
  {\bibfnamefont {T.}~\bibnamefont {Moitra}}, \bibinfo {author} {\bibfnamefont
  {Y.}~\bibnamefont {Os}}, \bibinfo {author} {\bibfnamefont {R.}~\bibnamefont
  {Paul}}, \bibinfo {author} {\bibfnamefont {J.}~\bibnamefont {Pedersen}},
  \bibinfo {author} {\bibfnamefont {M.}~\bibnamefont {Rinaldi}}, \bibinfo
  {author} {\bibfnamefont {R.~R.}\ \bibnamefont {Riso}}, \bibinfo {author}
  {\bibfnamefont {S.}~\bibnamefont {Roet}}, \bibinfo {author} {\bibfnamefont
  {E.}~\bibnamefont {Ronca}}, \bibinfo {author} {\bibfnamefont
  {F.}~\bibnamefont {Rossi}}, \bibinfo {author} {\bibfnamefont {B.~S.}\
  \bibnamefont {Sannes}}, \bibinfo {author} {\bibfnamefont {A.~K.}\
  \bibnamefont {Schnack-Petersen}}, \bibinfo {author} {\bibfnamefont {A.~S.}\
  \bibnamefont {Skeidsvoll}}, \bibinfo {author} {\bibfnamefont
  {L.}~\bibnamefont {Stoll}}, \bibinfo {author} {\bibfnamefont
  {G.}~\bibnamefont {Thiam}}, \bibinfo {author} {\bibfnamefont {J.~H.~M.}\
  \bibnamefont {Trabski}},\ and\ \bibinfo {author} {\bibfnamefont
  {H.}~\bibnamefont {Koch}},\ }\href {https://doi.org/10.1063/5.0309334}
  {\bibfield  {journal} {\bibinfo  {journal} {J. Chem. Phys.}\ }\textbf
  {\bibinfo {volume} {164}},\ \bibinfo {pages} {132501} (\bibinfo {year}
  {2026})}\BibitemShut {NoStop}%
\bibitem [{\citenamefont {Haugland}\ \emph {et~al.}(2020)\citenamefont
  {Haugland}, \citenamefont {Ronca}, \citenamefont {Kjønstad}, \citenamefont
  {Rubio},\ and\ \citenamefont {Koch}}]{Haugland2020-xh}%
  \BibitemOpen
  \bibfield  {author} {\bibinfo {author} {\bibfnamefont {T.~S.}\ \bibnamefont
  {Haugland}}, \bibinfo {author} {\bibfnamefont {E.}~\bibnamefont {Ronca}},
  \bibinfo {author} {\bibfnamefont {E.~F.}\ \bibnamefont {Kjønstad}}, \bibinfo
  {author} {\bibfnamefont {A.}~\bibnamefont {Rubio}},\ and\ \bibinfo {author}
  {\bibfnamefont {H.}~\bibnamefont {Koch}},\ }\href
  {https://doi.org/10.1103/PhysRevX.10.041043} {\bibfield  {journal} {\bibinfo
  {journal} {Phys. Rev. X}\ }\textbf {\bibinfo {volume} {10}},\ \bibinfo
  {pages} {041043} (\bibinfo {year} {2020})}\BibitemShut {NoStop}%
\bibitem [{\citenamefont {Angelico}\ \emph {et~al.}(2023)\citenamefont
  {Angelico}, \citenamefont {Haugland}, \citenamefont {Ronca},\ and\
  \citenamefont {Koch}}]{Angelico2023-jh}%
  \BibitemOpen
  \bibfield  {author} {\bibinfo {author} {\bibfnamefont {S.}~\bibnamefont
  {Angelico}}, \bibinfo {author} {\bibfnamefont {T.~S.}\ \bibnamefont
  {Haugland}}, \bibinfo {author} {\bibfnamefont {E.}~\bibnamefont {Ronca}},\
  and\ \bibinfo {author} {\bibfnamefont {H.}~\bibnamefont {Koch}},\ }\href
  {https://doi.org/10.1063/5.0172764} {\bibfield  {journal} {\bibinfo
  {journal} {J. Chem. Phys.}\ }\textbf {\bibinfo {volume} {159}},\ \bibinfo
  {pages} {214112} (\bibinfo {year} {2023})}\BibitemShut {NoStop}%
\bibitem [{\citenamefont {Schnappinger}\ \emph {et~al.}(2023)\citenamefont
  {Schnappinger}, \citenamefont {Sidler}, \citenamefont {Ruggenthaler},
  \citenamefont {Rubio},\ and\ \citenamefont
  {Kowalewski}}]{Schnappinger2023-cbohf}%
  \BibitemOpen
  \bibfield  {author} {\bibinfo {author} {\bibfnamefont {T.}~\bibnamefont
  {Schnappinger}}, \bibinfo {author} {\bibfnamefont {D.}~\bibnamefont
  {Sidler}}, \bibinfo {author} {\bibfnamefont {M.}~\bibnamefont
  {Ruggenthaler}}, \bibinfo {author} {\bibfnamefont {A.}~\bibnamefont
  {Rubio}},\ and\ \bibinfo {author} {\bibfnamefont {M.}~\bibnamefont
  {Kowalewski}},\ }\href {https://doi.org/10.1021/acs.jpclett.3c01842}
  {\bibfield  {journal} {\bibinfo  {journal} {J. Phys. Chem. Lett.}\ }\textbf
  {\bibinfo {volume} {14}},\ \bibinfo {pages} {8024} (\bibinfo {year}
  {2023})}\BibitemShut {NoStop}%
\bibitem [{\citenamefont {Smith}\ \emph {et~al.}(2020)\citenamefont {Smith},
  \citenamefont {Burns}, \citenamefont {Simmonett}, \citenamefont {Parrish},
  \citenamefont {Schieber}, \citenamefont {Galvelis}, \citenamefont {Kraus},
  \citenamefont {Kruse}, \citenamefont {Di~Remigio}, \citenamefont {Alenaizan},
  \citenamefont {James}, \citenamefont {Lehtola}, \citenamefont {Misiewicz},
  \citenamefont {Scheurer}, \citenamefont {Shaw}, \citenamefont {Schriber},
  \citenamefont {Xie}, \citenamefont {Glick}, \citenamefont {Sirianni},
  \citenamefont {O'Brien}, \citenamefont {Waldrop}, \citenamefont {Kumar},
  \citenamefont {Hohenstein}, \citenamefont {Pritchard}, \citenamefont
  {Brooks}, \citenamefont {Schaefer}, \citenamefont {Sokolov}, \citenamefont
  {Patkowski}, \citenamefont {DePrince}, \citenamefont {Bozkaya}, \citenamefont
  {King}, \citenamefont {Evangelista}, \citenamefont {Turney}, \citenamefont
  {Crawford},\ and\ \citenamefont {Sherrill}}]{Smith2020-kq}%
  \BibitemOpen
  \bibfield  {author} {\bibinfo {author} {\bibfnamefont {D.~G.~A.}\
  \bibnamefont {Smith}}, \bibinfo {author} {\bibfnamefont {L.~A.}\ \bibnamefont
  {Burns}}, \bibinfo {author} {\bibfnamefont {A.~C.}\ \bibnamefont
  {Simmonett}}, \bibinfo {author} {\bibfnamefont {R.~M.}\ \bibnamefont
  {Parrish}}, \bibinfo {author} {\bibfnamefont {M.~C.}\ \bibnamefont
  {Schieber}}, \bibinfo {author} {\bibfnamefont {R.}~\bibnamefont {Galvelis}},
  \bibinfo {author} {\bibfnamefont {P.}~\bibnamefont {Kraus}}, \bibinfo
  {author} {\bibfnamefont {H.}~\bibnamefont {Kruse}}, \bibinfo {author}
  {\bibfnamefont {R.}~\bibnamefont {Di~Remigio}}, \bibinfo {author}
  {\bibfnamefont {A.}~\bibnamefont {Alenaizan}}, \bibinfo {author}
  {\bibfnamefont {A.~M.}\ \bibnamefont {James}}, \bibinfo {author}
  {\bibfnamefont {S.}~\bibnamefont {Lehtola}}, \bibinfo {author} {\bibfnamefont
  {J.~P.}\ \bibnamefont {Misiewicz}}, \bibinfo {author} {\bibfnamefont
  {M.}~\bibnamefont {Scheurer}}, \bibinfo {author} {\bibfnamefont {R.~A.}\
  \bibnamefont {Shaw}}, \bibinfo {author} {\bibfnamefont {J.~B.}\ \bibnamefont
  {Schriber}}, \bibinfo {author} {\bibfnamefont {Y.}~\bibnamefont {Xie}},
  \bibinfo {author} {\bibfnamefont {Z.~L.}\ \bibnamefont {Glick}}, \bibinfo
  {author} {\bibfnamefont {D.~A.}\ \bibnamefont {Sirianni}}, \bibinfo {author}
  {\bibfnamefont {J.~S.}\ \bibnamefont {O'Brien}}, \bibinfo {author}
  {\bibfnamefont {J.~M.}\ \bibnamefont {Waldrop}}, \bibinfo {author}
  {\bibfnamefont {A.}~\bibnamefont {Kumar}}, \bibinfo {author} {\bibfnamefont
  {E.~G.}\ \bibnamefont {Hohenstein}}, \bibinfo {author} {\bibfnamefont
  {B.~P.}\ \bibnamefont {Pritchard}}, \bibinfo {author} {\bibfnamefont {B.~R.}\
  \bibnamefont {Brooks}}, \bibinfo {author} {\bibfnamefont {H.~F.}\
  \bibnamefont {Schaefer}, \bibfnamefont {3rd}}, \bibinfo {author}
  {\bibfnamefont {A.~Y.}\ \bibnamefont {Sokolov}}, \bibinfo {author}
  {\bibfnamefont {K.}~\bibnamefont {Patkowski}}, \bibinfo {author}
  {\bibfnamefont {A.~E.}\ \bibnamefont {DePrince}, \bibfnamefont {3rd}},
  \bibinfo {author} {\bibfnamefont {U.}~\bibnamefont {Bozkaya}}, \bibinfo
  {author} {\bibfnamefont {R.~A.}\ \bibnamefont {King}}, \bibinfo {author}
  {\bibfnamefont {F.~A.}\ \bibnamefont {Evangelista}}, \bibinfo {author}
  {\bibfnamefont {J.~M.}\ \bibnamefont {Turney}}, \bibinfo {author}
  {\bibfnamefont {T.~D.}\ \bibnamefont {Crawford}},\ and\ \bibinfo {author}
  {\bibfnamefont {C.~D.}\ \bibnamefont {Sherrill}},\ }\href
  {https://doi.org/10.1063/5.0006002} {\bibfield  {journal} {\bibinfo
  {journal} {J. Chem. Phys.}\ }\textbf {\bibinfo {volume} {152}},\ \bibinfo
  {pages} {184108} (\bibinfo {year} {2020})}\BibitemShut {NoStop}%
\bibitem [{\citenamefont {Smith}\ \emph {et~al.}(2018)\citenamefont {Smith},
  \citenamefont {Burns}, \citenamefont {Sirianni}, \citenamefont {Nascimento},
  \citenamefont {Kumar}, \citenamefont {James}, \citenamefont {Schriber},
  \citenamefont {Zhang}, \citenamefont {Zhang}, \citenamefont {Abbott},
  \citenamefont {Berquist}, \citenamefont {Lechner}, \citenamefont {Cunha},
  \citenamefont {Heide}, \citenamefont {Waldrop}, \citenamefont {Takeshita},
  \citenamefont {Alenaizan}, \citenamefont {Neuhauser}, \citenamefont {King},
  \citenamefont {Simmonett}, \citenamefont {Turney}, \citenamefont {Schaefer},
  \citenamefont {Evangelista}, \citenamefont {DePrince}, \citenamefont
  {Crawford}, \citenamefont {Patkowski},\ and\ \citenamefont
  {Sherrill}}]{Smith2018-tu}%
  \BibitemOpen
  \bibfield  {author} {\bibinfo {author} {\bibfnamefont {D.~G.~A.}\
  \bibnamefont {Smith}}, \bibinfo {author} {\bibfnamefont {L.~A.}\ \bibnamefont
  {Burns}}, \bibinfo {author} {\bibfnamefont {D.~A.}\ \bibnamefont {Sirianni}},
  \bibinfo {author} {\bibfnamefont {D.~R.}\ \bibnamefont {Nascimento}},
  \bibinfo {author} {\bibfnamefont {A.}~\bibnamefont {Kumar}}, \bibinfo
  {author} {\bibfnamefont {A.~M.}\ \bibnamefont {James}}, \bibinfo {author}
  {\bibfnamefont {J.~B.}\ \bibnamefont {Schriber}}, \bibinfo {author}
  {\bibfnamefont {T.}~\bibnamefont {Zhang}}, \bibinfo {author} {\bibfnamefont
  {B.}~\bibnamefont {Zhang}}, \bibinfo {author} {\bibfnamefont {A.~S.}\
  \bibnamefont {Abbott}}, \bibinfo {author} {\bibfnamefont {E.~J.}\
  \bibnamefont {Berquist}}, \bibinfo {author} {\bibfnamefont {M.~H.}\
  \bibnamefont {Lechner}}, \bibinfo {author} {\bibfnamefont {L.~A.}\
  \bibnamefont {Cunha}}, \bibinfo {author} {\bibfnamefont {A.~G.}\ \bibnamefont
  {Heide}}, \bibinfo {author} {\bibfnamefont {J.~M.}\ \bibnamefont {Waldrop}},
  \bibinfo {author} {\bibfnamefont {T.~Y.}\ \bibnamefont {Takeshita}}, \bibinfo
  {author} {\bibfnamefont {A.}~\bibnamefont {Alenaizan}}, \bibinfo {author}
  {\bibfnamefont {D.}~\bibnamefont {Neuhauser}}, \bibinfo {author}
  {\bibfnamefont {R.~A.}\ \bibnamefont {King}}, \bibinfo {author}
  {\bibfnamefont {A.~C.}\ \bibnamefont {Simmonett}}, \bibinfo {author}
  {\bibfnamefont {J.~M.}\ \bibnamefont {Turney}}, \bibinfo {author}
  {\bibfnamefont {H.~F.}\ \bibnamefont {Schaefer}}, \bibinfo {author}
  {\bibfnamefont {F.~A.}\ \bibnamefont {Evangelista}}, \bibinfo {author}
  {\bibfnamefont {A.~E.}\ \bibnamefont {DePrince}, \bibfnamefont {3rd}},
  \bibinfo {author} {\bibfnamefont {T.~D.}\ \bibnamefont {Crawford}}, \bibinfo
  {author} {\bibfnamefont {K.}~\bibnamefont {Patkowski}},\ and\ \bibinfo
  {author} {\bibfnamefont {C.~D.}\ \bibnamefont {Sherrill}},\ }\href
  {https://doi.org/10.1021/acs.jctc.8b00286} {\bibfield  {journal} {\bibinfo
  {journal} {J. Chem. Theory Comput.}\ }\textbf {\bibinfo {volume} {14}},\
  \bibinfo {pages} {3504} (\bibinfo {year} {2018})}\BibitemShut {NoStop}%
\bibitem [{\citenamefont {Schnappinger}\ and\ \citenamefont
  {Kowalewski}(2023)}]{Schnappinger2023-wp}%
  \BibitemOpen
  \bibfield  {author} {\bibinfo {author} {\bibfnamefont {T.}~\bibnamefont
  {Schnappinger}}\ and\ \bibinfo {author} {\bibfnamefont {M.}~\bibnamefont
  {Kowalewski}},\ }\href {https://doi.org/10.1021/acs.jctc.3c01135} {\bibfield
  {journal} {\bibinfo  {journal} {J. Chem. Theory Comput.}\ }\textbf {\bibinfo
  {volume} {19}},\ \bibinfo {pages} {9278} (\bibinfo {year}
  {2023})}\BibitemShut {NoStop}%
\bibitem [{\citenamefont {Schnappinger}\ and\ \citenamefont
  {Kowalewski}(2024)}]{Schnappinger2024-vt}%
  \BibitemOpen
  \bibfield  {author} {\bibinfo {author} {\bibfnamefont {T.}~\bibnamefont
  {Schnappinger}}\ and\ \bibinfo {author} {\bibfnamefont {M.}~\bibnamefont
  {Kowalewski}},\ }\href {https://doi.org/10.1021/acs.jpclett.4c01810}
  {\bibfield  {journal} {\bibinfo  {journal} {J. Phys. Chem. Lett.}\ }\textbf
  {\bibinfo {volume} {15}},\ \bibinfo {pages} {7700} (\bibinfo {year}
  {2024})}\BibitemShut {NoStop}%
\bibitem [{\citenamefont {Schnappinger}\ and\ \citenamefont
  {Kowalewski}(2025)}]{Schnappinger2025-fx}%
  \BibitemOpen
  \bibfield  {author} {\bibinfo {author} {\bibfnamefont {T.}~\bibnamefont
  {Schnappinger}}\ and\ \bibinfo {author} {\bibfnamefont {M.}~\bibnamefont
  {Kowalewski}},\ }\bibfield  {journal} {\bibinfo  {journal} {J. Chem. Theory
  Comput.}\ }\href {https://doi.org/10.1021/acs.jctc.5c00461}
  {10.1021/acs.jctc.5c00461} (\bibinfo {year} {2025})\BibitemShut {NoStop}%
\bibitem [{\citenamefont {Schnappinger}\ \emph {et~al.}(2024)\citenamefont
  {Schnappinger}, \citenamefont {Falvo},\ and\ \citenamefont
  {Kowalewski}}]{Schnappinger2024-rz}%
  \BibitemOpen
  \bibfield  {author} {\bibinfo {author} {\bibfnamefont {T.}~\bibnamefont
  {Schnappinger}}, \bibinfo {author} {\bibfnamefont {C.}~\bibnamefont
  {Falvo}},\ and\ \bibinfo {author} {\bibfnamefont {M.}~\bibnamefont
  {Kowalewski}},\ }\bibfield  {journal} {\bibinfo  {journal} {J. Chem. Phys.}\
  }\textbf {\bibinfo {volume} {161}},\ \href
  {https://doi.org/10.1063/5.0239877} {10.1063/5.0239877} (\bibinfo {year}
  {2024})\BibitemShut {NoStop}%
\bibitem [{\citenamefont {Glauber}(1963)}]{Glauber1963-coh}%
  \BibitemOpen
  \bibfield  {author} {\bibinfo {author} {\bibfnamefont {R.~J.}\ \bibnamefont
  {Glauber}},\ }\href {https://doi.org/10.1103/PhysRev.131.2766} {\bibfield
  {journal} {\bibinfo  {journal} {Phys. Rev.}\ }\textbf {\bibinfo {volume}
  {131}},\ \bibinfo {pages} {2766} (\bibinfo {year} {1963})}\BibitemShut
  {NoStop}%
\bibitem [{\citenamefont {Liebenthal}\ \emph {et~al.}(2023)\citenamefont
  {Liebenthal}, \citenamefont {Vu},\ and\ \citenamefont
  {DePrince}}]{Liebenthal2023-ra}%
  \BibitemOpen
  \bibfield  {author} {\bibinfo {author} {\bibfnamefont {M.~D.}\ \bibnamefont
  {Liebenthal}}, \bibinfo {author} {\bibfnamefont {N.}~\bibnamefont {Vu}},\
  and\ \bibinfo {author} {\bibfnamefont {A.~E.}\ \bibnamefont {DePrince},
  \bibfnamefont {3rd}},\ }\bibfield  {journal} {\bibinfo  {journal} {J. Phys.
  Chem. A}\ }\href {https://doi.org/10.1021/acs.jpca.3c01842}
  {10.1021/acs.jpca.3c01842} (\bibinfo {year} {2023})\BibitemShut {NoStop}%
\bibitem [{\citenamefont {Bonini}\ and\ \citenamefont
  {Flick}(2021)}]{Bonini2021}%
  \BibitemOpen
  \bibfield  {author} {\bibinfo {author} {\bibfnamefont {M.}~\bibnamefont
  {Bonini}}\ and\ \bibinfo {author} {\bibfnamefont {J.}~\bibnamefont {Flick}},\
  }\href@noop {} {\bibfield  {journal} {\bibinfo  {journal} {arXiv
  [physics.chem-ph]}\ } (\bibinfo {year} {2021})},\ \Eprint
  {https://arxiv.org/abs/2108.11564} {arXiv:2108.11564} \BibitemShut {NoStop}%
\bibitem [{\citenamefont {Haugland}\ \emph {et~al.}(2025)\citenamefont
  {Haugland}, \citenamefont {Philbin}, \citenamefont {Ghosh}, \citenamefont
  {Chen}, \citenamefont {Koch},\ and\ \citenamefont
  {Narang}}]{Haugland2025-uk}%
  \BibitemOpen
  \bibfield  {author} {\bibinfo {author} {\bibfnamefont {T.~S.}\ \bibnamefont
  {Haugland}}, \bibinfo {author} {\bibfnamefont {J.~P.}\ \bibnamefont
  {Philbin}}, \bibinfo {author} {\bibfnamefont {T.~K.}\ \bibnamefont {Ghosh}},
  \bibinfo {author} {\bibfnamefont {M.}~\bibnamefont {Chen}}, \bibinfo {author}
  {\bibfnamefont {H.}~\bibnamefont {Koch}},\ and\ \bibinfo {author}
  {\bibfnamefont {P.}~\bibnamefont {Narang}},\ }\href
  {https://doi.org/10.1063/5.0258935} {\bibfield  {journal} {\bibinfo
  {journal} {J. Chem. Phys.}\ }\textbf {\bibinfo {volume} {162}},\ \bibinfo
  {pages} {194106} (\bibinfo {year} {2025})}\BibitemShut {NoStop}%
\bibitem [{\citenamefont {Vu}\ \emph {et~al.}(2024)\citenamefont {Vu},
  \citenamefont {Mejia-Rodriguez}, \citenamefont {Bauman}, \citenamefont
  {Panyala}, \citenamefont {Mutlu}, \citenamefont {Govind},\ and\ \citenamefont
  {Foley}}]{Vu2024-qh}%
  \BibitemOpen
  \bibfield  {author} {\bibinfo {author} {\bibfnamefont {N.}~\bibnamefont
  {Vu}}, \bibinfo {author} {\bibfnamefont {D.}~\bibnamefont {Mejia-Rodriguez}},
  \bibinfo {author} {\bibfnamefont {N.~P.}\ \bibnamefont {Bauman}}, \bibinfo
  {author} {\bibfnamefont {A.}~\bibnamefont {Panyala}}, \bibinfo {author}
  {\bibfnamefont {E.}~\bibnamefont {Mutlu}}, \bibinfo {author} {\bibfnamefont
  {N.}~\bibnamefont {Govind}},\ and\ \bibinfo {author} {\bibfnamefont {J.~J.}\
  \bibnamefont {Foley}, \bibfnamefont {4th}},\ }\href
  {https://doi.org/10.1021/acs.jctc.3c01207} {\bibfield  {journal} {\bibinfo
  {journal} {J. Chem. Theory Comput.}\ }\textbf {\bibinfo {volume} {20}},\
  \bibinfo {pages} {1214} (\bibinfo {year} {2024})}\BibitemShut {NoStop}%
\bibitem [{\citenamefont {McTague}\ and\ \citenamefont
  {Foley}(2022)}]{McTague2022-xl}%
  \BibitemOpen
  \bibfield  {author} {\bibinfo {author} {\bibfnamefont {J.}~\bibnamefont
  {McTague}}\ and\ \bibinfo {author} {\bibfnamefont {J.~J.}\ \bibnamefont
  {Foley}, \bibfnamefont {4th}},\ }\href {https://doi.org/10.1063/5.0091953}
  {\bibfield  {journal} {\bibinfo  {journal} {J. Chem. Phys.}\ }\textbf
  {\bibinfo {volume} {156}},\ \bibinfo {pages} {154103} (\bibinfo {year}
  {2022})}\BibitemShut {NoStop}%
\bibitem [{\citenamefont {Rokaj}\ \emph {et~al.}(2018)\citenamefont {Rokaj},
  \citenamefont {Welakuh}, \citenamefont {Ruggenthaler},\ and\ \citenamefont
  {Rubio}}]{Rokaj2018-ww}%
  \BibitemOpen
  \bibfield  {author} {\bibinfo {author} {\bibfnamefont {V.}~\bibnamefont
  {Rokaj}}, \bibinfo {author} {\bibfnamefont {D.~M.}\ \bibnamefont {Welakuh}},
  \bibinfo {author} {\bibfnamefont {M.}~\bibnamefont {Ruggenthaler}},\ and\
  \bibinfo {author} {\bibfnamefont {A.}~\bibnamefont {Rubio}},\ }\href
  {https://doi.org/10.1088/1361-6455/aa9c99} {\bibfield  {journal} {\bibinfo
  {journal} {J. Phys. B At. Mol. Opt. Phys.}\ }\textbf {\bibinfo {volume}
  {51}},\ \bibinfo {pages} {034005} (\bibinfo {year} {2018})}\BibitemShut
  {NoStop}%
\bibitem [{\citenamefont {Sch{\"a}fer}\ \emph {et~al.}(2020)\citenamefont
  {Sch{\"a}fer}, \citenamefont {Ruggenthaler}, \citenamefont {Rokaj},\ and\
  \citenamefont {Rubio}}]{Schafer2020-cb}%
  \BibitemOpen
  \bibfield  {author} {\bibinfo {author} {\bibfnamefont {C.}~\bibnamefont
  {Sch{\"a}fer}}, \bibinfo {author} {\bibfnamefont {M.}~\bibnamefont
  {Ruggenthaler}}, \bibinfo {author} {\bibfnamefont {V.}~\bibnamefont
  {Rokaj}},\ and\ \bibinfo {author} {\bibfnamefont {A.}~\bibnamefont {Rubio}},\
  }\href {https://doi.org/10.1021/acsphotonics.9b01649} {\bibfield  {journal}
  {\bibinfo  {journal} {ACS Photonics}\ }\textbf {\bibinfo {volume} {7}},\
  \bibinfo {pages} {975} (\bibinfo {year} {2020})}\BibitemShut {NoStop}%
\bibitem [{\citenamefont {Sidler}\ \emph {et~al.}(2024)\citenamefont {Sidler},
  \citenamefont {Schnappinger}, \citenamefont {Obzhirov}, \citenamefont
  {Ruggenthaler}, \citenamefont {Kowalewski},\ and\ \citenamefont
  {Rubio}}]{Sidler2024-tw}%
  \BibitemOpen
  \bibfield  {author} {\bibinfo {author} {\bibfnamefont {D.}~\bibnamefont
  {Sidler}}, \bibinfo {author} {\bibfnamefont {T.}~\bibnamefont
  {Schnappinger}}, \bibinfo {author} {\bibfnamefont {A.}~\bibnamefont
  {Obzhirov}}, \bibinfo {author} {\bibfnamefont {M.}~\bibnamefont
  {Ruggenthaler}}, \bibinfo {author} {\bibfnamefont {M.}~\bibnamefont
  {Kowalewski}},\ and\ \bibinfo {author} {\bibfnamefont {A.}~\bibnamefont
  {Rubio}},\ }\href {https://doi.org/10.1021/acs.jpclett.4c00913} {\bibfield
  {journal} {\bibinfo  {journal} {J. Phys. Chem. Lett.}\ }\textbf {\bibinfo
  {volume} {15}},\ \bibinfo {pages} {5208} (\bibinfo {year}
  {2024})}\BibitemShut {NoStop}%
\bibitem [{\citenamefont {{\v C}{\'\i}{\v z}ek}\ and\ \citenamefont
  {Paldus}(1967)}]{Cizek1967-stability}%
  \BibitemOpen
  \bibfield  {author} {\bibinfo {author} {\bibfnamefont {J.}~\bibnamefont {{\v
  C}{\'\i}{\v z}ek}}\ and\ \bibinfo {author} {\bibfnamefont {J.}~\bibnamefont
  {Paldus}},\ }\href {https://doi.org/10.1063/1.1701562} {\bibfield  {journal}
  {\bibinfo  {journal} {J. Chem. Phys.}\ }\textbf {\bibinfo {volume} {47}},\
  \bibinfo {pages} {3976} (\bibinfo {year} {1967})}\BibitemShut {NoStop}%
\bibitem [{\citenamefont {Seeger}\ and\ \citenamefont
  {Pople}(1977)}]{Seeger1977-stability}%
  \BibitemOpen
  \bibfield  {author} {\bibinfo {author} {\bibfnamefont {R.}~\bibnamefont
  {Seeger}}\ and\ \bibinfo {author} {\bibfnamefont {J.~A.}\ \bibnamefont
  {Pople}},\ }\href {https://doi.org/10.1063/1.434318} {\bibfield  {journal}
  {\bibinfo  {journal} {J. Chem. Phys.}\ }\textbf {\bibinfo {volume} {66}},\
  \bibinfo {pages} {3045} (\bibinfo {year} {1977})}\BibitemShut {NoStop}%
\bibitem [{\citenamefont {Bauernschmitt}\ and\ \citenamefont
  {Ahlrichs}(1996)}]{Bauernschmitt1996-stability}%
  \BibitemOpen
  \bibfield  {author} {\bibinfo {author} {\bibfnamefont {R.}~\bibnamefont
  {Bauernschmitt}}\ and\ \bibinfo {author} {\bibfnamefont {R.}~\bibnamefont
  {Ahlrichs}},\ }\href {https://doi.org/10.1063/1.471637} {\bibfield  {journal}
  {\bibinfo  {journal} {J. Chem. Phys.}\ }\textbf {\bibinfo {volume} {104}},\
  \bibinfo {pages} {9047} (\bibinfo {year} {1996})}\BibitemShut {NoStop}%
\bibitem [{\citenamefont {Lee}\ and\ \citenamefont
  {Taylor}(1989)}]{Lee1989-t1diag}%
  \BibitemOpen
  \bibfield  {author} {\bibinfo {author} {\bibfnamefont {T.~J.}\ \bibnamefont
  {Lee}}\ and\ \bibinfo {author} {\bibfnamefont {P.~R.}\ \bibnamefont
  {Taylor}},\ }\href {https://doi.org/10.1002/qua.560360824} {\bibfield
  {journal} {\bibinfo  {journal} {Int. J. Quantum Chem.}\ }\textbf {\bibinfo
  {volume} {36}},\ \bibinfo {pages} {199} (\bibinfo {year} {1989})}\BibitemShut
  {NoStop}%
\bibitem [{\citenamefont {Whitten}(1973)}]{Whitten1973-df}%
  \BibitemOpen
  \bibfield  {author} {\bibinfo {author} {\bibfnamefont {J.~L.}\ \bibnamefont
  {Whitten}},\ }\href {https://doi.org/10.1063/1.1679012} {\bibfield  {journal}
  {\bibinfo  {journal} {J. Chem. Phys.}\ }\textbf {\bibinfo {volume} {58}},\
  \bibinfo {pages} {4496} (\bibinfo {year} {1973})}\BibitemShut {NoStop}%
\bibitem [{\citenamefont {Dunlap}\ \emph {et~al.}(1979)\citenamefont {Dunlap},
  \citenamefont {Connolly},\ and\ \citenamefont {Sabin}}]{Dunlap1979-df}%
  \BibitemOpen
  \bibfield  {author} {\bibinfo {author} {\bibfnamefont {B.~I.}\ \bibnamefont
  {Dunlap}}, \bibinfo {author} {\bibfnamefont {J.~W.~D.}\ \bibnamefont
  {Connolly}},\ and\ \bibinfo {author} {\bibfnamefont {J.~R.}\ \bibnamefont
  {Sabin}},\ }\href {https://doi.org/10.1063/1.438728} {\bibfield  {journal}
  {\bibinfo  {journal} {J. Chem. Phys.}\ }\textbf {\bibinfo {volume} {71}},\
  \bibinfo {pages} {3396} (\bibinfo {year} {1979})}\BibitemShut {NoStop}%
\bibitem [{\citenamefont {Vahtras}\ \emph {et~al.}(1993)\citenamefont
  {Vahtras}, \citenamefont {Alml{\"o}f},\ and\ \citenamefont
  {Feyereisen}}]{Vahtras1993-ri}%
  \BibitemOpen
  \bibfield  {author} {\bibinfo {author} {\bibfnamefont {O.}~\bibnamefont
  {Vahtras}}, \bibinfo {author} {\bibfnamefont {J.}~\bibnamefont
  {Alml{\"o}f}},\ and\ \bibinfo {author} {\bibfnamefont {M.~W.}\ \bibnamefont
  {Feyereisen}},\ }\href {https://doi.org/10.1016/0009-2614(93)89151-7}
  {\bibfield  {journal} {\bibinfo  {journal} {Chem. Phys. Lett.}\ }\textbf
  {\bibinfo {volume} {213}},\ \bibinfo {pages} {514} (\bibinfo {year}
  {1993})}\BibitemShut {NoStop}%
\bibitem [{\citenamefont {Weigend}(2002)}]{Weigend2002-rihf}%
  \BibitemOpen
  \bibfield  {author} {\bibinfo {author} {\bibfnamefont {F.}~\bibnamefont
  {Weigend}},\ }\href {https://doi.org/10.1039/b204199p} {\bibfield  {journal}
  {\bibinfo  {journal} {Phys. Chem. Chem. Phys.}\ }\textbf {\bibinfo {volume}
  {4}},\ \bibinfo {pages} {4285} (\bibinfo {year} {2002})}\BibitemShut
  {NoStop}%
\bibitem [{\citenamefont {Hestenes}\ and\ \citenamefont
  {Stiefel}(1952)}]{Hestenes1952-cg}%
  \BibitemOpen
  \bibfield  {author} {\bibinfo {author} {\bibfnamefont {M.~R.}\ \bibnamefont
  {Hestenes}}\ and\ \bibinfo {author} {\bibfnamefont {E.}~\bibnamefont
  {Stiefel}},\ }\href {https://doi.org/10.6028/jres.049.044} {\bibfield
  {journal} {\bibinfo  {journal} {J. Res. Natl. Bur. Stand.}\ }\textbf
  {\bibinfo {volume} {49}},\ \bibinfo {pages} {409} (\bibinfo {year}
  {1952})}\BibitemShut {NoStop}%
\bibitem [{\citenamefont {Dunning}(1989)}]{Dunning1989-xc}%
  \BibitemOpen
  \bibfield  {author} {\bibinfo {author} {\bibfnamefont {T.~H.}\ \bibnamefont
  {Dunning}, \bibfnamefont {Jr}},\ }\href {https://doi.org/10.1063/1.456153}
  {\bibfield  {journal} {\bibinfo  {journal} {J. Chem. Phys.}\ }\textbf
  {\bibinfo {volume} {90}},\ \bibinfo {pages} {1007} (\bibinfo {year}
  {1989})}\BibitemShut {NoStop}%
\bibitem [{\citenamefont {Kendall}\ \emph {et~al.}(1992)\citenamefont
  {Kendall}, \citenamefont {Dunning},\ and\ \citenamefont
  {Harrison}}]{Kendall1992-wu}%
  \BibitemOpen
  \bibfield  {author} {\bibinfo {author} {\bibfnamefont {R.~A.}\ \bibnamefont
  {Kendall}}, \bibinfo {author} {\bibfnamefont {T.~H.}\ \bibnamefont
  {Dunning}},\ and\ \bibinfo {author} {\bibfnamefont {R.~J.}\ \bibnamefont
  {Harrison}},\ }\href {https://doi.org/10.1063/1.462569} {\bibfield  {journal}
  {\bibinfo  {journal} {J. Chem. Phys.}\ }\textbf {\bibinfo {volume} {96}},\
  \bibinfo {pages} {6796} (\bibinfo {year} {1992})}\BibitemShut {NoStop}%
\bibitem [{\citenamefont {Hehre}\ \emph {et~al.}(1969)\citenamefont {Hehre},
  \citenamefont {Stewart},\ and\ \citenamefont {Pople}}]{Hehre1969-xx}%
  \BibitemOpen
  \bibfield  {author} {\bibinfo {author} {\bibfnamefont {W.~J.}\ \bibnamefont
  {Hehre}}, \bibinfo {author} {\bibfnamefont {R.~F.}\ \bibnamefont {Stewart}},\
  and\ \bibinfo {author} {\bibfnamefont {J.~A.}\ \bibnamefont {Pople}},\ }\href
  {https://doi.org/10.1063/1.1672392} {\bibfield  {journal} {\bibinfo
  {journal} {J. Chem. Phys.}\ }\textbf {\bibinfo {volume} {51}},\ \bibinfo
  {pages} {2657} (\bibinfo {year} {1969})},\ \bibinfo {note} {sTO-3G minimal
  basis}\BibitemShut {NoStop}%
\bibitem [{\citenamefont {Pulay}(1980)}]{Pulay1980-diis}%
  \BibitemOpen
  \bibfield  {author} {\bibinfo {author} {\bibfnamefont {P.}~\bibnamefont
  {Pulay}},\ }\href@noop {} {\bibfield  {journal} {\bibinfo  {journal} {Chem.
  Phys. Lett.}\ }\textbf {\bibinfo {volume} {73}},\ \bibinfo {pages} {393}
  (\bibinfo {year} {1980})}\BibitemShut {NoStop}%
\bibitem [{\citenamefont {Pulay}(1982)}]{Pulay1982-diis}%
  \BibitemOpen
  \bibfield  {author} {\bibinfo {author} {\bibfnamefont {P.}~\bibnamefont
  {Pulay}},\ }\href@noop {} {\bibfield  {journal} {\bibinfo  {journal} {J.
  Comput. Chem.}\ }\textbf {\bibinfo {volume} {3}},\ \bibinfo {pages} {556}
  (\bibinfo {year} {1982})}\BibitemShut {NoStop}%
\bibitem [{\citenamefont {Broyden}(1970)}]{Broyden1970-bfgs}%
  \BibitemOpen
  \bibfield  {author} {\bibinfo {author} {\bibfnamefont {C.~G.}\ \bibnamefont
  {Broyden}},\ }\href@noop {} {\bibfield  {journal} {\bibinfo  {journal} {IMA
  J. Appl. Math.}\ }\textbf {\bibinfo {volume} {6}},\ \bibinfo {pages} {76}
  (\bibinfo {year} {1970})}\BibitemShut {NoStop}%
\bibitem [{\citenamefont {Fletcher}(1970)}]{Fletcher1970-bfgs}%
  \BibitemOpen
  \bibfield  {author} {\bibinfo {author} {\bibfnamefont {R.}~\bibnamefont
  {Fletcher}},\ }\href@noop {} {\bibfield  {journal} {\bibinfo  {journal}
  {Comput. J.}\ }\textbf {\bibinfo {volume} {13}},\ \bibinfo {pages} {317}
  (\bibinfo {year} {1970})}\BibitemShut {NoStop}%
\bibitem [{\citenamefont {Goldfarb}(1970)}]{Goldfarb1970-bfgs}%
  \BibitemOpen
  \bibfield  {author} {\bibinfo {author} {\bibfnamefont {D.}~\bibnamefont
  {Goldfarb}},\ }\href@noop {} {\bibfield  {journal} {\bibinfo  {journal}
  {Math. Comput.}\ }\textbf {\bibinfo {volume} {24}},\ \bibinfo {pages} {23}
  (\bibinfo {year} {1970})}\BibitemShut {NoStop}%
\bibitem [{\citenamefont {Shanno}(1970)}]{Shanno1970-bfgs}%
  \BibitemOpen
  \bibfield  {author} {\bibinfo {author} {\bibfnamefont {D.~F.}\ \bibnamefont
  {Shanno}},\ }\href@noop {} {\bibfield  {journal} {\bibinfo  {journal} {Math.
  Comput.}\ }\textbf {\bibinfo {volume} {24}},\ \bibinfo {pages} {647}
  (\bibinfo {year} {1970})}\BibitemShut {NoStop}%
\bibitem [{\citenamefont {Barlini}\ \emph {et~al.}(2025)\citenamefont
  {Barlini}, \citenamefont {Bianchi}, \citenamefont {Trabski}, \citenamefont
  {Bloino},\ and\ \citenamefont {Koch}}]{Barlini2025}%
  \BibitemOpen
  \bibfield  {author} {\bibinfo {author} {\bibfnamefont {A.}~\bibnamefont
  {Barlini}}, \bibinfo {author} {\bibfnamefont {A.}~\bibnamefont {Bianchi}},
  \bibinfo {author} {\bibfnamefont {J.~H.~M.}\ \bibnamefont {Trabski}},
  \bibinfo {author} {\bibfnamefont {J.}~\bibnamefont {Bloino}},\ and\ \bibinfo
  {author} {\bibfnamefont {H.}~\bibnamefont {Koch}},\ }\href
  {https://doi.org/10.1021/acs.jctc.5c00680} {\bibfield  {journal} {\bibinfo
  {journal} {J. Chem. Theory Comput.}\ }\textbf {\bibinfo {volume} {21}},\
  \bibinfo {pages} {9323} (\bibinfo {year} {2025})}\BibitemShut {NoStop}%
\bibitem [{\citenamefont {Monzel}\ and\ \citenamefont
  {Stopkowicz}(2026)}]{Monzel2026-qedcc-sym}%
  \BibitemOpen
  \bibfield  {author} {\bibinfo {author} {\bibfnamefont {L.}~\bibnamefont
  {Monzel}}\ and\ \bibinfo {author} {\bibfnamefont {S.}~\bibnamefont
  {Stopkowicz}},\ }\href {https://doi.org/10.1021/acs.jctc.5c01343} {\bibfield
  {journal} {\bibinfo  {journal} {J. Chem. Theory Comput.}\ }\textbf {\bibinfo
  {volume} {22}},\ \bibinfo {pages} {5991} (\bibinfo {year}
  {2026})}\BibitemShut {NoStop}%
\bibitem [{\citenamefont {Zabrodsky}\ \emph {et~al.}(1992)\citenamefont
  {Zabrodsky}, \citenamefont {Peleg},\ and\ \citenamefont
  {Avnir}}]{Zabrodsky1992-csm}%
  \BibitemOpen
  \bibfield  {author} {\bibinfo {author} {\bibfnamefont {H.}~\bibnamefont
  {Zabrodsky}}, \bibinfo {author} {\bibfnamefont {S.}~\bibnamefont {Peleg}},\
  and\ \bibinfo {author} {\bibfnamefont {D.}~\bibnamefont {Avnir}},\ }\href
  {https://doi.org/10.1021/ja00046a033} {\bibfield  {journal} {\bibinfo
  {journal} {J. Am. Chem. Soc.}\ }\textbf {\bibinfo {volume} {114}},\ \bibinfo
  {pages} {7843} (\bibinfo {year} {1992})}\BibitemShut {NoStop}%
\bibitem [{\citenamefont {Kowalewski}\ and\ \citenamefont
  {Seeber}(2022)}]{Kowalewski2022-nix}%
  \BibitemOpen
  \bibfield  {author} {\bibinfo {author} {\bibfnamefont {M.}~\bibnamefont
  {Kowalewski}}\ and\ \bibinfo {author} {\bibfnamefont {P.}~\bibnamefont
  {Seeber}},\ }\href {https://doi.org/10.1002/qua.26872} {\bibfield  {journal}
  {\bibinfo  {journal} {Int. J. Quantum Chem.}\ }\textbf {\bibinfo {volume}
  {122}},\ \bibinfo {pages} {e26872} (\bibinfo {year} {2022})}\BibitemShut
  {NoStop}%
\bibitem [{\citenamefont {Fischer}(2025)}]{Fischer2025}%
  \BibitemOpen
  \bibfield  {author} {\bibinfo {author} {\bibfnamefont {E.~W.}\ \bibnamefont
  {Fischer}},\ }\href@noop {} {\bibfield  {journal} {\bibinfo  {journal} {J.
  Chem. Theory Comput.}\ }\textbf {\bibinfo {volume} {21}},\ \bibinfo {pages}
  {12081} (\bibinfo {year} {2025})}\BibitemShut {NoStop}%
\bibitem [{\citenamefont {Fetherolf}\ \emph {et~al.}(2026)\citenamefont
  {Fetherolf}, \citenamefont {Duong}, \citenamefont {Li},\ and\ \citenamefont
  {Hammes-Schiffer}}]{Fetherolf2026-dc}%
  \BibitemOpen
  \bibfield  {author} {\bibinfo {author} {\bibfnamefont {J.~H.}\ \bibnamefont
  {Fetherolf}}, \bibinfo {author} {\bibfnamefont {T.}~\bibnamefont {Duong}},
  \bibinfo {author} {\bibfnamefont {T.~E.}\ \bibnamefont {Li}},\ and\ \bibinfo
  {author} {\bibfnamefont {S.}~\bibnamefont {Hammes-Schiffer}},\ }\bibfield
  {journal} {\bibinfo  {journal} {ACS Photonics}\ }\href
  {https://doi.org/10.1021/acsphotonics.5c03076} {10.1021/acsphotonics.5c03076}
  (\bibinfo {year} {2026})\BibitemShut {NoStop}%
\bibitem [{\citenamefont {Gu}\ \emph {et~al.}(2023)\citenamefont {Gu},
  \citenamefont {Si}, \citenamefont {Li}, \citenamefont {Gao}, \citenamefont
  {Wang},\ and\ \citenamefont {Zhang}}]{Gu2023-uq}%
  \BibitemOpen
  \bibfield  {author} {\bibinfo {author} {\bibfnamefont {K.}~\bibnamefont
  {Gu}}, \bibinfo {author} {\bibfnamefont {Q.}~\bibnamefont {Si}}, \bibinfo
  {author} {\bibfnamefont {N.}~\bibnamefont {Li}}, \bibinfo {author}
  {\bibfnamefont {F.}~\bibnamefont {Gao}}, \bibinfo {author} {\bibfnamefont
  {L.}~\bibnamefont {Wang}},\ and\ \bibinfo {author} {\bibfnamefont
  {F.}~\bibnamefont {Zhang}},\ }\bibfield  {journal} {\bibinfo  {journal} {ACS
  Photonics}\ }\href {https://doi.org/10.1021/acsphotonics.3c00243}
  {10.1021/acsphotonics.3c00243} (\bibinfo {year} {2023})\BibitemShut {NoStop}%
\end{thebibliography}
\end{document}


\title{Supporting Information: \\ Enabling Polaritonic Electronic Structure Theory for Psi4:\\ The \texttt{psi4-cbo} package}

\author{Thomas Schnappinger}
\email{thomas.schnappinger@fysik.su.se}
\affiliation{Department of Physics, Stockholm University, AlbaNova University Center, SE-106 91 Stockholm, Sweden}

\author{Markus Kowalewski}
\affiliation{Department of Physics, Stockholm University, AlbaNova University Center, SE-106 91 Stockholm, Sweden}
\date{\today}%

\begin{abstract}

\end{abstract}

\maketitle

\tableofcontents

\clearpage

\section{Numerical verification of the \glsentryshort{cbohf} and \glsentryshort{qedhf} identity}

Table~\ref{tab:identity} compares two independent \gls{scf} solutions, testing the identity~\cite{Angelico2023-jh} given in Eq.~(9) of the manuscript: \gls{cbohf} with the photon coordinate optimized (\texttt{cbo\_rhf} with \texttt{qopt=True}) and the coherent-state \gls{qedhf} of Eq.~(7) (\texttt{cbo\_qedhf}), which optimizes the orbitals and the displacement amplitudes $z_m$ together.
Both use the same geometry, basis and \gls{scf} thresholds (\texttt{scf\_type pk}, $E_{\rm conv}=10^{-12}\,E_h$, $D_{\rm conv}=10^{-10}$), and a single $z$-polarized mode at $\omega_c=\SI{0.5}{\au}$.
The residual never exceeds $2.6\times10^{-13}$~$E_h$ across five basis sets, three coupling strengths and two molecules, the densities agree to $1.2\times10^{-9}$, and $z_m=\sqrt{\omega_m/2}\,q_m^{\rm opt}$ holds to $2.0\times10^{-9}$.

\begin{table}[h]
\centering
\caption{Absolute residual $\big|E_{\rm \gls{qedhf}} - E_{\rm \gls{cbohf}}(q^{\rm opt})
- \tfrac12\omega_c\big|$, in units of $10^{-13}\,E_h$, for a single $z$-polarized mode at
$\omega_c=0.5$~a.u. Entries of $0$ are exact zeros in
double precision.}
\begin{tabular}{lcccccc}
\hline
& \multicolumn{3}{c}{\ce{HF}} & \multicolumn{3}{c}{\ce{H2O}} \\
& \multicolumn{3}{c}{$\lambda_c$} & \multicolumn{3}{c}{$\lambda_c$} \\
\cmidrule(lr){2-4}\cmidrule(lr){5-7}
basis & $0.00$ & $0.05$ & $0.10$ & $0.00$ & $0.05$ & $0.10$ \\
\hline
cc-pVDZ & $0$ & $0.14$ & $0.14$ & $0$ & $0.14$ & $0.43$ \\
aug-cc-pVDZ & $0$ & $0.99$ & $0.57$ & $0$ & $0.28$ & $1.14$ \\
cc-pVTZ & $0$ & $2.27$ & $0.57$ & $0$ & $0.28$ & $0.14$ \\
aug-cc-pVTZ & $0$ & $2.13$ & $1.14$ & $0$ & $0.14$ & $0.99$ \\
cc-pVQZ & $0$ & $0.99$ & $0.71$ & $0$ & $2.56$ & $0.71$ \\
\hline
\end{tabular}
\label{tab:identity}
\end{table}

\clearpage

\section{AO-level working equations for \glsentrylong{cbohf}}

Section~II states the \gls{cboa} Hamiltonian (Eq.~(5)) at the operator level.
The energy actually evaluated is, in the AO basis with the spin-summed density matrix $D$,
\begin{equation}
\langle E_{\rm \gls{cbohf}}\rangle = \sum_{\alpha\beta}D_{\alpha\beta}\langle\alpha|\tilde h|\beta\rangle
+ \frac12\sum_{\alpha\beta\gamma\delta}D_{\alpha\beta}D_{\gamma\delta}\langle\alpha\beta|\tilde g|\gamma\delta\rangle
+ \tilde E_{n,c} ,
\end{equation}
with the cavity-dressed one- and two-electron integrals
\begin{equation}
\label{eq:1e_2e_ao}
\begin{split}
\langle\alpha|\tilde h|\beta\rangle &= \langle\alpha|\hat h|\beta\rangle + \omega_c q_c
\langle\alpha|\boldsymbol\lambda_c\cdot\hat{\mathbf r}|\beta\rangle -
\boldsymbol\lambda_c\cdot\boldsymbol\mu_{\rm nuc}\,\langle\alpha|\boldsymbol\lambda_c\cdot\hat{\mathbf r}|\beta\rangle
+ \tfrac12\langle\alpha|(\boldsymbol\lambda_c\cdot\hat{\mathbf r})^2|\beta\rangle , \\
\langle\alpha\beta|\tilde g|\gamma\delta\rangle &= \langle\alpha\beta|g|\gamma\delta\rangle - \tfrac12\langle\alpha\delta|g|\gamma\beta\rangle
+ \langle\alpha|\boldsymbol\lambda_c\cdot\hat{\mathbf r}|\beta\rangle\langle\gamma|\boldsymbol\lambda_c\cdot\hat{\mathbf r}|\delta\rangle
- \tfrac12\langle\alpha|\boldsymbol\lambda_c\cdot\hat{\mathbf r}|\delta\rangle\langle\gamma|\boldsymbol\lambda_c\cdot\hat{\mathbf r}|\beta\rangle ,
\end{split}
\end{equation}
where $\hat h$ and $g$ are the ordinary integrals, $\hat{\mathbf r}$ is the electron position operator (the electronic dipole operator is $-\hat{\mathbf r}$), $q_c$ the converged photon coordinate, and $\tilde E_{n,c}$ collects the nuclear and constant cavity terms.
The one-electron piece carries the linear light-matter term and half the \gls{dse} self-term.
The two-electron piece is an ordinary Coulomb/exchange integral plus the Hartree-like and exchange-like dipole-dipole contractions of the \gls{dse}~\cite{Schnappinger2023-cbohf}.

\clearpage

\section{Correlation and stability across systems and basis sets}

Figure~1 in the main manuscript resolves $\varepsilon_{ee}$ and $\varepsilon_{ep}$ in $(\lambda,\omega_c)$ for one system.
Tables~\ref{tab:eps_systems} and~\ref{tab:eps_basis} give complementary cuts at fixed $\omega_c$, for four systems and for five basis sets of \ce{H2}.

\begin{table}[htb]
\centering
\caption{Three-way correlation split across systems, STO-3G, single $z$-polarized mode at
$\omega_c=\SI{3550}{\per\centi\meter}$. Energies in $E_h$; the last column is the ratio that
Fig.~1 maps. The $\lambda=0$ row of \ce{H2} confirms that $\varepsilon_{ep}$ vanishes
identically there (values are numerical noise).}
\begin{tabular}{lccccc}
\hline
System & $\lambda$ & $\varepsilon_{ee}$ & $\varepsilon_{ep}$ (bare) & $\varepsilon_{ep}$ (dressed) & $|\varepsilon_{ep}|/|\varepsilon_{ee}|$ \\
\hline
\ce{H2}  & 0.00 & $-2.0525\times10^{-2}$ & $+1.6\times10^{-15}$ & $+1.6\times10^{-15}$ & $8\times10^{-14}$ \\
\ce{H2}  & 0.02 & $-2.0599\times10^{-2}$ & $-4.412\times10^{-6}$ & $+5.457\times10^{-7}$ & $2.1\times10^{-4}$ \\
\ce{H2}  & 0.05 & $-2.0989\times10^{-2}$ & $-2.746\times10^{-5}$ & $+3.451\times10^{-6}$ & $1.3\times10^{-3}$ \\
\ce{H2}  & 0.10 & $-2.2409\times10^{-2}$ & $-1.083\times10^{-4}$ & $+1.438\times10^{-5}$ & $4.8\times10^{-3}$ \\
\hline
\ce{LiH} & 0.02 & $-2.0415\times10^{-2}$ & $-1.493\times10^{-5}$ & $-5.369\times10^{-6}$ & $7.3\times10^{-4}$ \\
\ce{LiH} & 0.05 & $-2.0619\times10^{-2}$ & $-8.403\times10^{-5}$ & $-2.835\times10^{-5}$ & $4.1\times10^{-3}$ \\
\ce{LiH} & 0.10 & $-2.1452\times10^{-2}$ & $-2.530\times10^{-4}$ & $-6.479\times10^{-5}$ & $1.2\times10^{-2}$ \\
\hline
\ce{HF}  & 0.02 & $-2.5906\times10^{-2}$ & $-4.866\times10^{-6}$ & $+2.089\times10^{-7}$ & $1.9\times10^{-4}$ \\
\ce{HF}  & 0.05 & $-2.6233\times10^{-2}$ & $-3.028\times10^{-5}$ & $+1.280\times10^{-6}$ & $1.2\times10^{-3}$ \\
\ce{HF}  & 0.10 & $-2.7404\times10^{-2}$ & $-1.193\times10^{-4}$ & $+4.750\times10^{-6}$ & $4.4\times10^{-3}$ \\
\hline
\ce{H2O} & 0.02 & $-4.9611\times10^{-2}$ & $-3.368\times10^{-6}$ & $+6.928\times10^{-8}$ & $6.8\times10^{-5}$ \\
\ce{H2O} & 0.05 & $-4.9903\times10^{-2}$ & $-2.096\times10^{-5}$ & $+4.241\times10^{-7}$ & $4.2\times10^{-4}$ \\
\ce{H2O} & 0.10 & $-5.0957\times10^{-2}$ & $-8.255\times10^{-5}$ & $+1.576\times10^{-6}$ & $1.6\times10^{-3}$ \\
\hline
\end{tabular}
\label{tab:eps_systems}
\end{table}

\begin{table}[htb]
\centering
\caption{Basis-set dependence of the same split, \ce{H2} at $\lambda=0.10$,
$\omega_c=\SI{3550}{\per\centi\meter}$. Energies in $E_h$.}
\begin{tabular}{lcccc}
\hline
Basis & $\varepsilon_{ee}$ & $\varepsilon_{ep}$ (bare) & $\varepsilon_{ep}$ (dressed) & $|\varepsilon_{ep}|/|\varepsilon_{ee}|$ \\
\hline
STO-3G      & $-2.2409\times10^{-2}$ & $-1.083\times10^{-4}$ & $+1.438\times10^{-5}$ & $4.8\times10^{-3}$ \\
6-31G       & $-2.6908\times10^{-2}$ & $-2.271\times10^{-4}$ & $+2.341\times10^{-5}$ & $8.4\times10^{-3}$ \\
cc-pVDZ     & $-3.6280\times10^{-2}$ & $-2.338\times10^{-4}$ & $+1.034\times10^{-5}$ & $6.4\times10^{-3}$ \\
cc-pVTZ     & $-4.0968\times10^{-2}$ & $-2.386\times10^{-4}$ & $+8.328\times10^{-6}$ & $5.8\times10^{-3}$ \\
aug-cc-pVDZ & $-3.7318\times10^{-2}$ & $-2.346\times10^{-4}$ & $+4.838\times10^{-6}$ & $6.3\times10^{-3}$ \\
\hline
\end{tabular}
\label{tab:eps_basis}
\end{table}

$\varepsilon_{ee}$ nearly doubles from STO-3G to cc-pVTZ while $\varepsilon_{ep}$ is almost basis-insensitive above the minimal basis, so the ratio drifts downward as $\varepsilon_{ee}$ converges; STO-3G, which gives about half the $\varepsilon_{ep}$ of the larger bases, yields the smallest ratio and thus understates electron-photon correlation.
The system-to-system spread is a factor of 7 to 11 at fixed $(\lambda,\omega_c)$, ordered by the coupling operator, with \ce{LiH} highest and \ce{H2O} lowest.

The no-go factor $\omega_{\rm eff}^2/\omega_c^2=1-\boldsymbol\lambda^\top\boldsymbol\alpha\boldsymbol\lambda$, that is $T_3$ of Sec.~III\,C in the main manuscript in units of $\omega_c^2$, inherits the basis-set dependence of the polarizability, which Table~\ref{tab:nogo_basis_conv} follows for two monomers up the plain and augmented cc-pV$X$Z ladders.
The polarizability is strongly diffuse-function-sensitive, rising for \ce{H2O} from $4.99$ to $8.42$~a.u., so the photon diagnostic should be evaluated in an augmented basis.
Along the augmented series $\alpha_{\rm iso}$ changes by $0.9\%$ (\ce{H2O}) and $2.4\%$ (\ce{HF}) between triple- and quadruple-zeta, and the no-go factor settles at $0.9867$ (\ce{H2O}) and $0.9910$ (\ce{HF}).

\begin{table}[htb]
\centering
\caption{Basis-set convergence of the isotropic polarizability $\alpha_{\rm iso}$ (a.u.) and the
no-go factor $\omega_{\rm eff}^2/\omega_c^2$, single $z$-polarized mode, $\lambda=0.04$, electronic-scale
$\omega_c$. Both quantities converge along the augmented series.}
\begin{tabular}{lcccc}
\hline
& \multicolumn{2}{c}{\ce{H2O}} & \multicolumn{2}{c}{HF} \\
\cmidrule(lr){2-3}\cmidrule(lr){4-5}
Basis & $\alpha_{\rm iso}$ & $\omega_{\rm eff}^2/\omega_c^2$ & $\alpha_{\rm iso}$ & $\omega_{\rm eff}^2/\omega_c^2$ \\
\hline
cc-pVDZ     & 4.991 & 0.99194 & 2.354 & 0.99363 \\
cc-pVTZ     & 6.578 & 0.98938 & 3.334 & 0.99220 \\
cc-pVQZ     & 7.309 & 0.98824 & 3.909 & 0.99160 \\
aug-cc-pVDZ & 8.064 & 0.98738 & 4.354 & 0.99120 \\
aug-cc-pVTZ & 8.344 & 0.98688 & 4.714 & 0.99100 \\
aug-cc-pVQZ & 8.422 & 0.98674 & 4.829 & 0.99095 \\
\hline
\end{tabular}
\label{tab:nogo_basis_conv}
\end{table}

\clearpage

\section{Analytic-versus-numeric Hessian crossover}

Table~\ref{tab:hess_crossover} benchmarks the analytic joint $(R,q)$ Hessian against the numeric fallback on random water clusters, measured as cgroup peak memory; for the numeric route, the increase above the baseline process after importing Psi4 is given.
The numeric route is accelerated by density fitting~\cite{Whitten1973-df,Dunlap1979-df,Vahtras1993-ri,Weigend2002-rihf}, which replaces the four-index two-electron integrals by three-index quantities contracted with an auxiliary basis, as implemented in PSI4~\cite{Parrish2017-psi4,Smith2020-kq}.
The density-fitted \gls{scf} and gradient use the \texttt{-JKFIT} auxiliary sets matched to the orbital basis~\cite{Weigend2002-rihf}; density-fitted response builds, as in the analytic Hessian and the stability analysis, use the corresponding \texttt{-RIFIT} sets.

\begin{table}[htb]
\centering
\caption{Analytic against numeric joint Hessian, random water clusters, cc-pVDZ, both with
\texttt{density\_fitting=True}, which for the analytic route fits only the response build. Memory is
the cgroup-measured peak, for the numeric route the increase above the baseline after importing
Psi4; times are relative to the fastest run in the table, the analytic build at $N=1$. The numeric
timing at $N=7$ is an outlier of the benchmark run and is omitted.}
\begin{tabular}{cccccc}
\hline
& & \multicolumn{2}{c}{peak memory} & \multicolumn{2}{c}{relative time} \\
\cmidrule(lr){3-4}\cmidrule(lr){5-6}
$N_{\rm H_2O}$ & $n_{\rm bf}$ & analytic & numeric & analytic & numeric \\
\hline
1 & 24  & 0.71~GB & 199~MB & $1.0$    & $1.9$   \\
2 & 48  & 1.21~GB & 217~MB & $10$     & $8.4$   \\
3 & 72  & 3.00~GB & 215~MB & $52$     & $11$    \\
4 & 96  & 7.67~GB & 215~MB & $185$    & $22$    \\
5 & 120 & 17.4~GB & 222~MB & $502$    & $42$    \\
6 & 144 & 34.0~GB & 228~MB & $1227$   & $90$    \\
7 & 168 & 61.5~GB & 233~MB & $2653$   & ---     \\
8 & 192 & 104~GB  & 245~MB & $4799$   & $208$   \\
9 & 216 & 165~GB  & 363~MB & $9410$   & $329$   \\
\hline
\end{tabular}
\label{tab:hess_crossover}
\end{table}

Analytic peak memory scales as $\sim n_{\rm bf}^4$, the second-derivative two-electron integral signature, and walls at $N\approx9$--$10$ water molecules: an $N=10$ attempt was killed by the OOM handler at \SI{201}{\giga\byte} on the way to a predicted \SI{253}{\giga\byte}.
Density fitting cannot rescue the analytic route there, because PSI4~\cite{Parrish2017-psi4,Smith2020-kq} exposes fitted derivative integrals only at first order, so it accelerates the response build alone and not the two-electron derivative integrals that dominate memory.
The memory of the numeric route instead stays nearly flat in $N$, between \SI{199}{\mega\byte} and \SI{363}{\mega\byte} above the baseline, against \SI{165}{\giga\byte} for the analytic route at $N=9$, and it is the faster of the two from $N=2$ onwards.

\clearpage

\section{Symmetry tools: selection-rule labeling and the continuous symmetry measure}

Table~\ref{tab:sym_descent} illustrates both symmetry diagnostics on water ($C_{2v}$), rotating the cavity polarization axis by $\theta$ away from the molecular $C_2$ axis in the molecular plane, at fixed $\lambda=0.1$.
The selection-rule diagnostic decomposes the polarization direction into the point group's irreducible representations, and a vibrational mode couples linearly to the cavity if and only if its irrep appears in that decomposition.

The continuous symmetry measure $\mathrm{CSM}(R)$ is defined in Eq.~(18) of the main text; $1$ means that the density respects the operation $R$ exactly.

\begin{table}[htb]
\centering
\caption{Water, $C_{2v}$, $\lambda=0.1$, $\omega_c=\SI{1600}{\per\centi\meter}$: cavity
polarization rotated by $\theta$ from the $C_2$ axis in-plane. ``Residual'' is the largest
subgroup of $C_{2v}$ the polarization direction respects; CSM($C_2$) is the continuous symmetry
measure of the $C_2$ operation on the coupled density; polarization irrep weights are the
decomposition of $\hat{\mathbf e}$ onto $C_{2v}$'s irreps; $n_{\rm bright}$ is the number of modes
predicted (and confirmed) to couple.}
\begin{tabular}{cccccc}
\hline
$\theta$ (deg) & Residual group & CSM($C_2$) & $w(A_1)$ & $w(B_2)$ & $n_{\rm bright}$ \\
\hline
0  & $C_{2v}$ (order 4) & 1.0000 & 1.000 & 0.000 & 2 \\
15 & $C_s$ (order 2)    & 0.9986 & 0.933 & 0.067 & 3 \\
30 & $C_s$ (order 2)    & 0.9976 & 0.750 & 0.250 & 3 \\
45 & $C_s$ (order 2)    & 0.9972 & 0.500 & 0.500 & 3 \\
60 & $C_s$ (order 2)    & 0.9975 & 0.250 & 0.750 & 3 \\
75 & $C_s$ (order 2)    & 0.9986 & 0.067 & 0.933 & 3 \\
90 & $C_s$ (order 2)    & 1.0000 & 0.000 & 1.000 & 1 \\
\hline
\end{tabular}
\label{tab:sym_descent}
\end{table}

\begin{figure}[htb]
\centering
\includegraphics[width=0.65\linewidth]{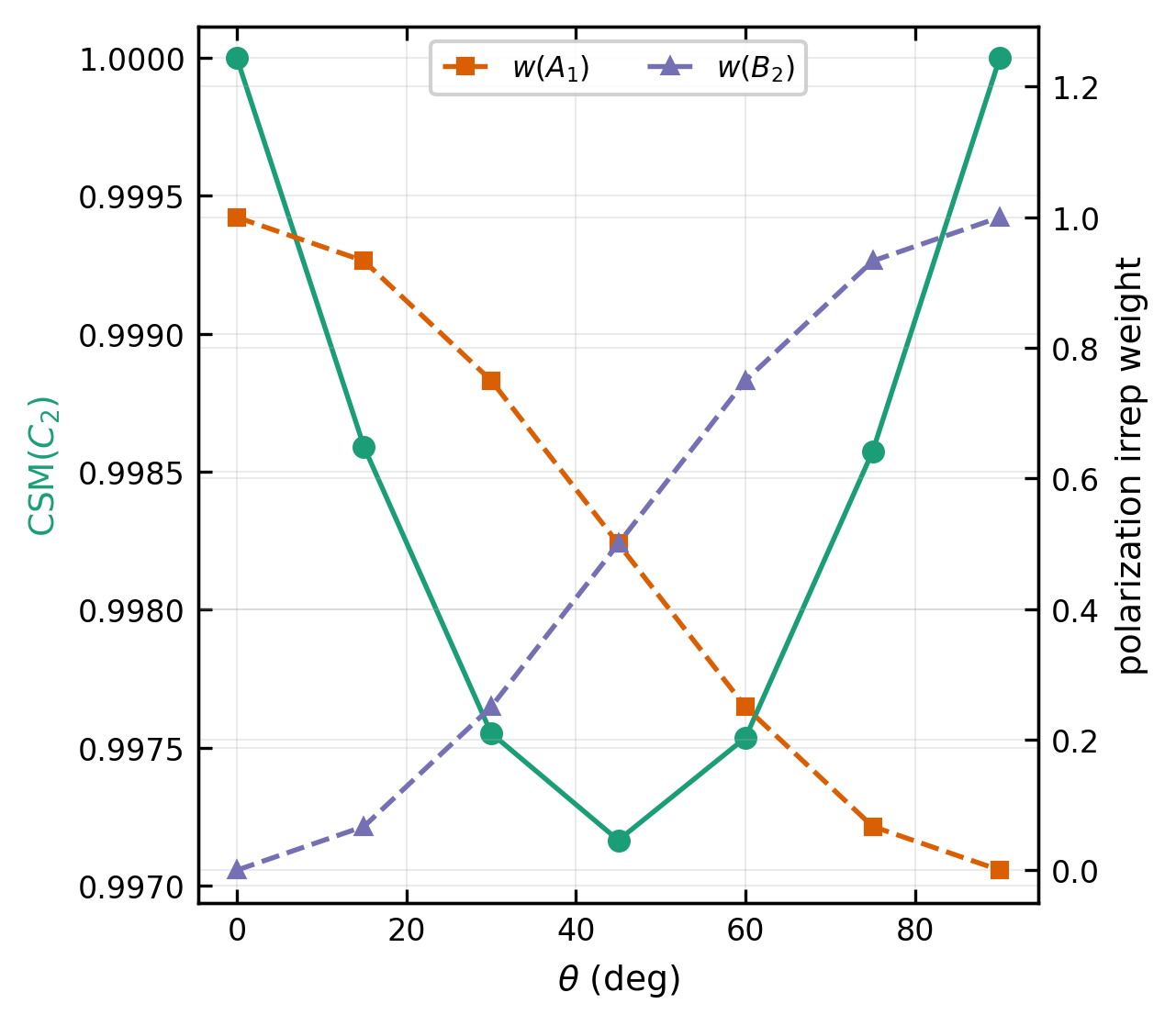}
\caption{$\mathrm{CSM}(C_2)$ (left axis) and the $A_1$/$B_2$ polarization-irrep weights (right
axis) against the polarization-rotation angle $\theta$, water, $C_{2v}$, $\lambda=0.1$, the same
seven points as Table~\ref{tab:sym_descent}. The two symmetry-adapted angles ($\theta=0^\circ,90^\circ$)
are marked distinctly from the five intermediate ones.}
\label{fig:sym_descent}
\end{figure}

\clearpage

\section{Matrix-free polarizabilities: the conjugate-gradient solver}

The $(A+B)$ response matrix~\cite{Pople1979-cphf} was explicitly formed in Ref.~\cite{Schnappinger2025-fx}, but it does not need to.
For a trial vector $X$ of dimension $n_{\rm occ}\times n_{\rm vir}$ with trial density $D^X=C_{\rm occ}XC_{\rm vir}^{\!\top}$, a single Coulomb/exchange build gives the matrix-vector product directly,
\begin{equation}
\label{eq:sigma_ab}
\big[(A+B)X\big]_{ia} = (\epsilon_a-\epsilon_i)X_{ia}
+ \big[C_{\rm occ}^{\!\top}\big(4J[D^X]-K[D^X]-K[D^X]^{\!\top}\big)C_{\rm vir}\big]_{ia}
+ \big[(A+B)^{\rm DSE}X\big]_{ia} ,
\end{equation}
where the \gls{dse} contribution of each mode, a rank-1 Coulomb-like term and two exchange-like terms, is built from the same trial density.
This $\sigma$-vector is used as the \texttt{matvec} of a \texttt{scipy} \texttt{LinearOperator} and solved by preconditioned conjugate gradients~\cite{Hestenes1952-cg} to $\mathrm{atol}=10^{-8}$, which replaces storage of the $n_{\rm ov}\times n_{\rm ov}$ matrix by one JK build per iteration, in the spirit of the direct \gls{scf} approach~\cite{Almlof1982-direct} where integrals are recomputed rather than stored.
The JK build follows the Psi4 option \texttt{scf\_type} and is density fitted~\cite{Whitten1973-df,Dunlap1979-df,Vahtras1993-ri,Weigend2002-rihf} with \texttt{scf\_type df}, using PSI4's fitted JK infrastructure~\cite{Parrish2017-psi4}.

Table~\ref{tab:cg_polar} benchmarks the solver against forming $(A+B)$ explicitly, for aniline across a Dunning sweep with cgroup-measured peak memory.
The explicit route solves the stored matrix by LU decomposition for $n_{\rm ov}<500$ and by conjugate gradients above; for aniline, $n_{\rm ov}=2700$ already at cc-pVDZ.
This is a performance benchmark rather than a physics result, run with two degenerate modes polarized along $x$ and $y$ at $\omega_c=\SI{22000}{\per\centi\meter}$ and dipole of aniline perpendicular to both.

\begin{table}[htb]
\centering
\caption{CG polarizability solve, aniline, $n_{\rm occ}=25$. ``CG peak'' is the memory increase
above the converged \gls{scf}; times are relative to the CG solve at cc-pVDZ.}
\begin{tabular}{ccccc}
\hline
Basis & $n_{\rm bf}$ & $(A+B)$ size & CG peak & CG time \\
\hline
cc-pVDZ     & 133 & 58~MB   & 48~MB   & $1.0$  \\
aug-cc-pVDZ & 224 & 198~MB  & 173~MB  & $3.0$  \\
cc-pVTZ     & 308 & 400~MB  & 289~MB  & $4.2$  \\
aug-cc-pVTZ & 483 & 1049~MB & 972~MB  & $15$   \\
cc-pVQZ     & 595 & 1624~MB & 1425~MB & $22$   \\
aug-cc-pVQZ & 882 & 3672~MB & 4611~MB & $82$   \\
\hline
\end{tabular}
\label{tab:cg_polar}
\end{table}

The explicit route, run at cc-pVDZ only, needs \SI{561}{\mega\byte} and $111$ times the CG time; larger bases were skipped.
The CG increment is dominated by the JK build and exceeds the stored $(A+B)$ matrix at aug-cc-pVQZ, so the gain of the matrix-free route is time and the absence of the $n_{\rm ov}^2$ build, not memory at every size.

\clearpage

\section{The Hessian rescale law: proof and validation}

\subsection{Proof}
\label{sec:rescale_proof}

The coupling vectors $\boldsymbol\lambda_m$ are held fixed while the mode frequencies change, and the photon displacement coordinates are classical parameters of the electronic problem.
In the \gls{cboa} Hamiltonian,
\begin{equation*}
\hat H_{\rm CBO}\big(\mathbf R;\mathbf q;\boldsymbol\omega\big) = \hat H_{\rm el}(\mathbf R)
  + \sum_{m=1}^{N_{\rm mod}}\Big[\tfrac12\omega_m^2 q_m^2 - \omega_m q_m\hat d_m + \tfrac12\hat d_m^2\Big],
\qquad \hat d_m=\boldsymbol\lambda_m\cdot\hat{\boldsymbol\mu},
\end{equation*}
coordinates and frequencies enter only through $x_m=\omega_m q_m$:
\begin{equation*}
\hat H_{\rm CBO}\big(\mathbf R;\mathbf q;\boldsymbol\omega\big) = \tilde H(\mathbf R;\mathbf x)
 = \hat H_{\rm el}(\mathbf R) + \sum_{m=1}^{N_{\rm mod}}\Big[\tfrac12 x_m^2 - x_m\hat d_m + \tfrac12\hat d_m^2\Big].
\end{equation*}
$\tilde H$ contains no frequency.
Every method that treats $\mathbf q$ as parameters of the electronic problem, \gls{cbohf}, CBO-DFT or a correlated method built on the same Hamiltonian, therefore gives an energy $E(\mathbf R,\mathbf q;\boldsymbol\omega)=\tilde E(\mathbf R,\mathbf x)$ with $\mathbf x=\boldsymbol\Omega\mathbf q$, $\boldsymbol\Omega={\rm diag}(\omega_1,\dots,\omega_{N_{\rm mod}})$, and $\tilde E$ independent of $\boldsymbol\omega$.
The same holds for the orbitals, the density and all expectation values, in particular the dipole moment, $\boldsymbol\mu(\mathbf R,\mathbf q;\boldsymbol\omega)=\tilde{\boldsymbol\mu}(\mathbf R,\mathbf x)$.

With $\boldsymbol\Omega$ constant, the chain rule gives $\partial E/\partial\mathbf R=\partial\tilde E/\partial\mathbf R$ and $\partial E/\partial q_m=\omega_m\,\partial\tilde E/\partial x_m$, and for the second derivatives
\begin{equation*}
\mathbf H_{\mathbf R\mathbf R} = \tilde{\mathbf H}_{\mathbf R\mathbf R} ,\qquad
\mathbf H_{\mathbf R q_m} = \omega_m\,\tilde{\mathbf H}_{\mathbf R x_m} ,\qquad
H_{q_m q_n} = \omega_m\omega_n\,\tilde H_{x_m x_n} ,
\end{equation*}
that is, with $\mathbf D(\boldsymbol\omega)={\rm diag}(\mathbf 1_{3N_{\rm at}},\boldsymbol\Omega)$,
\begin{equation*}
\mathbf H(\boldsymbol\omega) = \mathbf D(\boldsymbol\omega)\,\tilde{\mathbf H}\,\mathbf D(\boldsymbol\omega) ,
\end{equation*}
where $\tilde{\mathbf H}$ is the Hessian of $\tilde E$ with respect to $(\mathbf R,\mathbf x)$.

The photon coordinates are stationary when $\partial E/\partial q_m=0$, i.e.\ when $\partial\tilde E/\partial x_m=0$.
This condition contains only $\tilde E$, so its solution $\mathbf x^{\rm opt}(\mathbf R)$ does not depend on the frequencies, and $\mathbf q^{\rm opt}(\boldsymbol\omega)=\boldsymbol\Omega^{-1}\mathbf x^{\rm opt}$; at the mean-field level $x^{\rm opt}_m=\langle\hat d_m\rangle$.
At $\mathbf q^{\rm opt}$ the derivatives of $\tilde E$ are therefore taken at the same point $(\mathbf R,\mathbf x^{\rm opt})$ for every set of frequencies, and $\tilde{\mathbf H}$ is the same matrix.
For a reference set $\boldsymbol\omega^{\rm ref}$ and a target set $\boldsymbol\omega$ this gives
\begin{equation*}
\mathbf H(\boldsymbol\omega) = \mathbf S\,\mathbf H(\boldsymbol\omega^{\rm ref})\,\mathbf S ,\qquad
\mathbf S = \mathbf D(\boldsymbol\omega)\,\mathbf D(\boldsymbol\omega^{\rm ref})^{-1}
 = {\rm diag}\big(\mathbf 1_{3N_{\rm at}},\,\omega_1/\omega_1^{\rm ref},\dots,\omega_{N_{\rm mod}}/\omega_{N_{\rm mod}}^{\rm ref}\big) ,
\end{equation*}
which is Eq.~(10) of the main text: $\mathbf H_{\mathbf R\mathbf R}$ is unchanged, $\mathbf H_{\mathbf R q_m}$ scales with $\omega_m/\omega_m^{\rm ref}$ and $H_{q_mq_n}$ with $\omega_m\omega_n/(\omega_m^{\rm ref}\omega_n^{\rm ref})$.

The same argument gives three corollaries.
First, the energy minimized over the photon coordinates, $E^{\rm opt}(\mathbf R)=\tilde E(\mathbf R,\mathbf x^{\rm opt}(\mathbf R))$, does not depend on the frequencies, and neither does a geometry relaxed jointly in $\mathbf R$ and $\mathbf q$.
Second, the dipole derivatives obey $\partial\boldsymbol\mu/\partial\mathbf R=\partial\tilde{\boldsymbol\mu}/\partial\mathbf R$ and $\partial\boldsymbol\mu/\partial q_m=\omega_m\,\partial\tilde{\boldsymbol\mu}/\partial x_m$, so their photon components scale with $\omega_m/\omega_m^{\rm ref}$, and the polarizability derivatives behave the same way.
Since $-x_m\hat d_m$ acts on the electrons like a static field $x_m\boldsymbol\lambda_m$, $\partial\tilde{\boldsymbol\mu}/\partial x_m=\boldsymbol\alpha\boldsymbol\lambda_m$ with the polarizability $\boldsymbol\alpha$ of the coupled system, which gives Eq.~\eqref{eq:dmudq}.
Third, the dipole self-energy $\tfrac12\hat d_m^2$ contains no frequency, so the law holds with and without it and for any number of modes.

The derivation also shows where the rescaling fails.
(i) If the reference Hessian is not built at $\mathbf q^{\rm opt}$, $\mathbf S\mathbf H\mathbf S$ is the exact Hessian at $\mathbf x^{\rm ref}=\boldsymbol\Omega^{\rm ref}\mathbf q^{\rm ref}$, but not at the stationary photon coordinate of the target frequencies.
(ii) If the coupling changes with the frequency, as for a mode defined by a fixed field amplitude, where $\lambda\propto\omega^{-1/2}$, $\tilde H$ depends on $\boldsymbol\omega$ and the first step no longer holds.
(iii) For a quantized photon field, the photon energy $\omega_m(\hat b_m^\dagger\hat b_m+\tfrac12)$ and the bilinear coupling cannot be written in terms of $x_m$ alone, so the argument does not extend to electron-photon correlation; at the mean-field level the two descriptions differ only by the constant $\tfrac12\sum_m\omega_m$.
For water (STO-3G, $\lambda=0.05$), a reference with a residual photon gradient of $2\times10^{-4}$ shifts the rescaled frequencies by $0.9$~\si{\per\centi\meter} at twice the reference frequency, and a coupling that follows the frequency as in (ii) by $3.1$~\si{\per\centi\meter}; at $\mathbf q^{\rm opt}$ with fixed $\boldsymbol\lambda$ the rescaled and rebuilt frequencies agree to machine precision, with and without the dipole self-energy.

\subsection{Validation}

Eq.~(10) of the main text rescales a joint Hessian built at $q=q^{\rm opt}(\omega_{\rm ref})$ to any target frequency, with $H_{RR}$ unchanged, $H_{Rq}$ scaled by $\omega/\omega_{\rm ref}$ and $H_{qq}$ by $(\omega/\omega_{\rm ref})^2$.

Table~\ref{tab:omega_rescale} compares rescaled joint Hessians against fully rebuilt references, at all 21 target frequencies for the analytic \ce{H2CO} route and at three of them for the other two routes, for detuning scans on acrolein and formaldehyde.
Both couple to two degenerate modes polarized along $x$ and $z$, as in Sec.~VI, and their dipoles sit at opposite extremes relative to those axes.

\begin{table}[htb]
\centering
\caption{Rescaled versus freshly-rebuilt joint Hessians, maximum deviation over the rebuilt
points and over every frequency, aug-cc-pVDZ (acrolein) / cc-pVDZ (H$_2$CO),
$\lambda=0.05$, 21-point scans. The last column is the cost of rebuilding all 21 points relative to
the rescaled route, one reference build plus 21 rescales, taken as unity. For acrolein and the
numeric H$_2$CO route the rebuild cost is measured on 3 of the 21 points and scaled to the full
scan; for the analytic H$_2$CO route all 21 were rebuilt.}
\begin{tabular}{lcccc}
\hline
System (Hessian route) & window (\si{\per\centi\meter}) & max $|\Delta H|$ ($E_h$) & max $|\Delta\nu|$ (\si{\per\centi\meter}) & rebuild cost \\
\hline
Acrolein (analytic)   & 1000--2000 & $4\times10^{-13}$ & $5\times10^{-10}$ & $21.0$ \\
H$_2$CO (analytic)    & 1200--2600 & $6\times10^{-14}$ & $5\times10^{-11}$ & $21.8$ \\
H$_2$CO (numeric, DF) & 1200--2600 & $7\times10^{-8}$  & $0.0025$          & $20.8$ \\
\hline
\end{tabular}
\label{tab:omega_rescale}
\end{table}

\clearpage

\section{Infrared intensity of the photon coordinate}

Section~VI attributes the intensity of the nearly pure photon mode to its own dipole derivative rather than to brightness borrowed from the coupled band.
The mechanism follows from the coupling term $-\omega_c q\,\hat d$ with $\hat d=\boldsymbol\lambda\cdot\hat{\boldsymbol\mu}$.
Displacing $q$ imposes a static field $\omega_c\boldsymbol\lambda\,\delta q$ on the molecule, which responds through its polarizability, so that (Sec.~\ref{sec:rescale_proof})
\begin{equation}
\label{eq:dmudq}
\frac{\partial\boldsymbol\mu}{\partial q}\;=\;\omega_c\,\boldsymbol\alpha\,\boldsymbol\lambda ,
\qquad
I \;=\; \Big|\frac{\partial\boldsymbol\mu}{\partial Q}\Big|^2 ,
\end{equation}
with $\boldsymbol\alpha$ the polarizability of the coupled molecule.
For a pure photon mode, $Q=q$ and $I=\omega_c^2|\boldsymbol\alpha\boldsymbol\lambda|^2$, which grows as $\lambda^2$ as long as $\boldsymbol\alpha$ does not change with the coupling.
The intensity therefore scales with the polarizability rather than with any vibrational transition dipole, which is why a cavity mode that couples to no vibration is nonetheless bright.

Table~\ref{tab:photon_intensity} tests this directly.
A single water molecule is coupled to one $z$-polarized mode with a frequency of \SI{800}{\per\centi\meter}, far below the bend and the stretches, so the photon mode stays pure at every coupling (photon fraction $>0.996$) and its intensity is entirely its own.
The geometry is that of \ce{H2O} in Table~\ref{tab:geom_small}, held fixed, so nothing but $\lambda$ changes between rows.
$I/\lambda^2$ falls by $41\%$ between $\lambda=0.005$ and $0.15$, because the \gls{dse} screens the polarizability of the coupled molecule; mixing with the vibrations changes the intensity by less than $2\%$.

\begin{table}[htb]
\centering
\caption{Infrared intensity of an isolated cavity mode against coupling strength. Single \ce{H2O},
aug-cc-pVDZ, one $z$-polarized mode at $\omega_c=\SI{800}{\per\centi\meter}$, fixed geometry of
Table~\ref{tab:geom_small}. $I=|\partial\boldsymbol\mu/\partial Q|^2$ in the same arbitrary units used for the
spectra of Sec.~VI, scaled by $10^{3}$. The photon fraction confirms the mode
stays pure, so $I$ is entirely the photon coordinate's own contribution.}
\begin{tabular}{lcccc}
\hline
$\lambda$ & $\omega$ (\si{\per\centi\meter}) & photon fraction & $10^{3}I$ & $I/\lambda^2$ \\
\hline
0.005 & 799.9 & 1.0000 & 0.049 & 1.972 \\
0.01 & 799.6 & 1.0000 & 0.197 & 1.968 \\
0.02 & 798.6 & 0.9999 & 0.781 & 1.951 \\
0.03 & 796.8 & 0.9998 & 1.732 & 1.924 \\
0.05 & 791.2 & 0.9994 & 4.604 & 1.842 \\
0.075 & 780.9 & 0.9988 & 9.546 & 1.697 \\
0.1 & 767.6 & 0.9981 & 15.25 & 1.525 \\
0.15 & 734.8 & 0.9968 & 26.27 & 1.168 \\
\hline
\end{tabular}
\label{tab:photon_intensity}
\end{table}

\clearpage

\section{Reference geometries}

Cartesian coordinates in \AA, in the lab frame each system's cavity modes are defined in (\texttt{nocom}/\texttt{noreorient}, no Psi4 recentering).
Table~\ref{tab:geom_small} collects the small systems.
\ce{H2}, \ce{HF}, \ce{LiH} and \ce{H2O} are the correlation-map systems of Sec.~V, each placed with its dipole along the lab-frame $z$ axis.
The last entry is the water geometry of the symmetry analysis of Table~\ref{tab:sym_descent}, independent of the \ce{H2O} above, with its $C_2$ axis along $z$.

\begin{table}[htb]
\centering
\caption{Small systems: correlation map (Sec.~V) and symmetry analysis
(Table~\ref{tab:sym_descent}).}
\begin{tabular}{lrrr}
\hline
Atom & $x$ & $y$ & $z$ \\
\hline
\multicolumn{4}{l}{\textbf{H$_2$}, $n=2$} \\
H   & 0.000000 & 0.000000 & 0.000000 \\
H   & 0.000000 & 0.000000 & 0.740000 \\
\multicolumn{4}{l}{\textbf{HF}, $n=2$} \\
H   & 0.000000 & 0.000000 & 0.000000 \\
F   & 0.000000 & 0.000000 & 0.917000 \\
\multicolumn{4}{l}{\textbf{LiH}, $n=2$} \\
Li  & 0.000000 & 0.000000 & 0.000000 \\
H   & 0.000000 & 0.000000 & 1.595000 \\
\multicolumn{4}{l}{\textbf{H$_2$O}, $n=3$} \\
O   & 0.000000 & 0.000000 & 0.117300 \\
H   & 0.000000 & 0.757200 & -0.469200 \\
H   & 0.000000 & -0.757200 & -0.469200 \\
\multicolumn{4}{l}{\textbf{Water, symmetry analysis}, $n=3$} \\
O   & 0.000000 & 0.000000 & -0.068516 \\
H   & 0.000000 & -0.790690 & 0.543702 \\
H   & 0.000000 & 0.790690 & 0.543702 \\
\hline
\end{tabular}
\label{tab:geom_small}
\end{table}

Table~\ref{tab:geom_acrolein_h2co} gives acrolein and formaldehyde, used for the example of Sec.~VI and the rescale validation above.
The first acrolein structure is the aug-cc-pVDZ cavity-free optimized structure from which every acrolein result of Sec.~VI starts, and each fixed-$\lambda$ result re-optimizes further in the presence of both cavity modes; the rescale validation of Table~\ref{tab:omega_rescale} uses the second, optimized at the RHF/cc-pVDZ level.
Table~\ref{tab:geom_aniline} gives aniline, used for the conjugate-gradient benchmark of Table~\ref{tab:cg_polar}.

\begin{table}[htb]
\centering
\caption{Acrolein and formaldehyde: showcase of Sec.~VI (first acrolein structure) and rescale
validation (Table~\ref{tab:omega_rescale}; second acrolein structure and H$_2$CO).}
\begin{tabular}{lrrr}
\hline
Atom & $x$ & $y$ & $z$ \\
\hline
\multicolumn{4}{l}{\textbf{Acrolein}, $n=8$} \\
O   & -0.603468 & 1.674206 & 0.001807 \\
C   & -0.596898 & 0.483819 & -0.000314 \\
C   & 0.636214 & -0.334449 & 0.000662 \\
C   & 0.577017 & -1.659745 & -0.001835 \\
H   & -1.540263 & -0.084436 & -0.003201 \\
H   & 1.574382 & 0.203336 & 0.003487 \\
H   & -0.372534 & -2.179492 & -0.004651 \\
H   & 1.468153 & -2.270804 & -0.001166 \\
\multicolumn{4}{l}{\textbf{Acrolein, rescale validation (RHF/cc-pVDZ optimized)}, $n=8$} \\
O   & -0.576088 & 1.598368 & 0.000000 \\
C   & -0.576088 & 0.410981 & 0.000000 \\
C   & 0.652393 & -0.414200 & 0.000000 \\
C   & 0.591286 & -1.737039 & 0.000000 \\
H   & -1.521099 & -0.160223 & 0.000000 \\
H   & 1.593041 & 0.122883 & 0.000000 \\
H   & -0.360186 & -2.257674 & 0.000000 \\
H   & 1.482279 & -2.351297 & 0.000000 \\
\multicolumn{4}{l}{\textbf{H$_2$CO}, $n=4$} \\
C   & 0.000000 & 0.000000 & -0.529700 \\
O   & 0.000000 & 0.000000 & 0.673300 \\
H   & 0.000000 & 0.936700 & -1.109700 \\
H   & 0.000000 & -0.936700 & -1.109700 \\
\hline
\end{tabular}
\label{tab:geom_acrolein_h2co}
\end{table}

\begin{table}[htb]
\centering
\caption{Aniline, conjugate-gradient benchmark (Table~\ref{tab:cg_polar}).}
\begin{tabular}{lrrr}
\hline
Atom & $x$ & $y$ & $z$ \\
\hline
N   & 0.000000 & 0.000000 & 2.190000 \\
H   & 0.812000 & 0.000000 & 2.784000 \\
H   & -0.812000 & 0.000000 & 2.784000 \\
C   & 0.000000 & 0.000000 & 0.809000 \\
C   & 1.213000 & 0.000000 & 0.124000 \\
C   & -1.213000 & 0.000000 & 0.124000 \\
C   & 1.210000 & 0.000000 & -1.268000 \\
C   & -1.210000 & 0.000000 & -1.268000 \\
C   & 0.000000 & 0.000000 & -1.958000 \\
H   & 2.156000 & 0.000000 & 0.668000 \\
H   & -2.156000 & 0.000000 & 0.668000 \\
H   & 2.150000 & 0.000000 & -1.807000 \\
H   & -2.150000 & 0.000000 & -1.807000 \\
H   & 0.000000 & 0.000000 & -3.041000 \\
\hline
\end{tabular}
\label{tab:geom_aniline}
\end{table}